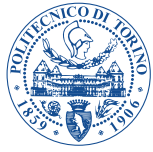 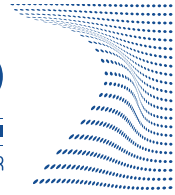

Doctoral Dissertation
Doctoral Program in Electrical, Electronics and Communications Engineering
(31st cycle)

# Big Data for Traffic Monitoring and Management

## Martino Trevisan
\* \* \* \* \* \*

**Supervisors**
Prof. Marco Mellia, Supervisor
Dr. Idilio Drago, Co-supervisor

**Doctoral Examination Committee:**
Prof. Anna Brunstrom, Referee, Karlstad University
Prof. Steve Uhlig, Referee, Queen Mary, University of London
Prof. Ana Paula Couto Da Silva, Universidade Federal de Minas Gerais
Prof. Sanjay Rao, Purdue University
Dr. Marco Fiore, Consiglio Nazionale delle Ricerche

Politecnico di Torino
February 27, 2019

I hereby declare that, the contents and organisation of this dissertation constitute my own original work and does not compromise in any way the rights of third parties, including those relating to the security of personal data.

......................................
Martino Trevisan
Turin, February 27, 2019

# Acknowledgements

The three main chapters of this thesis are named *What*, *Awesome* and *Pain*, respectively. There is no particular reason behind this, as these acronyms where invented at the time the respective papers were submitted. However, if you read them together, you come up with the sentence *What awesome pain.* After three years as a PhD student, I can say that this is a good definition for PhD. Hard work, rewarded by great satisfaction. As such, I really must thank all people around me over the last years. My parents deserve a great thank, as well as my girlfriend Francesca. I must thank my friends Matteo, Andrea, Leonardo and Giorgio for these last years of youth, and my flatmates that are still young. Finally, thank you to my collegues for the great times in the *experts'* lab, to my supervisor Marco and to Idilio and Maurizio.

# Abstract

The last two decades witnessed tremendous advances in the Information and Communications Technologies. Beside improvements in computational power and storage capacity, communication networks carry nowadays an amount of data which was not envisaged only few years ago. Together with their pervasiveness, network complexity increased at the same pace, leaving operators and researchers with few instruments to understand what happens in the networks, and, on the global scale, on the Internet.

Fortunately, recent advances in data science and machine learning come to the rescue of network analysts, and allow analyses with a level of complexity and spatial/temporal scope not possible only 10 years ago. In my thesis, I take the perspective of an Internet Service Provider (ISP), and illustrate challenges and possibilities of analyzing the traffic coming from modern operational networks. I make use of big data and machine learning algorithms, and apply them to datasets coming from passive measurements of ISP and University Campus networks. The marriage between data science and network measurements is complicated by the complexity of machine learning algorithms, and by the intrinsic multi-dimensionality and variability of this kind of data. As such, my work proposes and evaluates novel techniques, inspired from popular machine learning approaches, but carefully tailored to operate with network traffic.

In this thesis, I first provide a thorough characterization of the Internet traffic from 2013 to 2018. I show the most important trends in the composition of traffic and users' habits across the last 5 years, and describe how the network infrastructure of Internet big players changed in order to support faster and larger traffic. Then, I show the challenges in classifying network traffic, with particular attention to encryption and to the convergence of Internet around few big players. To overcome the limitations of classical approaches, I propose novel algorithms for traffic classification and management leveraging machine learning techniques, and, in particular, big data approaches. Exploiting temporal correlation among network events, and benefiting from large datasets of operational traffic, my algorithms learn common traffic patterns of web services, and use them for (i) traffic classification and (ii) fine-grained traffic management. My proposals are always validated in experimental environments, and, then, deployed in real operational networks, from which I report the most interesting findings I obtain. I also focus on the Quality of Experience (QoE) of web users, as their satisfaction represents the final objective of computer networks. Again, I show that using big data approaches, the network can achieve visibility on the quality of web browsing of users. In general, the algorithms I propose help ISPs have a detailed view of traffic that flows in their network, allowing fine-grained traffic classification and management, and real-time monitoring of users QoE.

# Contents

# List of Tables

VI

# List of Figures

# Chapter 1

# Introduction

In the last two decades Internet has become a fundamental infrastructure for the industry and the preferred means for entertainment. Born to interconnect universities and research laboratories, Internet nowadays permeates the globe, and allows billions of people to communicate and access multimedia content. The amount of traffic carried by Internet has become huge, and it is indeed expected to keep increasing in the next decade. According to the Cisco Visual Networking Index [55] the annual global IP traffic will reach 3.3 ZB per year by 2021 growing from the 1.2 ZB per year registered in 2016.

In this scenario, network measurements are a fundamental instrument to understand how the Internet evolves and to identify potential issues, like anomalous behavior of users, impairment of network devices or malfunctions of servers and content providers. Network devices, user equipments and monitoring systems produce a deluge of data that contain unique knowledge about both mere technological aspects and anthropological issues. However, data alone is not sufficient to achieve knowledge, as useful information is typically buried among endless sequences of uninteresting records. Moreover, the recent trend towards encryption makes the life of network analysts harder, as a significant fraction of information is now carried by the network in an encrypted form, and, thus, not available to network operators. In this picture, the so called *data science* becomes necessary to analyze an always bigger amount of data in which knowledge is more and more hidden, and often can be obtained only when large datasets are processed in an aggregated fashion.

In this thesis, I describe my work in extracting knowledge from network data. In particular, I focus on *passive* measurements, a technique in which a monitoring infrastructure collects data regarding the activity of a population of users connected to the Internet. To this end, particular devices called *network probes* are deployed in an operational network, and run a suitable software to collect the desired statistics as the network packets flow. This approach is opposite to the so-called *active* measurements, in which network traffic is generated in a controlled environment. The latter allows larger freedom, as the observed network traffic is (almost) under the control of the experimenter. Nevertheless, active measurements pose some limitations when studying the behavior of network devices and users on the large Internet, where a bigger effort is required to mimic a real scenario. As such, passive measurement are considered a useful means to understand what happens on *real* networks, where real users utilize in-operation network devices. However, measurements are only the first step of the complicated processes behind data science applications. My thesis focuses on such processes, and proposes methodologies to obtain valuable knowledge from raw collected measurements. Most of the chapters are taken from papers already published in international conferences and journals, and, at the beginning of each, I report the venue in which the content has been presented.

In Chapter 2, I first describe the employed datasets and the methodology I followed to gather them. I make use of 3 datasets coming from passive monitoring ISP subscribers and campus users for a period of 5 years. As they are used throughout this thesis, I summarize datasets here, and specify on each chapter which (and which part) I employ.

The first work presented in this thesis is contained in Chapter 3, and focuses on Internet traffic characterization. Indeed, knowing traffic is crucial for operating the network, understanding users' need, and ultimately improving applications. In the chapter, I provide an in-depth longitudinal view of Internet traffic. I take the point of the view of a national-wide ISP and analyze 5 years of flow-level measurements to pinpoint and quantify trends. I show that an ordinary broadband subscriber nowadays downloads more than twice as much as they used to do 5 years ago. Bandwidth hungry video services drive this change, while social messaging applications boom (and vanish) at incredible pace. I study how protocols and service infrastructures evolve over time, highlighting unpredictable events that may hamper traffic management policies.

Next, in Chapter 4 I illustrate the challenges of passive monitoring in the current Internet, with particular attention to traffic classification. The widespread deployment of encryption and the convergence of the web services towards HTTP/HTTPS challenge traditional classification techniques. Algorithms to classify traffic are left with little information, such as server IP addresses, flow characteristics and queries performed at the DNS. Moreover, due to the usage of Content Delivery Networks and cloud infrastructure, it is unclear whether such coarse metadata is sufficient to differentiate the traffic. In this chapter, I study to what extent basic information visible at flow-level measurements is useful for traffic classification on the web. By analyzing a large dataset of flow measurements, I quantify how often the same server IP address is used by different services, and how services use hostnames. A very simple classifier that relies only on server IP addresses and on lists of hostnames can distinguish up to 55% of the traffic volume. This testifies the challenges behind passive measurements, and calls for more sophisticated techniques able to extract meaningful information. I the next chapters, I will explore such challenges and show how passive measurements become meaningful only when aggregating large datasets and processing them with suitable Data Science techniques.

Motivated by the aforementioned reasons, in Chapter 5 I propose new algorithms for traffic analysis and classification, with particular attention to web services. Indeed, nowadays HTTP(S) is the main means to access the Internet, but traditional solutions for traffic classification and metering fall short in providing visibility in users' activities. In the chapter, I present the Web Helper Accounting Tool (WHAT) that overcomes these challenges by (*i*) identifying the main domain name representative of the *service* being accessed, and (*ii*) grouping together traffic due to the access to such main service. WHAT is a completely unsupervised system, that relies only on passive measurements, assuming that the domain names associated to network flows is still visible.[1] The grouping of all flows enables accurate accountability per service. I provide an extensive evaluation and case studies to demonstrate WHAT effectiveness, thus enabling an accurate accounting of the traffic associated to each user action.

Then, my thesis focuses on traffic management. In Chapter 6, I show that the algorithms I propose for traffic classification can be successfully used for traffic management too, where classification is only one of the building blocks that allow innovative traffic routing. The most promising technique for traffic management is called Software Defined Network (SDN), which enables programmable management in computer networks, and aims at providing a homogeneous paradigm to provide all network devices. However, the complexity of modern Internet traffic challenges the standard SDN approach, based on simple per-flow management. I propose a new approach based on a "per service" management concept, which allows to identify and prioritize all traffic of important web services, while segregating others, even if they are running on the same cloud platform, or served by the same CDN. In this chapter, I design and evaluate AWESoME, Automatic WEb Service Manager, a novel SDN application to address the above problem. On the one hand, it leverages big data algorithms to automatically build models describing the traffic of thousands of web services. On the other hand, it uses the models to install rules in SDN devices to steer all flows related to the originating services. It correctly disambiguates those cases in which the same CDN is used by multiple services by taking into account the *sequence* of servers contacted by the client.

Finally, Chapter 7 addresses the issue of Quality of Experience (QoE) of users that access web services. Indeed, understanding QoE of web browsing is key to optimize services and keep users' loyalty. This is crucial for both Content Providers and Internet Service Providers (ISPs). However, quality is intrinsically subjective, and the complexity of today's pages challenges its measurement. In this chapter, I propose PAIN (PAssive INdicator), an automatic system to monitor the performance of web pages from passive measurements. With unsupervised learning, PAIN automatically creates a machine learning model from the timeline of events generated by browsers to render web pages, and uses it to measure web performance in real-time. I compare PAIN to objective metrics based on in-browser instrumentation and find strong correlations between the approaches. I let PAIN run on an operational ISP network, and find that it is able to pinpoint performance variations across time and groups of users.

---

[1] It can still be retrieved from DNS traffic and TLS handshake.

# Chapter 2

# Measurements and data collection

In this chapter, I describe the datasets I use throughout the thesis and the measurement methodology at the basis of their collection. In this thesis, I build my analysis on data collected by the passive monitoring infrastructure of two operational networks in Italy. The measurement infrastructure captures and analyses in real-time traffic from vantage points located at the edge of the network. A schematic view of the infrastructure is depicted in Figure 2.1. It processes traffic directly in the Point-of-Presences (POPs). Exploiting router span ports or optical splitters (depending on the link rates), it mirrors the traffic to the monitoring probes. Both uplink and downlink streams generated by the users are exposed to the probes. Since probes are deployed in the first level of aggregation, no traffic sampling is performed. Users are assigned fixed IP addresses, that the probes immediately anonymize in a consistent way to keep users' privacy.

Each probe is equipped with multiple high-end network interfaces. Packets are captured using the Intel Data Plane Development Kit (DPDK) [56] that allows line-rate capture even for multiple 10 Gbit/s links. Traffic is then processed by a custom-made passive traffic analyzer, called Tstat [107].

Each probe exports only flow records, i.e., a single entry for each TCP/UDP stream with per-flow statistics.[1] Each record contains classical fields on flow monitoring [51], such as IP addresses, port numbers, timestamps, packet-wise and byte-wise counters. Advanced analyzers extract some fields from packet payloads, such as information seen in the Application-Layer Protocol Negotiation (ALPN) fields of TLS handshakes, which allows to identify HTTP/2 and SPDY flows, and fields from QUIC public headers. Tstat also exports the *domain name* of the contacted servers, exchanged in clear in HTTP `Host:` headers, or requested in the TLS Server Name Indication (SNI) within TLS Client Hello messages. For flows missing such information, Tstat exports the host name the client resolved via DNS queries prior to open the flow.[2] This mechanism, called DN-Hunter, is explained in details in [7]. DN-Hunter correlates DNS traffic to TCP flows as follows. Consider a client having IP address `1.1.1.1` accessing `www.acme.com`. The client first contacts the DNS server to resolve the server hostname into a list of IP addresses, i.e., `2.2.2.2` and `3.3.3.3`. Then, the client might contact one of the retrieved IP addresses. Tstat keeps a circular buffer in memory with all DNS responses and, when seeing a flow from the client `1.1.1.1` going to the server `2.2.2.2`, it assigns the domain `www.acme.com` to the flow.

---

[1]Streams are expired either by the observation of particular packets (e.g., TCP packets with RST flag set) or by timeouts. See http://tstat.polito.it/measure.shtml.

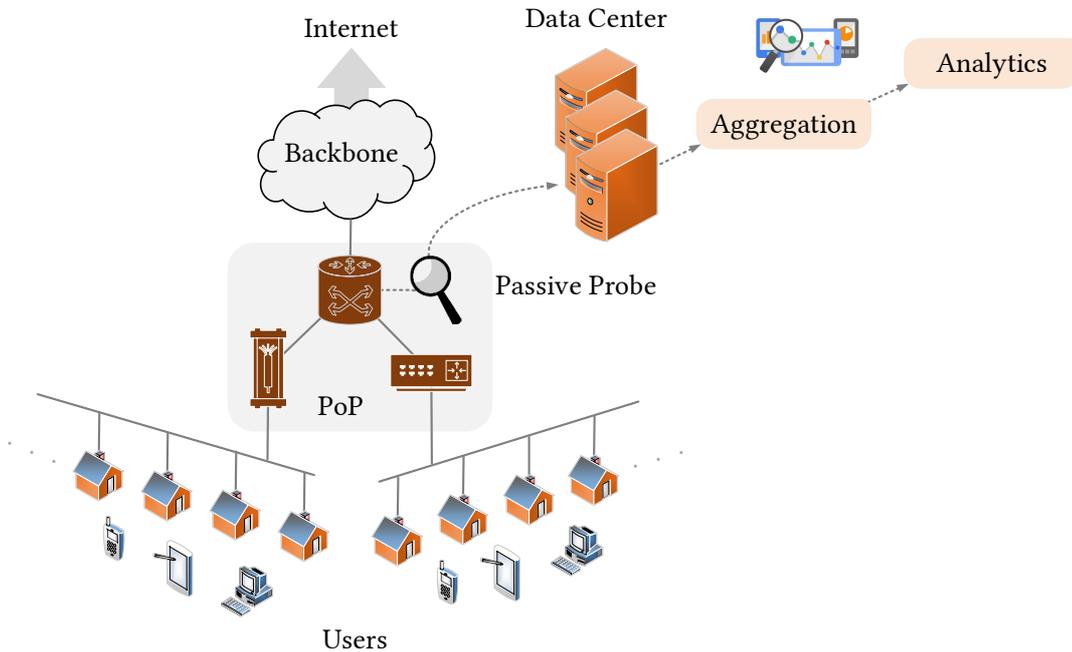

Figure 2.1: Measurement infrastructure and processing steps for a single probe. I deployed three probes collecting data for (i) ADSL subscribers, (ii) FTTH subscribers, (iii) University Campus users.

In the following chapters, I also perform more sophisticated analysis, relying on the estimation of RTT provided by Tstat for TCP flows [75]: It searches for acknowledged TCP segments, registering the time from the observation of the TCP segment and its acknowledgment. For each flow, Tstat exports the minimum, average and maximum RTT estimation, as well as the number of RTT samples. Notice that this metric represents only the RTT from the probe to servers, missing the delay from clients to the probes. In this deployment I miss thus the access delay, since probes are deployed at the first aggregation level (BRAS or edge router).

---

[2]The vantage points observe all DNS traffic directed to local resolvers.

Table 2.1: Overview of the datasets.

| Name | Flows | Users | Access Technology | Duration |
|------|-------|-------|-------------------|----------|
| ADSL | 141 G | 10 k | ADSL | 5 Years |
| FTTH | 44 G | 5 k | Fiber-To-The-Home | 5 Years |
| Campus | 163 G | 20 k | Wired and WiFi | 5 Years |

In this monitoring infrastructure, Tstat is installed in two distinct networks: (i) a University campus in Italy where $\approx$ 15 000 users are connected; and (ii) two PoPs of a nation-wide Internet Service Provider. The campus dataset includes traffic generated by students and professors using wired and WiFi networks during 5 years, from 2013 in 2018. I refer to this dataset with `Campus`. Regarding the ISP dataset, I consider the traffic of two PoPs, covering more than 10 000 ADSL and 5 000 Fiber-To-The-Home (FTTH) subscribers, all located in the same city in Italy, and active since 2013. I refer to these datasets with ADSL and FTTH. ADSL downlink capacity varies from 4 Mbit/s up to 20 Mbit/s, with uplink limited to 1 Mb/s. FTTH users enjoy 100 Mb/s downlink, and 10 Mbit/s uplink. Each subscription refers to an installation, where users' devices (PCs, smartphones, tablets, smart TVs etc) connect via WiFi and Ethernet through a home gateway. ADSL customers are almost totally residential customers (i.e., households), whereas a small but significant number of business customers exist among the FTTH customers. During the years of measurements, I observed a steady reduction on the number of active ADSL users and an increase in FTTH installations. The ISP has confirmed these trends are due to churning and technology upgrades. Datasets are summarized in Table 2.1.

Flow records are created, anonymized and stored on the local probe disks. Daily, logs are copied into a long-term storage in a centralized data center and discarded from the probes. Our data center has enough capacity to preserve historical data.

By the time of writing this thesis, the considered datasets covers 5 years of measurements, totaling 63.9 TB of compressed and anonymized flow logs (around 348 billion flow records). To process this deluge of data, I use a Hadoop-based cluster running Apache Spark. This structure allows both to update predefined analytics continuously, as well as to run specific queries on historical collections.

# Chapter 3

# Five Years at the Edge: Watching Internet from the ISP Network

The work I present in this chapter is mostly taken from my paper "*Five Years at the Edge: Watching Internet from the ISP Network*" presented in the 14[th] International Conference on emerging Networking EXperiments and Technologies (CoNEXT 2018) [105].

## 3.1   Introduction

Measurements have always been among the best ways to understand complex systems. Not surprisingly, measurements are the key means to gather information about the overall status of the Internet, identify eventual issues, and ultimately improve its design [116, 76, 96]. The Internet being an evolving system, novel measurement systems are continuously devised to extract information about applications, protocols, deployments, etc. However, having a long-term picture on how the Internet is evolving is a rather challenging task. Researchers often design new tools and approaches that focus on specific phenomena, which are observed and described in details for limited time intervals. It is rare to find works that offer a longitudinal view on systems over time.

In this chapter, I offer a view of the Internet in the past 5 years as seen from an operational network. I rely on a humongous amount of data collected from a nationwide Internet Service Provider (ISP) infrastructure. I focus on broadband Internet access via ADSL and FTTH technologies. I instrument some of the ISP aggregation links with passive monitoring probes. By observing packets flowing on the links, the probes extract detailed per flow information, that I collect and store on a centralized data lake. Keeping the pace with Internet evolution during 5 years is per se a challenging task. I rely on custom designed software probes that have been constantly updated during the monitoring period to account for and report information about new protocols and services.

Technically, I follow a well-established approach. Passive measurements are popular among researchers since early 2000 [5, 24], with current tools able to process several tens of Gb/s on commodity hardware [77, 107]. Extracting information from packets is possible thanks to Deep Packet Inspection (DPI) techniques [1], while the availability of big data solutions [28, 119] makes it possible to store and process large volumes of traffic with unprecedented parallelism.

Here, I dig into this data, depicting trends, highlighting sudden changes and observing sudden infrastructure upgrades. Instead of focusing on a specific angle, I aim at offering examples of general trends on the Internet evolution. The Internet indeed rapidly evolves: Services get popular and other get abandoned; Users change habits; New protocols change the way information is carried. Observing such trends is vital to understand the Internet, the users, and the systems.

First I give an overview of users' habits over 5 years, assessing the costs of broadband customers to the ISP in terms of traffic consumption. I observe for example that the traffic per broadband customer has increased at a constant rate over the years, with a growth of heavy users, i.e., those who exchange tens of GB per day. When comparing service usage between ADSL and FTTH customers, I see that the larger capacity offered to FTTH customers has a moderate impact on per customer data consumption.

Next, I turn my attention to the traffic loads imposed by web services to the ISP. I quantify the rise (and death) of services in terms of traffic volumes as well as popularity among customers. Here I confirm and precisely quantify some well-known trends, typically stated by content provided, but rarely measured from the network point of view: video content – no longer accessed via peer-to-peer systems – drives the bandwidth demand. Yet, users of modern social messaging systems such as Instagram (accessed from mobile phones) consume more and more traffic. Indeed, the traffic of each Instagram user is already comparable to the traffic per user of popular video-on-demand platforms, such as Netflix or YouTube.

Finally I study how changes in the infrastructure and protocols have impacted the ISP network. For example, I detail the (slow) migration of services to HTTPS and several (sudden) deployments of custom protocols by large companies that may hamper traffic engineering and troubleshooting of ISP networks. I testify the growth in the infrastructure of popular services, and show how services are more and more deployed close to users, with caches deployed at the first aggregation point at the ISP, in an effort to cut off the latency to reach the Internet contents.

Despite the dataset is limited to one country and focused on broadband Internet (thus missing mobile networks), I believe the information I offer is key to understand trends and inform researchers and practitioners about recent changes on Internet infrastructure and users' behavior. Our dataset includes traffic from more than 15,000 users, and it is collected in a central area of Europe. Even if we generally miss the traffic of mobile network, we catch smart-phone traffic of users that connect through WiFi when at home.

The chapter is organized as follows: Section 3.2 presents the monitoring infrastructure and the analyzed dataset. Section 3.3 investigates traffic demand of ISP customers, while Section 3.4 illustrates trends of services in terms of traffic volume and popularity. Section 3.5 analyses protocol usage and episodes of unpredictable traffic variations, whereas Section 3.6 shows notable trends in Big Players' infrastructure. Section 3.7 summarizes the related work. Finally, Section 3.8 concludes the chapter.

## 3.2   Measurement methodology

In this section, I now describe the measurement methodology and tools used to collect the data.

### 3.2.1   Measurement architecture

In this chapter, I build on data collected from a nation-wide ISP. Data collection methodology is described in Section. 2, and I make use of the ADSL and FTTH datasets. In short, I have instrumented a Point-of-Presence of the ISP to collect passive measurements by means of passive meter running Tstat [107], a tool that exports rich flow level logs containing hundreds of statistics. The dataset includes 5 years of anonymized traces, coming from monitoring of more than 10 000 ADSL and 5 000 Fiber-To-The-Home (FTTH) subscribers.

My analytics methodology follows a two-stage approach: firstly data is aggregated on a per day basis, secondly, advanced analytics and visualizations are computed. In the aggregation stage, queries compute per-day and per-subscription aggregates about traffic consumption, protocol usage, and contacted services. This round requires processing of millions of raw flow records.

Special attention is needed for identifying the services used by subscribers. Content providers are known to rely on large infrastructure and/or Content Delivery Networks (CDNs), which make the association between flow records and services tricky. For this step, I rely mostly on the server domain names. Examples of the association domain-service are provided in Table 3.1. Flexible matching based on regular expressions is allowed.[1] Along the years, my team and I have *continuously* monitored the most common server domain names seen in the network, maintaining the list of domains associated with the services of interest. For ambiguous cases [106], e.g., domains used by multiple services, I rely on heuristics, mostly based on traffic volumes, to decide whether a subscriber actually contacted a particular service (see Section 3.4.1). This methodology thus allows on-the-fly and historical classification of services. Once such aggregated dataset is available, flexible analytics perform the analysis and visualization of the data.

---

[1]The full list of rules to classify services can be found in https://smartdata.polito.it/five-years-at-the-edge-watching-internet-from-the-isp-network/.

Table 3.1: Examples of domain-to-service associations.

| Domain | Service |
|---|---|
| `facebook.com` | Facebook |
| `fbcdn.com` | Facebook |
| `^fbstatic-[a-z].akamaihd.net$` (RegExp) | Facebook |
| `netflix.com` | Netflix |
| `nflxvideo.net` | Netflix |

### 3.2.2 Challenges in long-term measurements

Several challenges arise when handling a large-scale measurement infrastructure. Network probes are the most likely point of failure, as they are subjected to a continuous and high workload. During the period considered in this chapter, probes suffered few outages, lasting from few hours up to some months (when severe hardware issues arose). As such, the results I present have missing data for those periods. The data is not available for clear privacy reasons, but I'm developing, as a follow-up of this work, a online graphical toolkit to visualize and play with aggregated statistics.

A second issue arises from the evolution of network protocols and service infrastructure. Large content providers have the power of suddenly deploying new protocols leaving passive monitors and ISPs with few or no documentation to handle them. I incurred several cases, and report the experience in addressing them.

Third, the domain-to-service associations definition needs to be continuously updated. Also in this case, there is no public information to support this operation, so that my team and I have to manually define and update rules, often by running active experiments to observe patterns.

At last, users' privacy must be preserved. For this, I carefully limit the collected information and always consider only aggregated statistics. Customers' IP addresses and server names are the most privacy-sensitive information being collected. The former gets immediately anonymized by probes, while the latter is used to derive aggregate statistics on per-service basis. Importantly, all data collection is approved and supervised by the responsible teams in the ISP.

## 3.3 The cost of a user

I first characterize the traffic consumed by subscribers in the last 5 years. This analysis is instrumental to understand costs of ISPs in terms of capacity and forecasting trends.

For the results that follow, I consider only *active subscribers*. A subscriber is considered *active* if she/he has generated at least 10 flows, downloaded more than 15 kB and

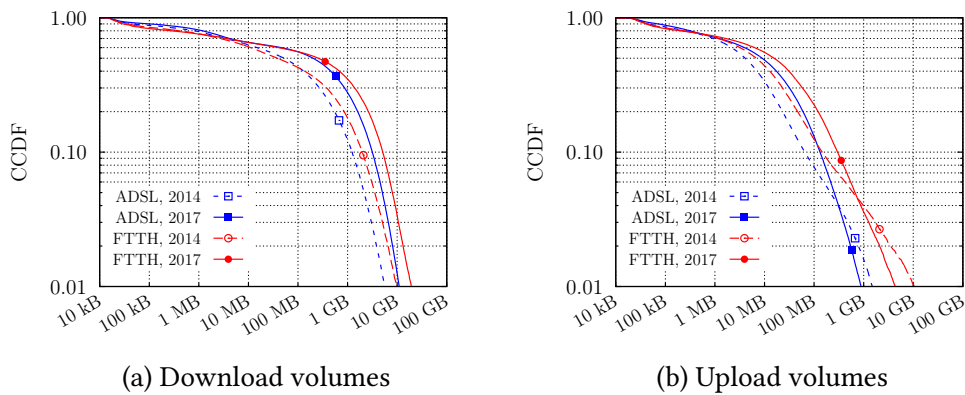

<div align="center">

(a) Download volumes        (b) Upload volumes

</div>

Figure 3.1: CCDF of per active subscriber daily traffic for April 2014 and 2017.

uploaded more than 5 kB.[2] This simple criterion lets me filter those cases where only background traffic is present, e.g., generated by the access gateway, or by incoming traffic (due to, e.g., port scans). On average I observe about 80% subscribers active each day, with respect to the total number of subscribers observed in the whole trace. Less than 0.01% of data, in terms of flow records, is discarded at this step.

Notice that these percentages are actually a lower-bound given churning (see Section 3.2.1). Notice also that smartphones contribute to make subscribers active in more days.

### 3.3.1 How much you eat: Consumption per day

Figure 3.1 depicts the empirical Complementary Cumulative Distribution Function (CCDF) of daily traffic consumption of active subscribers in the ISP. In other words, for each day, I compute the overall traffic each active subscriber exchanges. I report the CCDF of all measurements as seen in April 2014 and 2017. Figure 3.1 depicts CCDFs separately per access-link technology and down/up links. Log scales are used.

Observe the bimodal shape of the distribution. In about 50% of days, subscribers download (upload) less than 100 MB (10 MB) – i.e., days of light usage. However, a heavy tail is present. For more than 10% of the days, subscribers download (upload) more than 1 GB (100 MB) – i.e., days of heavy usage. Manual inspection shows that many different subscribers present days of heavy usage, often alternating between days of light and heavy usage.

Comparing 2014 (dashed lines) with 2017 (solid lines), I notice an increase in daily traffic consumption. The median values have increased by a factor 2 for both ADSL and FTTH installations, and for both upload and download. This behavior highlights an increasing trend in average per-subscriber traffic volume, that I examine more in depth later in this section.

---

[2]These thresholds have been determined by visually inspecting knee points in the distributions of daily traffic per user.

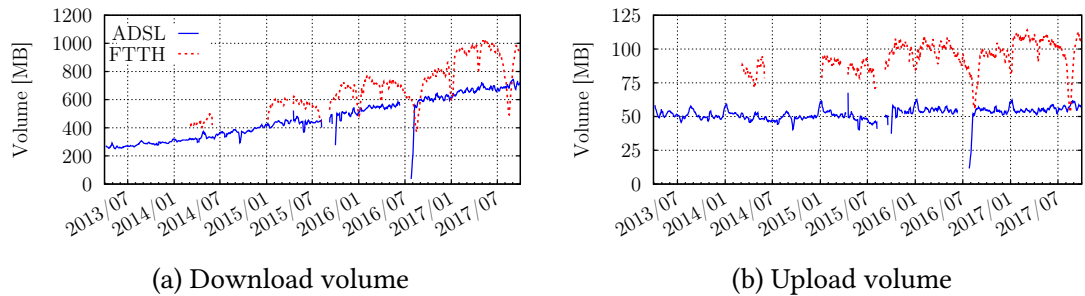

(a) Download volume                    (b) Upload volume

Figure 3.2: Average per-subscription daily traffic.

I observe no differences for the days of light usage when contrasting ADSL (blue curves) and FTTH installations (red curves). Instead, during heavy usage days, FTTH users download about 25% more data than ADSL users – a moderate increase given they enjoy 5-20 times higher capacity. The differences are higher considering upload traffic: ADSL users are indeed bottlenecked by the 1 Mb/s uplink, thus FTTH subscribers upload twice as much per day.

At last, I witness an interesting effect in uploaded traffic: Even if traffic volume increased in median between 2014 and 2017, the tail of the distributions in Figure 3.1b decreased. Notice the clearly visible bump in the tails present in 2014, which disappeared in 2017. This trend is rooted in the decline of Peer-To-Peer (P2P) traffic, both in volume and popularity, as I will show in Section 3.4.

### 3.3.2   Eager and Eager: Trends on traffic consumption

Figure 3.2 illustrates subscribers' traffic consumption over time. The *x*-axis spans over the 54 months of the dataset, *y*-axis shows the average byte consumption over monitored subscriptions, separately per access technology and down/up link. Curves in the figure contain interruptions caused by outages in monitoring probes, without affecting trends.[3]

Considering the average amount of data downloaded daily, illustrated in Figure 3.2a, a clear increasing trend emerges. For ADSL subscribers, average daily traffic increased at a constant rate – from 300 MB in 2013 up to 700 MB in late 2017. FTTH subscribers consume on average 25% more traffic as previously noticed, topping to 1 GB per day on average in 2017. Interesting, very similar slow increasing trends have been reported 10 years ago [20].

---

[3]FTTH figures are noisier than the ADSL ones due to the smaller numbers of FTTH customers. Some drops in FTTH curves are visible during summer and holiday breaks, thanks to the low number of customers and their profiles (e.g., business customers).

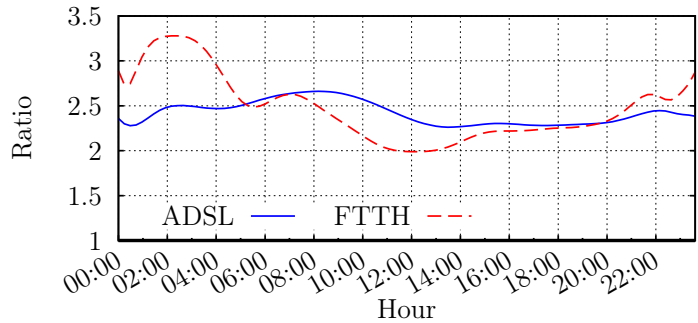

Figure 3.3: Ratio of traffic consumption between April 2017 and April 2014 for download.

When considering uploads (Figure 3.2b), I confirm that the higher uplink capacity lets FTTH users to upload more with respect to ADSL. The latter has been bottlenecked during the captures and thus the average amount of data remains constant. FTTH subscribers show a modest increase in average uploaded traffic over time. This modest increase is due to two factors. At the one hand, P2P uploads have decreased significantly in recent years. On the other hand, this decrease has been compensated by a significant increase in the upload of user-generated content to the cloud, including to cloud storage services (e.g., iCloud or Dropbox) as well as to social networks and video providers (e.g., YouTube and Instagram).

To check whether the increase observed in Figure 3.2 is homogeneous during the hours of the day, I consider the downloaded volume in each 10 minute-long time interval. I then average all values seen for the same time bin in all days of a month. At last I compute the ratio between April 2017 and April 2014. Figure 3.3 shows results (curves are smoothed using a Bezier interpolation). It confirms that the average amount of traffic consumed in 2017 is more than 2 times larger than 2014. Interestingly, the increase is higher during late night hours. Manual inspections reveals that this is due to software updates of operating systems (mainly Windows and Mac OS/iOS updates). FTTH users exhibit also a higher increase during prime time, which I confirm to be associated to the consumption of video streaming content.

## 3.4 The cost of services

### 3.4.1 Give me that: Service popularity

The changes in the per-subscriber traffic volume can be due to changes in the users' habits (e.g., people using different services), or changes in the services (e.g., high definition videos being automatically served). In this section, I analyse in details how popular and bandwidth demanding services evolved throughout years. I again focus on *active subscribers*, observing the fraction of them that accessed a given *service* on a daily basis.

Notice that selecting subscribers that contacted a service is not trivial. Indeed popular services may be *unintentionally* contacted by users. Consider for example Facebook. Its social buttons are embedded in websites and generate traffic to the same Facebook domains as an access to `facebook.com` services. To coarsely distinguish these cases, I have inspected the distribution of daily traffic per subscriber for each considered service. Not reported here for brevity, I manually set per-service thresholds to separate (i) subscribers with at least one visit to main services (moderate to large traffic volumes), and (ii) subscribers which unintentional contacted domains due third party objects (negligible volumes).

I start by providing a coarse picture about service popularity over time.[4] Figure 3.4a shows per-day percentage of active users that access popular services. I depict the ADSL data only, since FTTH results in similar figures. The multi-color palette highlights changes in the popularity of services, which are coarsely sorted by type. For instance, Google search engine is accessed regularly by about 60% of active users on a daily basis, and this pattern is rather constant over time.[5] On the contrary, Bing shows a constant growth, moving from less than 15% to about 45% of active users that contacted it at least one time per day in 2017. This pattern is likely a consequence of Windows telemetry which uses `bing.com` domains. Interestingly, DuckDuckGo, a privacy respecting search engine, is used only by few tens of users (less than 0.3% of population), unveiling a scarce interest for privacy in the monitored subscribers.

Figure 3.4b depicts a similar picture for the percentage of downloaded bytes for each service in the ISP traffic mix. The multi-color palette is limited to 10% to improve the visualization, since only YouTube is over this percentage in the studied ISP. One can observe how services have changed their contributions to the traffic mix during the monitored period. Notice, for instance, how services such as Facebook, Instagram, WhatsApp and Netflix have increased traffic share throughout the years. Others, such as SnapChat have gained momentum only during a limited period.

Overall, I observe a continuously changing picture, with services showing an increase in popularity and traffic share, some of which with remarkable growth, while others that struggle to gain grounds. Next, I dive into some interesting use cases.

---

[4]Data tables used to generate these figures, including popularity of services and bytes per user per day, can be downloaded from https://smartdata.polito.it/five-years-at-the-edge-watching-internet-from-the-isp-network/.

[5]Some fluctuations are due to changes in Google domains that have taken time to be identified and updated in probes.

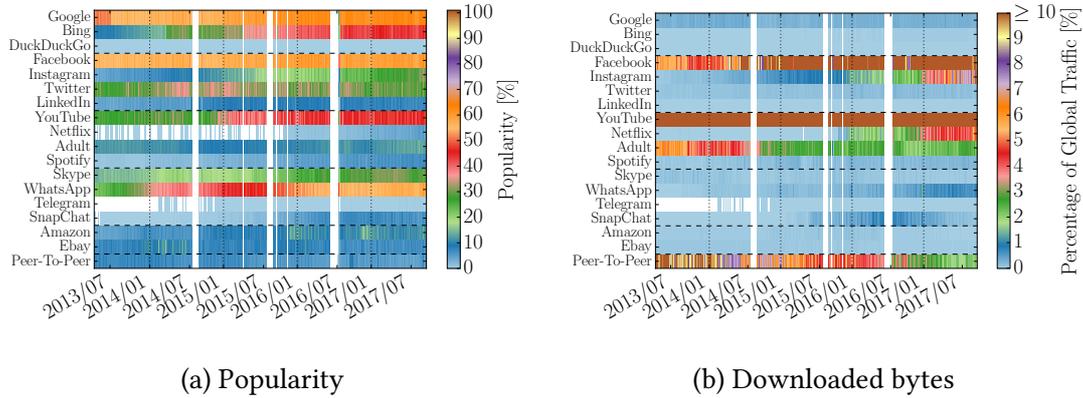

(a) Popularity        (b) Downloaded bytes

Figure 3.4: Popularity and percentage of downloaded bytes for selected services over time.

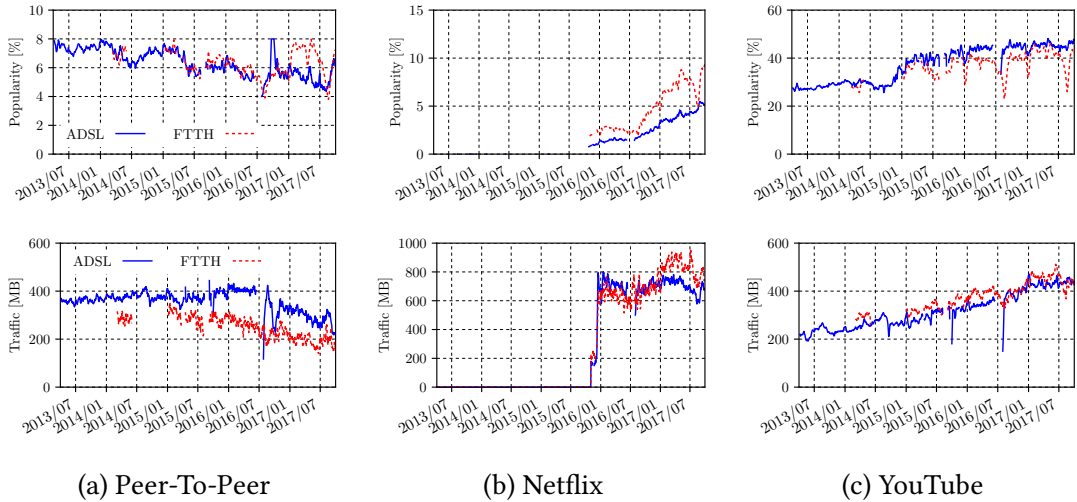

(a) Peer-To-Peer     (b) Netflix     (c) YouTube

Figure 3.5: Popularity (top) and volumes (bottom) for P2P and 2 popular video streaming services.

### 3.4.2 The downfall of Peer-To-Peer - finally

It is no news that P2P is no longer among the preferred means to download content. Here I quantify this phenomenon showing the popularity of P2P applications over the years. Figure 3.5a details the percentage of active users using a P2P service (Bittorrent, eMule and variants) (top plot) and the average P2P traffic volume per user (bottom plot). I still observe a hardcore group of users that exchange about 400 MB of P2P data daily. At end of 2016 the traffic volume they generate starts to decrease. Interestingly, FTTH subscribers start abandoning P2P applications earlier in terms of volume. Based on findings of previous studies [40, 71], a conjecture to explain this decline is that the availability of cheap, easy and legal platforms to access content is finally contributing to the downfall of P2P. In the following I explore this conjecture.

### 3.4.3   The usual suspects: YouTube and Netflix

I now consider popular video streaming services. Figure 3.5b shows the percentage of active users accessing Netflix (top) and the average per-user daily traffic (bottom). Netflix has gained momentum since the day it started operating in Italy. FTTH subscribers have been eager to adopt it, with about 10% of the active users using it on a daily basis at the end of 2017. Considering weekly statistics, I see that more than 18% (12%) of FTTH (ADSL) subscribers access Netflix at least once in 2017. Considering the amount of traffic they consume (bottom plot), I see no major differences between ADSL and FTTH subscribers up to end of 2016. Since October 2016, Netflix started offering Ultra HD content. This is reflected into each active FTTH subscriber downloading close to 1 GB of content on average per day. ADSL subscribers instead cannot enjoy it, or are not willing to pay the extra fee.

Next, I evaluate YouTube (Figure 3.5c). The figure shows a consolidated service, that is accessed regularly by users, who are consuming more and more content: more than 40% of active subscribers access it daily, and download more than 400 MB (about half of Netflix volume per subscriber). Interestingly, no differences are observed between ADSL and FTTH subscribers − hinting that YouTube video works similarly on FTTH and ADSL.

### 3.4.4   The new elephants in the room: Social messaging applications

I now study usage patterns for social messaging applications, namely SnapChat, WhatsApp and Instagram. All are popular applications accessed mostly on smartphones, whose traffic I observe once connected via WiFi from home. As before, I consider popularity and daily traffic consumption per active subscriber (recall Section 3.4.1), depicted in top and bottom plots in Figure 3.6.

Interesting trends emerge in the rise and fall of social networking apps. Observe first Snapchat (Figure 3.6a). It enjoyed a period of notoriety starting from 2015, topping in 2016 when it was adopted by around 10% of subscribers. Each active subscriber used to exchange up to 100 MB of data daily! Starting from 2017, the volume of data starts to decrease, with active subscribers that nowadays exchange less than 20 MB per day. Popularity is mostly unaffected, suggesting that people keep having the Snapchat app, but hardly use it.

The decline of SnapChat coincides with the growth of other social apps. See WhatsApp in Figure 3.6b: Its popularity is indisputable, with a steady growth in adopters that has almost reached saturation. Observe instead the growth in daily volume per active subscriber. Each subscriber exchanges around 10 MB daily, pointing to the intensive use of the app for sharing multimedia content. Note also the large peaks in the figure, corresponding to Christmas and New Year's Eve, when people exchange wishes using WhatsApp.

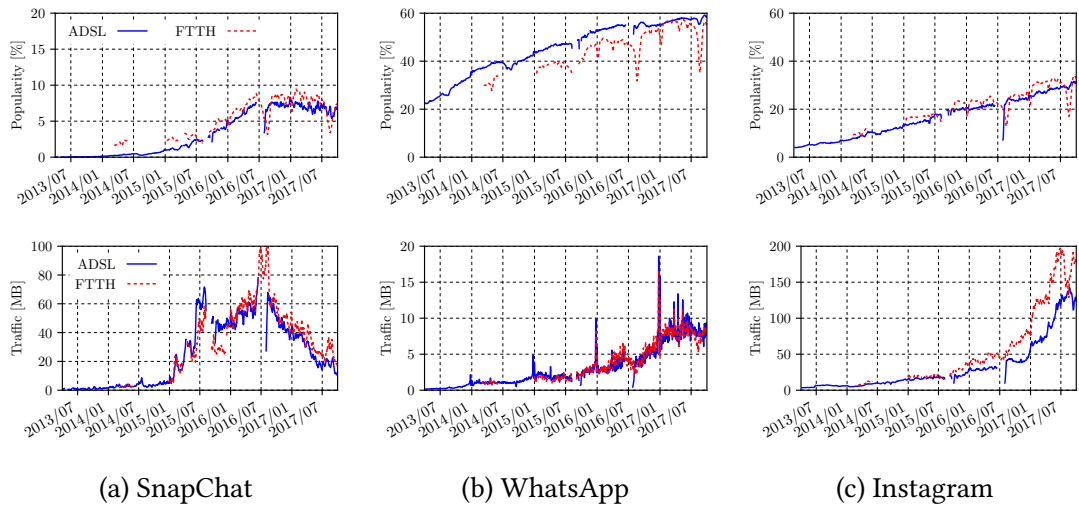

(a) SnapChat                    (b) WhatsApp                    (c) Instagram

Figure 3.6: Popularity (top) and volumes (bottom) for 3 popular social messaging services.

Finally, considering Figure 3.6c (Instagram), I see a constant growth in popularity and, more impressive, a massive growth in traffic volumes. Each active subscriber exchanges on average 200 MB and 120 MB per day, for FTTH and ADSL respectively. This is almost a quarter of the traffic of the active customers contacting Netflix! Recalling that Instagram, Snapchat and WhatsApp are predominantly used from mobile terminals, these figures point to a shift on traffic of broadband users, with mobile terminals taking a predominant role even when people are at home.

## 3.5    Web trends, and surprises

In this section, I study how web protocols usage varied across the last 5 years. I show in particular events associated with the slow migration of services towards newer standard web protocols, and sudden relevant changes on the traffic matrix caused by experiments of big players with custom protocols.

In its early life, the Web was predominantly plain Hyper Text Transfer Protocol (HTTP) traffic. It is by now known that most of the web traffic is running encrypted [80], first with the deployment of HTTPS, followed by the push towards HTTP/2 [6] (for which practical deployments rely on TLS) and more recently QUIC [68]. I here want

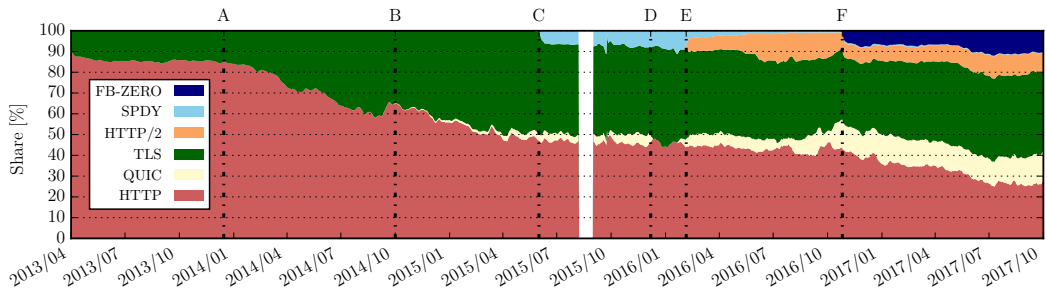

Figure 3.7: Web protocol breakdown over 5 years. Sudden changes and custom protocols deployment in the wild are highlighted.

to document to what extent these protocols have been used by the monitored ISP subscribers.

Figure 3.7 answers this question. It shows the traffic share of the several Web protocols observed in the network over time. Five years ago, in 2013, only the two "classic" web protocols were observed, with the majority of traffic served by clear-text HTTP, and only around 13% of the web traffic due to TLS/HTTPS. Then, several notable changes happened, which are marked with letters in the figure:[6]

A) January 2014: YouTube starts serving video streams over HTTPS. The migration has taken Google several months in 2014, in which one can see a steady change in the mix of HTTP and HTTPS traffic. HTTPS share tops to 40% at the end of 2014 already, and it is mainly driven by YouTube traffic.

B) October 2014: After announcing it in 2013, Google starts testing QUIC in the wild deploying its Chrome Web browser. Web traffic carried by QUIC (carried over UDP) starts growing steadily.

C) June 2015: I update the probes to explicitly report SPDY protocol (previously generically labeled as HTTPS). I discover 10% of traffic carried by an experimental protocol, reaching mainly Akamai and Google web servers.

D) December 2015: Google disables QUIC for security issues [68]. Suddenly 8% of the traffic falls back to TCP and HTTPS/SPDY. Around a month after, the bug is fixed and QUIC is suddenly back.

E) February 2016: Google migrates traffic from SPDY to HTTP/2, slowly followed by other players.

---

[6]These events have been confirmed manually throughout the years while upgrading the software of the probes to keep-up with protocols evolution.

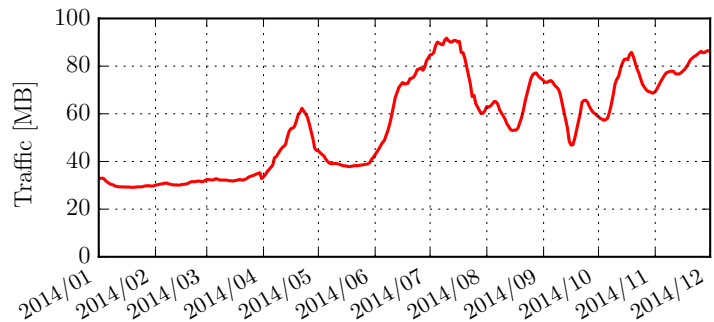

Figure 3.8: Facebook average daily per-user traffic before and after automatic video play.

F) November 2016: Facebook suddenly deploys "FB-Zero", a protocol with a custom 0-RTT modification of TLS used from the Facebook mobile app only.[7] Suddenly, 8% of web traffic moves to this new protocol. More than a half of Facebook traffic is now carried by Zero, showing that mobile app traffic surpassed website, even for fixed ADSL installations.

At the end of 2017, HTTP is down to 25%, with HTTP/2 that is slowly gaining momentum. QUIC and Zero together carry 20−25% of web traffic. Both are yet to be standardized protocols, showing how giants like Google and Facebook are free to deploy experiments on the web, since they own both server and client applications. Such experiments may create issues on ISP networks, e.g., making network proxies and firewalls suddenly inefficient, or creating issues with home gateway.

Finally, I illustrate in Figure 3.8 another interesting episode of sudden traffic changes. Around March/April 2014, Facebook started enabling video auto-play for its applications. The immediate effect on ISP traffic is striking. Figure 3.8 illustrates the daily average traffic per subscriber towards Facebook. Starting in March 2014 the average per-day traffic of a subscriber towards Facebook has grown from around 35 MB to around 70 MB in a month. After an apparent pause in the deployment of the feature during May, the service enabled video auto-play again. In July, the daily traffic per subscriber was around 90 MB on average, 2.5 times higher than the rate observed in March 2014!

This figure illustrate once more how the big players controlling key client software and servers can perform impactful changes in the Internet, complicating the planning and management of ISP networks.

---

[7]Zero protocol would be announced only in January 2017 − https://goo.gl/vuQ1Jy

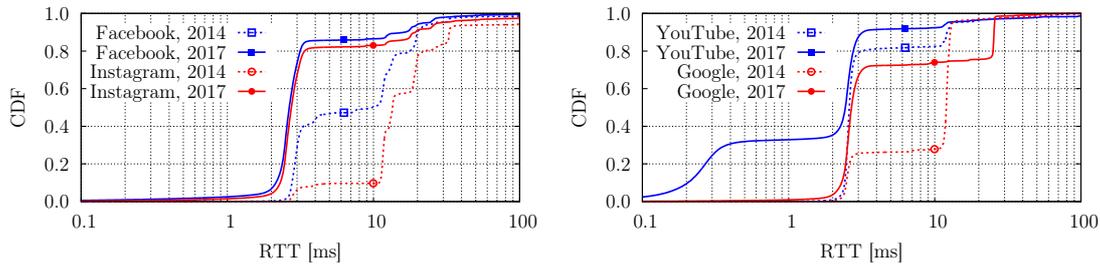

(a) CDF of RTT in 2014 and 2017 for Facebook and Instagram.

(b) CDF of RTT in 2014 and 2017 for YouTube and Google.

Figure 3.9: CDF of Round Trip time.

## 3.6  Where are my servers?

In the previous section I have shown both slow and sudden changes due to overall trends, and big players migration policies. Here I go deeper into showing the impact of big players infrastructure changes over the years.

### 3.6.1  The birth of the sub-millisecond Internet

CDNs were born in the '90s to reduce both the load on centralized server and the delay to access the content. Nowadays shared and private CDNs are making it possible to scale Internet content distribution, allowing users to fetch content from nearby surrogate servers. Being delay one of the main parameters affecting users' Quality of Experience, I focus my attention on how it changed over years.

I consider the Round Trip Time (RTT) as performance index. Remind that probes measure RTT by matching TCP segments sent by clients with corresponding TCP ACKs sent by servers. It represents the RTT from the probe to the server – excluding the access network delay. For all TCP connections to a given service, I extract the minimum per-flow RTT, and plot the corresponding CDF. By doing so for a long time interval and large sample of users, I can spot how the RTT distribution is composed. Thus, I focus on the body of the distribution of minimum per-flow RTT, ignoring samples in the tails of the distribution, which may be caused by queuing and processing delays.

Figure 3.9 shows the results contrasting measurements seen in April 2014 versus April 2017. I focus on Facebook and Google services as notable examples of big players that pay particular attention to speed up content delivery. Consider Instagram traffic (red curves) on Figure 3.9a. Dashed line refers to 2014 figures. At those time, there were already CDN surrogate nodes at just 3 ms RTT from the ISP PoP. However it served only  10% of flows. Other traffic was served by far away CDN nodes, with RTT of 10, 20 and 30 ms.[8] About 7% of flows was served by servers with RTT higher than 100 ms – a clear sign of intercontinental path. Facebook caches (blue curves) follows a very similar placement – with different share of traffic being served by different caches.

Consider now the 2017 CDF (solid lines). Results clearly show that many more requests are now served by close servers, with 80 % of both Instagram and Facebook traffic that is served by the 3 ms far CDN nodes. As I will show later, this change is due to two factors: i) Facebook that deployed its own CDN; And ii) Instagram infrastructure being integrated into Facebook one.

Look now at Figure 3.9b which depicts the RTT CDF for Google web search servers and YouTube streaming servers. In 2014, 80 % of YouTube traffic (blue curves) was already being served by nodes that were just 3 ms far away from the ISP PoP. This is to guarantee the high volume due to video traffic. In 2017, this already marginal figure decreased even more – with the YouTube video cache now breaking the sub millisecond RTT. That is, YouTube now directly places video servers inside the PoP, at the first level of aggregation, toward a very distributed and pervasive infrastructure. Interestingly, Google search engine web servers (red curves) have not yet reached such a fine grained penetration. This is because they have to handle less traffic, and perform more complicated processing than YouTube video caches.

I have confirmed these findings by directly contacting the ISP staff, who reported the deployment of third-party CDN and cache nodes at the ISP first aggregation point.

I repeated the analysis for other services – not reported for the sake of brevity. With the only notable exception of WhatsApp, whose servers are still following a centralized approach with RTT in the 100 ms range, all services are exhibiting the same trend, with more and more CDN surrogate servers being placed closer and closer to the edge of the network.

On the one hand, this proliferation of edge caches, and the delay of modern FTTH access network is leading us to the sub-millisecond Internet [97]. On the other hand, this poses new burdens on the ISPs, which have to host (and in some cases manage) infrastructure of different content and CDN providers inside their network. Network Function Virtualisation (NFV) would possibly help in reducing this burden [48], allowing ISPs to host virtual CDN surrogates into their infrastructure.

### 3.6.2 The Internet of few giants

I now analyse the infrastructure of large content providers. Indeed, during the last 5 years most web services incurred restructuring, replacing servers, deploying their own CDN, etc.

Figure 3.10 depicts the evolution over time of the infrastructure of Facebook (left plots), Instagram (center plots), and YouTube (right plots). Top plots show the server IP addresses being active in each day, for the considered service. The *y*-axis represents a single server IP address, sorted in order of appearance. A red dot is present if for that day, that IP address was being used only and only for traffic of the considered service. A blue dot is present if that IP addresses served also content for other services. Finally, no dot is present if for that day that IP address did not serve any content.

---

[8]Fraction changes by hour. Figures refer to statistics collected on the whole month.

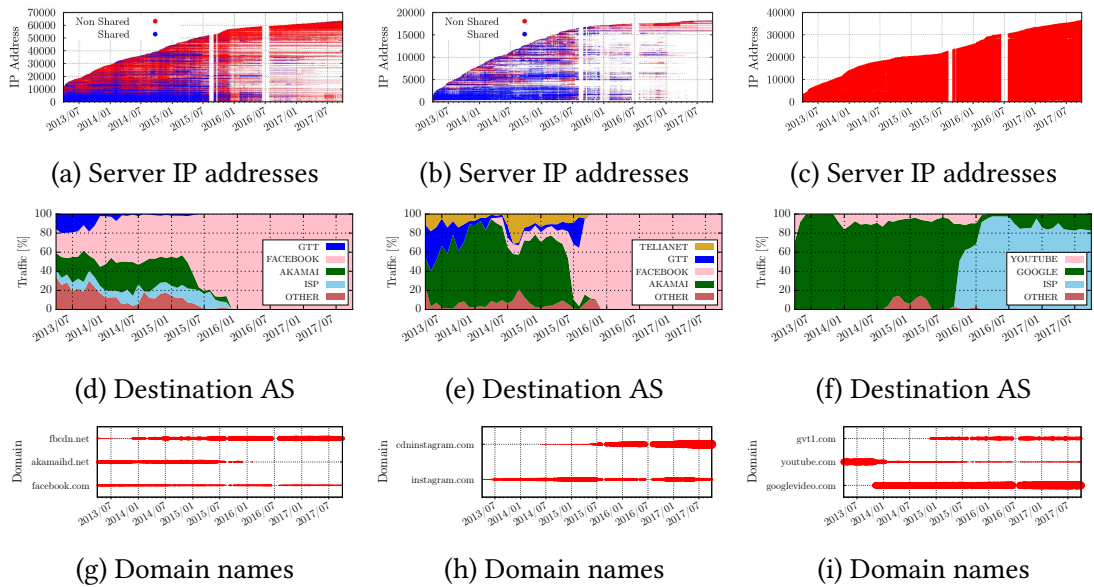

(a) Server IP addresses      (b) Server IP addresses      (c) Server IP addresses

(d) Destination AS      (e) Destination AS      (f) Destination AS

(g) Domain names      (h) Domain names      (i) Domain names

Figure 3.10: Facebook (left), Instagram (center) and YouTube (right) infrastructure evolution over time.

In all cases, I see that new IP addresses keep appearing over time, counting several tens of thousands unique IP addresses. Compare Facebook and Instagram in Figure 3.10a and Figure 3.10b, respectively. During 2013 and 2014, a good fraction of addresses were shared with other services. During the second half of 2015, I notice that both started having major changes, with i) a decrease in the number of servers being contacted, and ii) a specialization of servers that are not shared with any other services. In details, the total number of IP addresses used daily by Facebook dropped from 3 800 to less than 1 000, out of which 700 are still shared. Since July 2016, shared IP addresses drop to very few.

To better understand the reason behind this major change, I analyse to which Autonomous System Number (ASN) each IP addresses belonged.[9] Middle plots in Figure 3.10 show the breakdown of the per-day contacted IP addresses over major ASNs.

---

[9] I use the Routing Information Base (RIB) for each month from a major vantage point in the Route Views project to map IP addresses to ASNs

Figure 3.10d and Figure 3.10e show a migration from generic CDNs to the Facebook private CDN. In 2013, both services used third party CDNs, whose IP addresses where thus shared with other services. For Facebook, the migration started before 2013, and was completed by the end of 2015. For Instagram, the integration with Facebook infrastructure started in 2014 (Facebook acquired Instagram in April 2012), and was completed by end of 2015. This migration has two major effects: i) IP addresses are now dedicated to either Facebook, or Instagram; ii) the number of IP addresses contacted per day reduces. Indeed since 2016 only 1 000 IP addresses are used to serve Facebook traffic, and only 300 for Instagram. Contrasting these figures with Figure 3.9a, I notice that this change also benefited the RTT, which reduced significantly.

To better describe these changes, bottom plots in Figure 3.10 detail the traffic share served by most important second level domain names. The thicker is the line, the higher is the fraction of traffic served. For instance, Figure 3.10g confirms the migration from generic Akamai CDN to Facebook proprietary infrastructure. Even more evident is the migration for Instagram in Figure 3.10h.

Finally, I study the YouTube infrastructure evolution as a case of study of a very popular service with a massive infrastructure. From Figure 3.10c, it is already possible to see how different YouTube is with respect to the previous two cases. Indeed, YouTube always used a totally dedicated infrastructure to serve videos. Its infrastructure keeps growing until now, where 40 000 IP addresses are used daily. By looking at Figure 3.10f, I observe that starting from the end of 2015, the caches deployed in the ISP start serving most of YouTube traffic. This benefited RTT as previously shown. Regarding to the Domain names used by YouTube, Figure 3.10i shows three main changes: until January 2014, all the traffic was served by the *youtube.com* domain; In 2014 the *googlevideo.com* domain suddenly appeared, and immediately handled the majority of traffic; Finally, in 2015 YouTube introduced *gvt1.com*.

These results confirm the trend toward a consolidation of large services, which deploy their own infrastructure, in a more and more capillary way, reaching several tens of thousands of IP addresses. Furthermore, these infrastructure undergo sudden and undocumented changes that have impact on the traffic monitoring and management of IPSs and corporate networks.

## 3.7 Related Work

Several works measured Internet traffic from different points of views. Gebert *et al.* [41] characterized the observed traffic mixtures in an ISP network during 14 days. Liu *et al.* [69] designed a large scale measurement infrastructure and deployed it in the core network of a cellular operator. Their focus is on the architecture, not on measurements. Authors of [39] reported their experience on operating a monitoring infrastructure in ISP networks during 20 months in 2013, describing how protocols and services are typically consumed from such networks. Muhammad *et al.* [94] analyzed a week-long traffic trace collected from a tier-1 cellular network, showing how machine-to-machine traffic is different from human-generated traffic. All these works cover a relatively short period, which prevent them to evaluate how the identified phenomena have evolved over time.

Some works provided longitudinal views on Internet evolution. The authors of [29] analyzed a dataset of BGP measurements that covers 12 years, showing how the BGP ecosystem has evolved. Authors of [36] presented one of the first longitudinal studies of Internet traffic, covering the period of 1998–2003. Authors of [13] evaluated 7 years of MAWI traces, summarizing the evolution of Internet traffic in Japan. In [9] authors evaluated 23 months of data collected from 53 k broadband installations, highlighting for instance the relation between capacity and demand.

My work is similar to those efforts in terms of the employed methodology and general goals. Similar to [13, 36, 29, 9] I focus on long-term trends instead of exploring details of a measurement snapshot. I report statistics and trends about users' habits, usage of services and protocols, while also focusing on the infrastructure changes. More important, I show figures from a recent period, thus updating the knowledge about Internet usage.

Also in terms of methodology, I monitor close to end-users (e.g., similar to [38, 71, 13]) and not in the core (e.g., as in [90, 67]). This allows to provide a comprehensive picture of users' data consumption, which is particularly relevant for ISPs.

Regarding my conclusions, I highlight many interesting facts about the Internet traffic mix. A number of recent studies also reported on Internet traffic mix using different vantage points. Authors of [90] reported the traffic observed in an IXP in 2013, comparing their findings to other vantage points [71, 41]. Labovitz *et al.* [67] analyzed two years of network measurements collected from several Internet backbones, illustrating how core Internet traffic is converging around few big players.

My work updates these studies showing trends from 2013 onward. Similar to [43] and others, I present traffic mix focusing on services and the most popular application-layer protocols. Whereas my data would allow to drill down on per-protocol breakdowns (e.g., as in [26]), these details are left out for the sake of brevity.

As said above, many of my conclusions validate results already identified in previous works. Examples of known results that are confirmed or extended by my measurements include: (i) the slow increasing trend on traffic per user [20]; (ii) the predominance of video traffic [1, 33]; (iii) the fast increase in HTTPS deployment [34]; (iv) the decline of P2P [40, 71]; (v) the concentration of Internet traffic around few big players [67]; (vi) the deployment of experimental protocols resulting in sudden changes in the traffic mix due to bugs and private tests by large companies [91, 61, 68].

In some other cases, my results add more data points to complement previous findings. For example, I could not find a clear general relation between the capacity customers and their demands as in [9]. However, for customers relying on particular services (like Netflix) these conclusions seem to hold true. Besides that, I also shed light on new aspects of the Internet evolution, such as the costs of services to providers, the usage dynamics of new social network services such as Instagram and Snapchat, among others.

Finally, some companies such Cisco periodically report traffic trends and forecasts [23], including predictions on connected devices, Internet usage and traffic nature. By reporting detailed statistics from measurements collected in operational networks, my work complements such studies and can contribute in gaining a better understanding of Internet traffic.

## 3.8  Conclusion

In this chapter, I evaluated the evolution of the traffic during 5 years (2013–2017) for a large ISP network. By processing large scale and longitudinal measurements from a national ISP in Italy, I characterized the traffic consumption of broadband subscribers, and the infrastructure web services deploy to reach customers. I observed subscribers' daily traffic that more than doubled in the analyzed period. I studied the typical loads imposed by popular and bandwidth hungry services. I testified the death of P2P in exchange for legal, cheap and easy to use video providers, and the quick rise and death of social messaging applications typically accessed via mobile phones, able to generate massive amount of data.

I observed the concentration of services within few big Internet providers, each deploying its own infrastructure, unrolling custom protocols, and penetrating more and more network boundaries. In the rush to bring server closer and closer to users, I witnessed the birth of the sub-millisecond CDNs, where Internet giants like Google or Facebook are placing caches directly in the ISP PoPs. All such changes, and the unpredictability they are appearing, complicate the planning and management of the networks, possibly calling for closer integration between content providers and operators.

I believe the figures I presented in this chapter are vital to researchers, ISPs and even web service provider to better understand the liveliness of the Internet, which continuously changes, mixing slow and unpredictable changes.

# Chapter 4

# Towards Web Service Classification using Addresses and DNS

The work I present in this chapter is mostly taken from my paper "*Towards Web Service Classification using Addresses and DNS*" presented in the 7$^{th}$ International Workshop on TRaffic Analysis and Characterization (TRAC 2016) [106].

## 4.1   Introduction

Monitoring how web services are used and how they consume network resources is key to Internet Service Providers (ISP) when operating and planing the network. Similarly, companies have a vital need of monitoring their enterprise networks – e.g., to ensure usage of accredited services, or to control the access to unauthorized ones. Traffic classification has always taken a key role, and a variety of methods has been developed throughout the years. Initially focusing on protocol classification, e.g., HTTP vs FTP vs P2P, classification goals must now target the identification of "web services", e.g., YouTube vs Facebook vs Whatsapp. Indeed, HTTP is becoming the de-facto application layer protocol over which people access the large majority of Internet applications. Deep Packet Inspection (DPI), behavioral techniques [65, 15], have been used for traffic classification. These methods have been recognized so far as effective for several monitoring needs [109].

The convergence of web toward proprietary and encrypted protocols, however, challenges classification algorithms again. Indeed, I already observe a clear trend towards moving Internet services to protocols such HTTPS [80], with HTTP 2.0 behind the corner and TLS encryption by default. While this trend is well-justified by the urgency in improving end-users' privacy, it renders many traffic classification algorithms useless, since packet payload cannot be accessed anymore.

In addition, a handful of big players [42] is taking a prominent role in the Internet, where content is more and more being served from shared infrastructure, such as in Content Delivery Networks (CDNs) and cloud computing platforms. This further challenges behavioral classifiers [64], which rely on host profiling to determine the applications running on servers.

This chapter revisits the question of whether basic traffic features can be used to differentiate traffic of major web services. The ambitious goal is to understand how feasible would be the classification of web services traffic based only on server IP addresses and queries to the DNS, i.e., the few features that are likely going to remain visible. By relying on a large dataset containing flow-level measurements of user activity annotated with DNS queries, I first investigate to what extent server IP addresses provide enough evidences of the services used by people. I then evaluate the amount of traffic that can be distinguished by combining server hostname and addresses to create rules.[1] Finally, I discuss how stable such rules are in time.

Previous works have studied the importance of different features for traffic classification. In particular, a comprehensive survey on classification methods for encrypted traffic is presented in [110]. The authors of [60, 100] found that IP addresses are among the most informative features. I perform similar analysis to quantify how traffic of modern services can be classified using only addresses and hostnames. Authors of [85] are the first to claim the use of DNS to classify traffic. In contrast to the method proposed by authors of [85], I neglect well-known protocols (e.g., FTP or P2P). Instead, I focus on typical services that make the majority of encrypted web traffic nowadays, and characterize when hostnames are needed, and when only addresses would be sufficient for classification. More recently, authors of [101, 37] used Server Name Indication (SNI) strings found in TLS handshakes and DNS queries for classification. While authors concluded that hostnames alone are insufficient, they targeted protocol classification (e.g., SIP, HTTP, etc.), thus missing fine-grained identification of single web services. Other authors [7, 78] argue the usefulness of DNS for classifiers, but mostly focusing on how to label flows, missing a study of classification accuracy.

My work is a preliminary evaluation of web service classifiers in the modern Internet. My analysis provides the following main findings:

- Up to 65% of the IP addresses are associated to a single hostname. Those servers however are responsible for less than 15% of web traffic volume.

- Despite the simplicity, classification based solely on (group of equivalent) IP addresses can discern up to 55% of the web traffic. This can be achieved by uncovering and aggregating the various hostnames related to a given service, and then enumerating corresponding IP addresses.

- Lifetime of classification rules varies strongly, with some services requiring weekly updates and others showing stable names and addresses even after a year.

---

[1]In the remaining of the chapter, I omit the word "server" unless necessary.

Table 4.1: Overview of the datasets.

| Name | Flows | Server IPs | Period |
|------|-------|-----------|--------|
| `ADSL` | 13.25G | 49.25M | 1 year |
| `Campus` | 1.12G | 2.55M | 2 months |
| `Campus-DNS` | – | 1.13M | 2 weeks |

- Even when tagging flows with hostnames on-line using all DNS queries of each client (e.g., as in [7]), there can be complex scenarios when facing big cloud computing platforms (e.g., Amazon AWS)

These results are a first step towards classification algorithms that are able to work with minimal metadata. While these data will certainly not solve some identification problems (e.g., for network forensics and intrusion detection), I believe they represent a set of non-intrusive features to tackle common monitoring tasks, such as traffic accounting and engineering.

## 4.2 Datasets and Methodology

The aim of this chapter is to investigate whether IP addresses and DNS traffic provide enough information to design web service classifiers, targeting in particular those prominent services which adopt encryption, such as HTTPS, QUIC or SPDY. I take a data driven approach and look into real traces to run a feasibility check in this chapter, before going through a complete system design.

### 4.2.1 Datasets

I use two data sources in my analysis. First, I rely on the datasets described in Chapter 2. In short, I use Tstat [35] to perform passive measurements and collect data related to users' activity. I use two datasets coming from a University Campus and an ISP network both located in Italy, namely `Campus` and `ADSL`. The analysis of this chapter is limited to a shorter period, as the duration of datasets reaches 5 years: For `ADSL`, I consider one full year, while for `Campus`, I use a period of 2 months.

Second, in parallel to Tstat, I deploy PDNS[2] in one of the monitored links to get a deeper insight into the association between hostnames and server IP addresses. PDNS logs all DNS activity in the network independently from the resolver the client employs, including queries and responses with the returned addresses and the time-to-live found.

Table 4.1 summarizes the datasets employed in this chapter. The data from campus includes traffic generated by wired and WiFi networks during 2 months in 2015. PDNS was deployed in the campus for 2 weeks in Nov 2015. The residential dataset includes traffic of users' devices connected via Ethernet and/or WiFi at home during a full year (2015). In total, my datasets include statistics about more than 14 billion flows, and around 790 million records in DNS requests/responses.

---

[2]https://github.com/gamelinux/passivedns

### 4.2.2 Methodology

I study the association between IP addresses and hostnames to understand the role of addresses in modern traffic classification. I first assume hostnames provide sufficient means to distinguish services – i.e., different services use different hostnames. I will discuss later to what extent this assumption holds in practice. Hostnames coming either from SNIs or from DNS queries are the ground-truth in this scenario. I characterize how the relation between names and addresses evolves over time. In particular, I look for those IP addresses that serve only a single hostname, i.e., only one hostname is associated to a given IP address. I call this *singleton IP addresses*, or singleton in brief. I then quantify the percentage of traffic exchanged with singletons, to obtain an indication of the classification coverage that could be achieved using only the IP addresses as features.

Motivated by the low volume of traffic that could be discerned by such an approach, I study how to improve classification by enumerating the different hostnames (and addresses) used by services. I call the list of names of a service its *bag of domains*. I interactively build the bag of domains for a list of services by relying on SNIs and hostnames exported by Tstat. A graphical framework allows one to inspect names linked to IP addresses. I illustrate this procedure with examples in the next section. I focus on popular services running over HTTPS – e.g., Facebook, Google Video, Dropbox, Apple iCloud, Twitter etc. – since those services cause the greatest part of the encrypted traffic in the monitored networks.

## 4.3 Enumerating Names and Addresses of Services

In this section, I visually explore how hostnames and addresses are associated. I represent the associations as a graph, in which nodes are IP addresses and hostnames, and edges exist if a hostname has been resolved to an address. I initially search for hostnames containing terms of interest. For example, by searching for `whatsapp` in the data, I discover that Whatsapp services are offered from at least two second-level domains – i.e., `whatsapp.com` and `whatsapp.net`. I call those the *core domains*, and from them I explore linked IP addresses, and correlated hostnames.[3]

---

[3]Terms of interested could be obtained by active experimentation with target services in a testbed such as in [11].

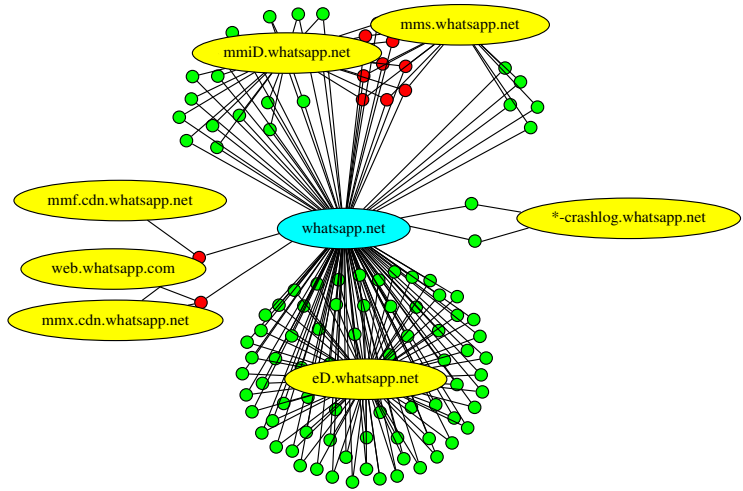

Figure 4.1: IP addresses and hostnames of Whatsapp. Most IP addresses are exclusively used by the service.

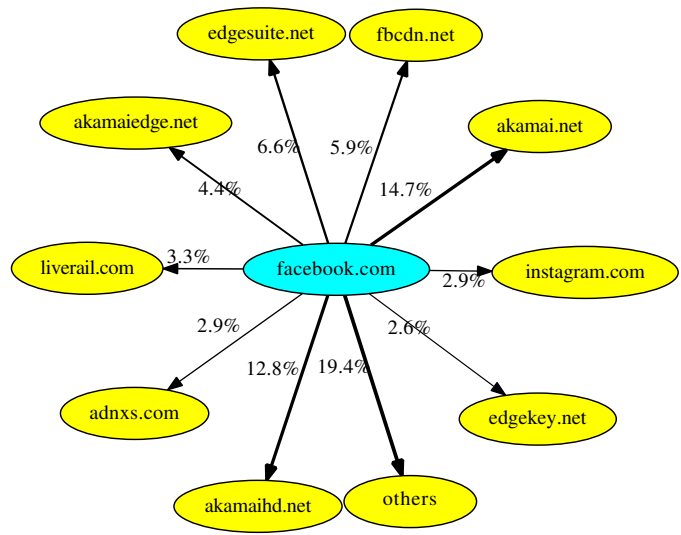

Figure 4.2: Hostnames sharing IP addresses with Facebook.

Figure 4.1 and Figure 4.2 provide examples. Figure 4.1 depicts how second-level domains are associated with `whatsapp.net`. For simplicity, the figure is built using a 5-minute sample of `Campus-DNS` trace. The core domain is shown as a central node; IP addresses are nodes colored either green (singletons, i.e., edge links them to only one `whatsapp.net` sub-domain), or red (not singletons, with multiple edges to multiple domains); and yellow nodes represent `whatsapp.net` sub-domains sharing IP addresses with each others.

I notice that Whatsapp IP addresses are not shared with other services. Therefore, once all addresses are enumerated, Whatsapp traffic can be identified without further

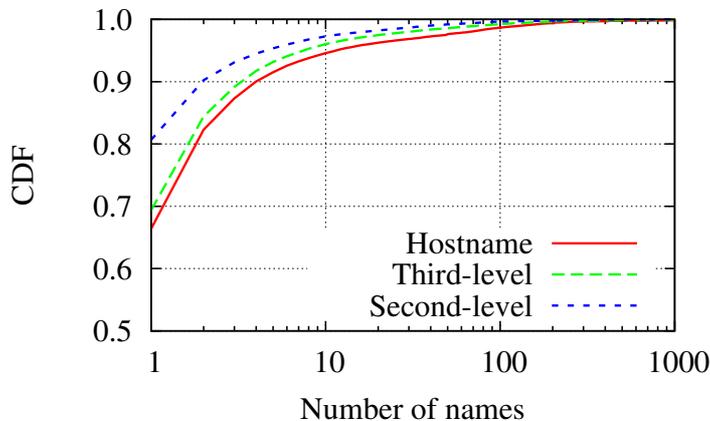

Figure 4.3: Distribution of names per IP address. `Campus-DNS`.

information.

Figure 4.2 shows more complicated scenarios emerging from `facebook.com`. To improve visualization, nodes representing IP addresses are replaced by edges labeled with the percentage of addresses connected to pairs of hostnames – e.g., 3.3% of the addresses seen as `facebook.com` are also seen as `liverail.com`. Besides sharing addresses with Facebook's affiliated services (e.g., Instagram), Facebook's usage of Akamai CDN results in thousands of hostnames unrelated to Facebook appearing in the graph as time progresses.

In summary, the association between IP addresses and hostnames brings information, but the presence of CDNs create conflicts and ambiguity. Next, I quantify how much traffic could be classified despite such ambiguities.

## 4.4   Classification Using IP Addresses

I first provide an overview on all IP addresses and hostnames in the 2-week long dataset of DNS traffic (i.e., `Campus-DNS`). I perform this analysis focusing on DNS *A* records. For each IP address returned in a DNS response, I collect all hostnames requested by clients.

Using `Campus-DNS` dataset, I count how many hostnames are linked to each IP address. Three levels of granularity are considered: (i) complete hostnames, e.g., `img.www.example.com`; (ii) third-level domains, e.g., `www.example.com`; (iii) second-level domains, e.g., `example.com`. Figure 4.3 reports the empirical Cumulative Distribution Function (CDF) of the number of names associated with each IP address.

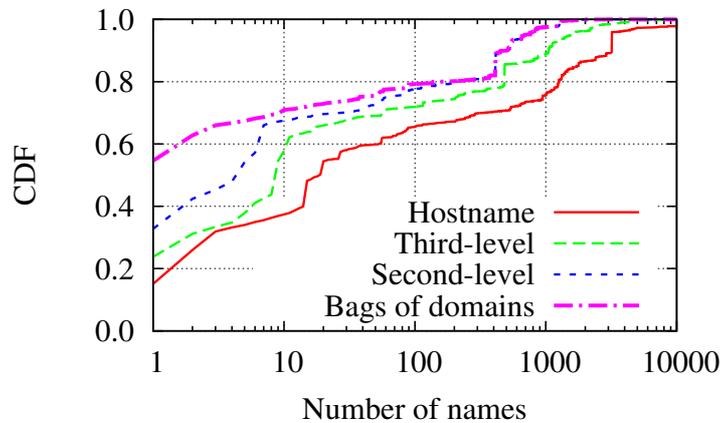

Figure 4.4: Distribution of the traffic related to IP addresses with different numbers of hostnames. `Campus`.

On the leftmost part, I notice that more than the 65% of the IP addresses are singletons. This percentage grows to 70% when considering third-level domains, and to 80% when considering second-level domains. Those results confirm previous observations (e.g., see [7, 16]) and, at first, suggest that a great part of the traffic could be easily classified by simply using server IP addresses.

A completely different picture however emerges when taking traffic volume into account. Although most IP addresses are singletons, such addresses are responsible for a small fraction of the traffic. I quantify this effect in Figure 4.4 using the `Campus` dataset. For each IP address, I count the amount of bytes it handles, and compute then the handled fraction. Figure 4.4 shows the resulting CDF. Remind that I include only HTTP and HTTPS traffic here. Less than 15% of the traffic is owing to singletons, even if those cases are 65% of the addresses. The picture does not change considerably when third- or second-level domains are used: Percentages are 25% and 33%, respectively. In a nutshell, a classifier that takes only IP addresses as input would identify up to 33% of the traffic without mistakes. Part of the remaining traffic would necessarily be misclassified, since many hostnames (and thus services) possibly run over the same address. I conclude that server IP addresses alone provide a very poor classification coverage for the web traffic.

## 4.5 Classification Using Bags of Domains

I repeat the analysis after creating *bags of domains*. A bag represents the set of domains a service uses to handle its content. I consider 25 coarsely defined groups of services, including e.g., *Google*, *Facebook* and *Dropbox*. For example, Facebook bag of domains includes `facebook.com` as well as Facebook's domains pointing to CDNs, such as `fbcdn.net` and `fbstatic-a.akamaihd.net`. So far, I manually group domain names that belong to a given service, since I observe that bags of domains are rather stable in the datasets.

Table 4.2: Popular services and classification precision.

| Core Domain | *facebook.com* | *google.com* | *googlevideo.com* | *whatsapp.net* | *twitter.com* | *dropbox.com* |
|---|---|---|---|---|---|---|
| Number of Addresses | 3,196 | 7,286 | 13,133 | 851 | 279 | 2,227 |
| Singletons (%) | 29.8 | 58.7 | 79.9 | 99.8 | 83.9 | 59.9 |
| Traffic to singl. (%) | 85.8 | 38.5 | 1.2 | 100.0 | 78.7 | 91.3 |
| Precision (%) | 59.1 | 33.8 | 77.2 | 100.0 | 96.1 | 99.1 |

Given a bag of domains, I extract IP addresses corresponding to any name in the bag. I next check if those addresses have been resolved also for hostnames not in the bag. Those IP addresses that create ambiguity are discarded. Those that correspond to hostnames in the bag only are *singletons for the service* and thus provide a good classification, i.e., traffic is uniquely linked to the targeted service.

Figure 4.4 reports the CDF of traffic according to singletons for the services. The bags of domains substantially increases the fraction of traffic that can be discerned. Three regions are visible in the figure. First, close to 55% of the traffic is related to IP addresses that are connected to a single bag of domains. Second, up to 10% of the traffic is caused by IP addresses shared by at most 10 names or bags. Part of these cases seems to occur because I have created bags only for few popular services, and thus names could be aggregated further. Third, about 20% of the traffic volume is caused by IP addresses shared by hundreds or thousands of names. Those cases are mostly servers in CDNs, and it is hard to discern services without full information about hostnames queried by clients. The intuition suggests that the bag of domains approach would be ineffective for this latter group. I will investigate these cases further in coming sections. I perform a similar calculation accounting flows instead of bytes, obtaining very similar results, not reported here for lack of space.

I conclude that a very simple classifier that relies on server IP addresses only could discern up to 55% of the web traffic. However, this is achievable only if service hostnames are aggregated, and their addresses are enumerated. Important, IP addresses in bags of domains can be learned by inspecting logs in DNS servers, or by actively querying the DNS system. Finally, the development of a methodology to automatically create bags of domains and to enumerate IP addresses is explored in the next chapters.

## 4.6   Use Cases

### 4.6.1   A Deeper Look into Popular Services

I now focus on six popular services and study in details how hostnames and addresses are used. I further estimate the precision of different classification approaches when applied to these services. Table 4.2 reports statistics about 6 services over two weeks of observations. I calculate statistics using the period in which `Campus` and `Campus-DNS` datasets are coincident. I focus on the most popular web services categories such as Social Networks, Search Engines and Cloud Storage. Thus, I take into account Facebook, Google, Whatsapp, Dropbox and Twitter, considering all traffic to their *bags of domains*.

Table 4.2 shows that the number of IP addresses hosting each service (2nd row) varies considerably,[4] as it varies the percentage of those addresses that are fully dedicated to the services (3rd row - singletons). For instance, while 99.8% of the IP addresses serving Whatsapp are singletons, more than 40% of the addresses of Google are observed in DNS queries related to non-`google.com` bag of domains. No address has been seen in more than one of the considered bags, except for Google and Googlevideo: all non-singletons of Google Video appear within Google's bag, and the 89.9% of Google's are in Google Video's, unveiling a shared infrastructure.

Next, I quantify the traffic related to singletons (4th row): Using the `Campus` trace, and using the DN-Hunter or SNI as ground truth to identify the service associated to a flow, I sum up all traffic for all hostnames in each bag of domains. I then compute the fraction of traffic that is associated with singletons for the same service. This number gives an estimation of the coverage if one relies only on the singletons to classify – i.e., the coverage when the classification provides 100% precision.

I can see that the percentage of traffic going to singletons is quite low for some services. Note for instance that only 1.2% of Google Video traffic goes to singletons, despite these being almost 80% of IP addresses of `googlevideo.com`. This happens since the traffic balance among the thousands of GoogleVideo servers is highly skewed toward a small subset of them, i.e., the most popular ones. Those addresses are also the ones for which hostnames of other bags of domains are found, and thus they are not singletons. For other services, singletons provide very high coverage: 100% of Whatsapp traffic goes to singletons (cfr. Figure 4.1), whereas percentages are relatively high also for Facebook (85%), Twitter (78%) and Dropbox (91%).

Finally, I estimate the precision of a classifier that marks *all* traffic related to addresses in the bags as belonging to the given services, being those singletons or not. That is, I estimate the precision of a classifier that have maximum coverage for the selected services. Since not all addresses are singletons (see 2nd row in Table 4.2), I expect to make classification mistakes.

The last row in Table 4.2 quantifies such mistakes. I can see that for three examples in the table – Whatsapp (100%), Twitter (96%) and Dropbox (99%) – the precision would be indeed very high. This means that only a minor amount of traffic not belonging to the services is mixed in their bags of domains. Google Video also presents a very high precision thanks to high traffic volume of the YouTube video service. Services that are mixed up with Google Video produce a low volume, even if they reach addresses in the Google Video bag. For Facebook, the classification precision is rather low, and this questions the applicability of the approach for such cases. This is because Facebook uses Akamai CDN, which hosts a multitude of alien services, which generate overall a large amount of traffic.

---

[4]The total number of addresses serving each service is likely higher since only *contacted* addresses are counted.

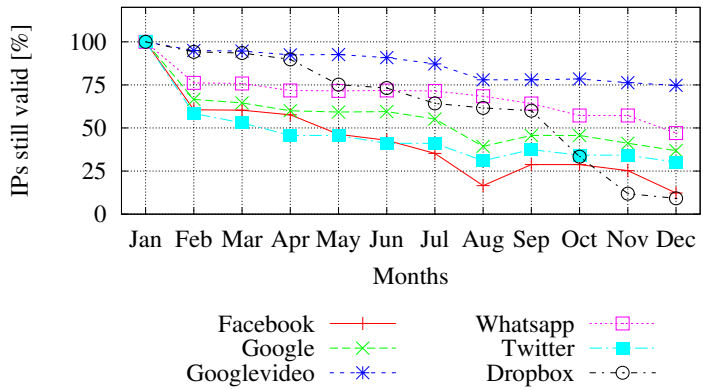

Figure 4.5: Persistence of addresses of popular services.

All in all, the classification based solely on addresses and bags of domains shows interesting potential. It enables the classification of a high share of traffic, with high coverage and precision for many popular services, while requiring minimal collection of data. Yet, a per-service assessment of precision and coverage is needed.

### 4.6.2  Names and Addresses over Time

In this section, I analyze how the associations between names and addresses evolve over time. In particular, I'm interested in knowing how stable the rules based on IP addresses and bags of domains are for popular services. I investigate such aspects using ADSL dataset, which covers a full year of a residential network. For each month of data, I create lists with all addresses used by popular services considering their bags of domains. I then track how the lists change throughout the year.

Figure 4.5 summarizes results by showing the percentage of addresses that is still on the lists, when compared to the first month of observation. One can see that all services present changes after Jan 2015, which is used as reference in the figure. Similar shapes emerge if other months are taken as reference. However, it is interesting to notice differences among services. Whereas the list of addresses for Google Video, for instance, is rather stable, as little as 15% of the Dropbox addresses seen in Jan 2015 remain in the list. Manual inspection suggests that addresses are passing for migration from US data-centers to EU data-centers; clients are now diverted to different data-centers than in previous months [5]. In several cases, such as for Twitter, almost 50% of the addresses already disappeared after a single month of observation.

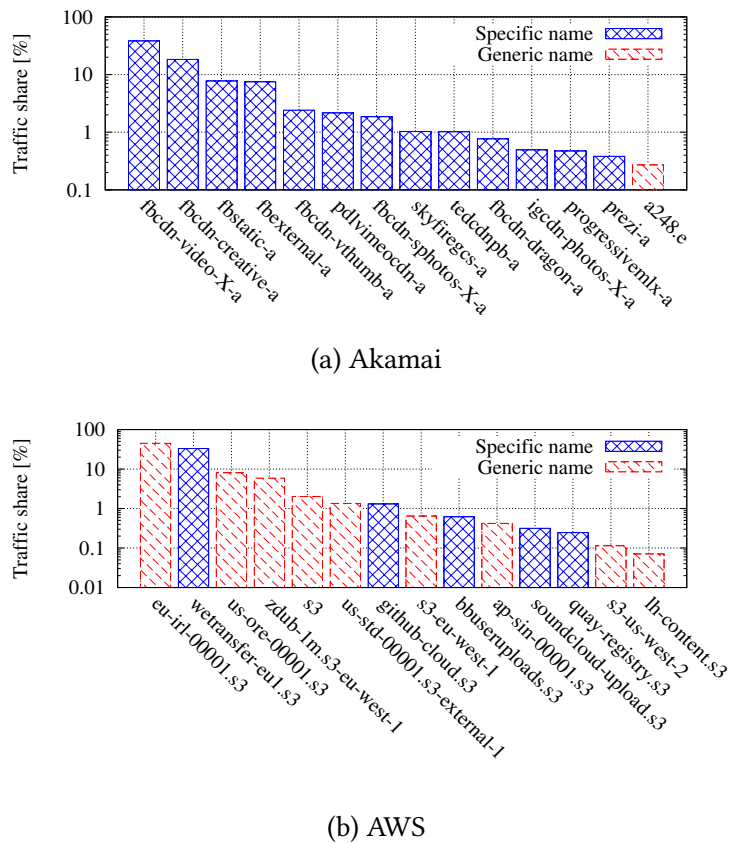

(a) Akamai

(b) AWS

Figure 4.6: Top sub-domains hosted by Amazon and Akamai.

Overall, I conclude that while the lists of addresses are stable in short intervals, they radically change in medium to long periods (Figure 4.5). Such intervals strongly depend on services and location of the monitored network. Classifiers relying on lists of addresses must deploy a methodology to constantly check and update their lists.

## 4.7 Traffic in Ambiguous Names

In the previous sections I evaluated the traffic related to IP addresses using the annotated hostnames as ground-truth. Now I investigate to what extent annotated traffic is reliable to the classification problem. I evaluate the case where each flow is associated to a hostname directly at the vantage point, as done by DN-Hunter or by extracting the SNI via DPI. Thus, the question is whether hostnames are unique to bags of domains of different services. I thus quantify how often hostnames are used by different services.

---

[5]See also https://www.dropbox.com/help/9063

I focus on two examples, *Amazon Web Services (AWS)* and *Akamai*, and enumerate all sub-domains of `amazonaws.com` and `akamaihd.net` contacted by clients in the datasets. Then, I manually try to identify the services relying on each sub-domain. In some case, hostnames give a clear hint about services – e.g., `fbcdn-sphotos-c-a.akamaihd.net` is used by Facebook, although generic names of the infra-structure providers are often observed as well – e.g., `eu-irl-00001.s3.amazonaws.com` is used by many services outsourcing to AWS.

Figure 4.6 highlights the top sub-domains of the providers according to their traffic share. Sub-domains are split into two groups: specific and generic. The first contains sub-domains that can be definitively associated to a service, whereas all other sub-domains are marked as generic.

When summing up all bytes related to specific sub-domains, I notice that 98% of the traffic related to Akamai can be distinguished. Therefore, classifiers can reach high coverage and precision when handling Akamai traffic, provided that information about hostnames requested by clients is available.

The scenario is different for AWS. Only 23% of the traffic related to Amazon has an informative sub-domain. One can see in Figure 4.6 that only 3 among the top-14 AWS sub-domains in the datasets provide hints on the service generating the traffic. Such cases without informative names are indeed hard to be discerned and will require a much more elaborated classification methodology. I explore in the next chapters classifiers that correlate names of distinct flows, including both temporal and spacial correlations among flows.

## 4.8   Conclusions

This chapter provided a first look into traffic classification for modern web services. I visually explored how hostnames and addresses are associated, and studied the role of IP addresses in classification. My results show that up to 55% of web traffic can be identified relying solely on addresses. This coverage is however achieved only if the several hostnames used by services are uncovered, and the respective addresses are enumerated. For some specific services, IP addresses can classify most of the traffic. Those results call for the development of novel classification methods, which will operate with minimal information collected from the network, thus respecting users' privacy.

Nevertheless, I also pointed out that the association between hostnames and addresses changes frequently. For instance, for a selection of services, more than half of the addresses were changed during one year of observations.

This chapter identified several directions for further work. Firstly, I showed that a number of services shares hostnames, in particular those services hosted at cloud environments. The identification of services is not possible in such cases, even when flows are tagged with client-requested hostnames. Methods to classify this traffic are needed, and I will pursue that in the future. This motivates the design and implementation of algorithms to automatically retrieve the list of hostnames associated to services (i.e., the bags of domains) as well as to detect changes and to update the list over time. This scenarios are explored in the following chapters.

# Chapter 5

# WHAT: Automatic Accounting of Modern Web Services

The work I present in this chapter is mostly taken from my paper "*WHAT: A Big Data Approach for Accounting of Modern Web Services*" presented in the 2016 IEEE International Conference on Big Data (Big Data 2016) [108].

## 5.1 Introduction

Monitoring how web services are used and how they consume network resources is key to Internet Service Providers (ISP) when operating and planing the network. Similarly, companies have a vital need of monitoring their enterprise traffic – e.g., to limit the consumption of bandwidth, to spot sudden growth in usage of services, and to enforce corporate polices on allowed applications and services. With as much as 40% [19] of traffic generated by a corporation that is directed to webservices offering "shadow IT" services, i.e., cloud or SaaS applications, network managers lack of tools to understand and control network usage.

Monitoring is the key to understand, and traffic classification plays a fundamental role in knowing what applications and services are being accessed by observing what protocols and servers are being used. A variety of methods have been developed in the past [65, 15]. Today a large and growing fraction of information exchanges happening over the Internet is based on the HTTP(S) protocol, i.e., in the Web. Whether users are browsing the web, accessing business or leisure applications, using mobile or desktop applications, sharing or accessing content, chances are HTTP(S) is used to support the communication. The clear trend towards encryption by default [80] leaves in-network monitors with mostly layer-3 and layer-4 information, eventually augmented with the name of the server as obtained via DNS [7] or TLS handshake parsing. As illustrated in the previous chapter, even the identity of the server to which traffic is directed cannot be leveraged to associate traffic to specific user activities because (i) Content Delivery

Networks (CDNs) and cloud computing platforms co-locate multiple services and applications and (ii) websites, services and mobile applications generate HTTP(S) flows to different servers because of dynamic content, ads, plugins, and trackers, etc.

In this chapter, I address the challenge of accounting traffic to web-services. Specifically, I answer the questions of *what is the service a user intends to visit?* and *what is the traffic associated to such visit?*

WHAT (Web Helper Accounting Tool), the system presented in this chapter, starts from a simple flow level trace annotated with the domain name of the servers. It addresses the first question by automatically singling out *core domains* as representative of the web services that users originally intended to visit. The second question is addressed by identifying the *support domains* that are subordinate to each core domain as a result of downloading pictures, plugins, videos, ads, and then triggering traffic to hosts that serve it, e.g., servers of CDNs, clouds, etc. At this point WHAT is able to tally the overall volume of traffic associated to user intentions.

The novel approach introduced by *WHAT* is based upon the following steps:

- *Automatic identification of core domains*, based on a machine learning classifier that, as demonstrated from the results presented in this chapter, achieves excellent accuracy (higher than 96% in our evaluation).

- *Bag of Domains (BoDs) creation*: the BoD provides a model of the traffic generated by accessing a web service that is based on the unordered set of all possible support domains that may be triggered by the core domain visit. Ingenuity is required to weight support domains and avoid background traffic to pollute BoDs. *WHAT* successfully adopts text processing approaches to obtain representative BoDs.

- *Classification*: newly observed traffic flows are uniquely associated to a specific BoD, as core or support domains. This task is not trivial since some domains can be core in a BoD and support in others. For example, the same YouTube video can be accessed from the YouTube page or embedded in any web page.

The final output is the set of flows annotated with the core domain of the BoD they were associated to. Analytics can then be run to present statistics on the corresponding traffic. My contributions are:

- A novel technique to identify the web services users intend to access, and model to associate traffic to them;

- A fully working system implementing such technique and capable of applying it on simple flow level traces;

- A thorough performance evaluation, done considering both a 1.5 year long dataset collected from a live ISP network, and synthetic traces generated from actual browsing history of 30 volunteers;

- The application of *WHAT* to case studies, investigating how people consume YouTube videos embedded in other webpages, or how CDN or tracking services traffic is split among services.

The remaining of the chapter is organized as follows: Section 5.2 explains the problem in detail by posing examples of possible scenarios. Section 5.3 describes *WHAT* internals while Section 5.4 introduces and characterizes the datasets. Section 5.5 then presents results of evaluation and parameter tuning, while Section 5.6 reports my experience in deploying *WHAT* in a real operational network. Section 5.7 presents related work on traffic classification and user activity discovery. I conclude in Section 5.8.

## 5.2 Scenario & Problem Statement

I assume that a passive network monitoring infrastructure is in place and exposes per-flow information records. Beside traditional information such as flow identifier, client identifier, volume, timestamp etc., I assume each flow is annotated with the domain name of the server being contacted.[1] *WHAT* aims at classifying traffic flows according to the website that triggers them. It targets primarily accounting applications.

Consider a user browsing the web and visiting two services C1 and C2, as sketched on the top plot of Figure 5.1. Her browser opens dozens of TCP connections (arrows in the timeline) to issue HTTP/HTTPS requests to tens of servers in different domains to retrieve elements of the page such as images, CSS files, third-party objects, or scripts that, once run, may generate even more requests.[2] Some of these domain names may be completely different from the one of the originally visited service. I am interested in accounting all such traffic as *triggered by* the original service.

I call the originally requested domain the *core domain*, and those automatically triggered by core domains the *support domains*. The top plot in Figure 5.1 shows a sequence of flows generated by the user visting C1 and C2, over time. The plot consists in a timeline with arrows marking the instant of time flows to a server are started. Taller arrows are the core domains, while arrows with the same color are flows triggered by the core domain. The terminal the user is using may be running background applications that in turn may generate traffic flows which are not linked to the actual web services she is accessing to. I call those *background domains*.

The goal of *WHAT* is to account all traffic generated by a user visiting a core domain. For this, it relies on the list of possible support domains that may serve objects that are part of the main service. I call this list the *Bag of Domains* (BoD) for the given core domain.

---

[1]Annotation can be done using DNS traffic or TLS handshake, as I describe later in 5.3.2.

[2]I use the term "domain" informally throughout the chapter, meaning Fully Qualified Domain Name (FQDN).

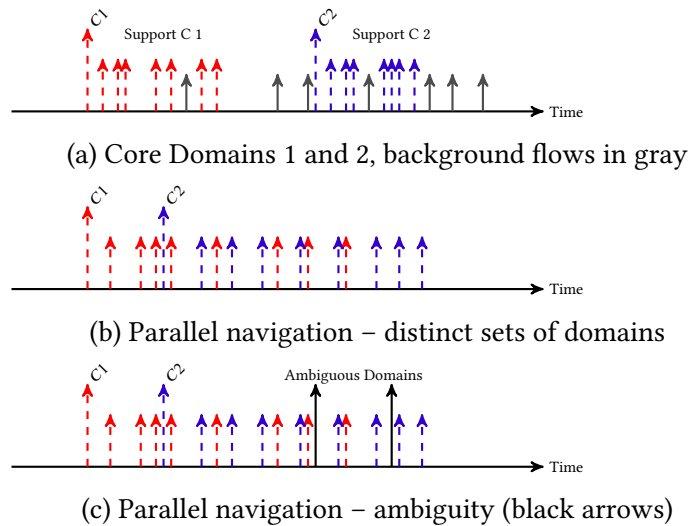

(a) Core Domains 1 and 2, background flows in gray

(b) Parallel navigation − distinct sets of domains

(c) Parallel navigation − ambiguity (black arrows)

Figure 5.1: Examples of entries (flows) in network traces and how *WHAT* behaves when labeling them.

## 5.2.1 Examples

Figure 5.1a depicts a simple case in which a user contacts two websites in two different moments, while some applications are generating background traffic − see red, blue and gray arrows. Flows to each core domain are seen first, followed by flows to support domains. In the hypothetical case of Figure 5.1a, there is no ambiguity among BoDs, and domains associated with background traffic. Every domain seen in the network belongs to exactly one BoD, while background domains are not part of any BoDs. Thus, *WHAT* annotates each flow whose domain is in the BoD of the core domain closest in time. I call these *best BoD*.

Figure 5.1b shows a scenario where parallel navigation takes place. For instance, the user may use multiple browsers or tabs to navigate through pages at the same time. Or multiple devices may share the same connection via, e.g., NAT. In these cases, the flow id information may not allow to identify which user/terminal/application/tab, i.e., client, is generating the traffic. Yet, if the BoDs of C1 and C2 have no common domain, *WHAT* can correctly associate flow to C1 and C2 by checking in which BoD they belong to.

Figure 5.1c depicts a scenario where flows with ambiguous names appear in the trace. Those are ambiguous domains (black arrows), and ingenuity must be used to link them to the correct core domain. I envision three major situations generating such cases, which are solved via heuristics.

- The ambiguous domain is a support domain, but it appears in several BoDs. *WHAT* assumes that the closest core domain is the right one − i.e., *C*2 is this example. The accuracy of this assumption when increasing the number of parallel navigations is evaluated later in this chapter.

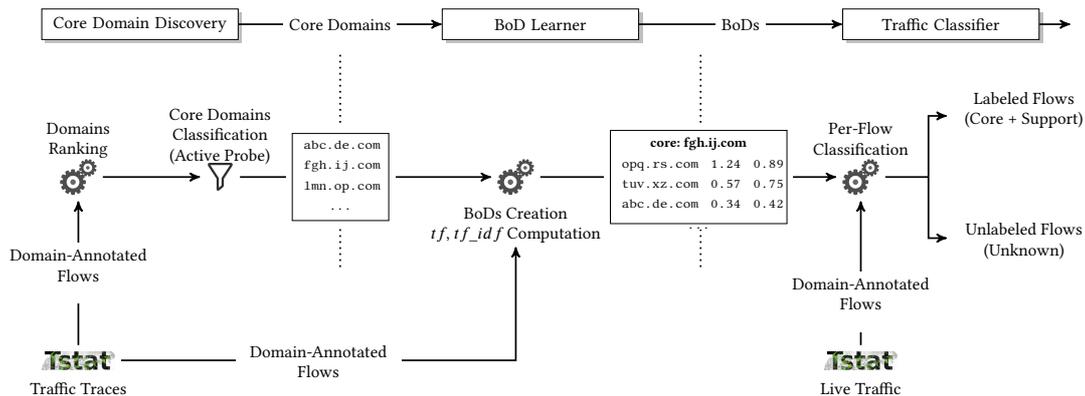

Figure 5.2: Architecture of *WHAT* to classify interactive web flows.

- A support domain is also a core domain. For instance www.facebook.com can be accessed directly by the user (as core domain), or as third-party service (e.g., a website embedding a www.facebook.com plugin). To disambiguate those cases, *WHAT* examines flows coming *before* and *after* the ambiguous entry. It calculates the chance for that flow to be a core domain that triggers a new independent navigation, or rather to be a support domain for the previous core domain.

- A domain is used by background and core services. This is the case of a background application accessing e.g., www.dropbox.com, or the user accessing it on the web. As before, *WHAT* compute a score to consider it support or background domain.

These simple examples clearly show how tangled the picture can be. Next, I describe the system design and how *WHAT* handles all these cases.

## 5.3 The WHAT System

### 5.3.1 Architecture Overview

*WHAT* is a supervised system. It first builds a model based on labeled data traces, and then uses the model to classify traffic online. It aims at defining models in an as much as possible automatic way, minimizing user intervention, and naturally adapting them to the usage scenario. Flow records exported by passive network measurement devices are the entities to be classified by *WHAT* (e.g., NetFlow, or logs collected by proxies). *WHAT* then assigns flows to the service (identified by its domain name) that most likely triggered them – i.e., with the mostly likely core domain.

Figure 5.2 summarizes *WHAT* architecture. It is composed of three modules: The *Core Domains Discoverer*, the *BoD Learner* and the *Traffic Classifier*. The former two modules use archived traces to train the system, whereas the latter module is deployed in the network to classify new flows from live networks. I next describe the expected input data format, followed by the working internals of each module in *WHAT*.

### 5.3.2 Input Data

*WHAT* is designed to receive ordinary flow records, such as exported by netflow. Given a flow $f$ (e.g., client and server IP addresses, ports and transport protocol), let $ts_f, te_f$ be the start and end timestamp, i.e., the time of the first and last packet. I assume that the initiator of the flow is the client, and the other end-point is the server. The knowledge of the network topology may ease this step if the direction of traffic is known *a priori* – i.e., all clients belong to known subnets. I assume that the flow record is enriched with information about the server FQDNs $d_f$ used by clients when obtaining the server IP address. Flow meters typically export information from the network and transport layers, missing the association between server IP addresses and FQDNs. Different methods can be used to annotate flow records with FQDNs. For example, DNS logs can be employed to extract queries/responses and annotate records in a post-processing phase. Equally, some flow meters export such information on-the-fly directly from the measurement point for popular protocols [51]. For instance, Deep Packet Inspection allows one to extract Server Name Identification (SNI) from encrypted TLS flows, or server `Host:` from plain HTTP flows.

In this chapter, I rely on Tstat [107] to collect data summarizing flows, using the monitoring infrastructure described in Section 2.

### 5.3.3 Core Domains Discoverer

The first task for training *WHAT* is to identify core domains, i.e., those domains $C = \{c_1, \dots, c_k\}$ that the users directly access to, and support domains, i.e., those domains $S = \{s_1, \dots, s_j\}$ which the application generates to fetch all objects that are part of the page. This is a classic classification problem, i.e., given a domain $d$, return if it is a core or support domain.

Instead of building a custom heuristic to solve the problem, *WHAT* classifies domains by means of a decision tree classifier [47]. The training of the decision tree is performed using a labeled dataset, in which a list of core and support domains is given, and features are extracted to characterize each of them. During training, the classifier builds an internal model (a decision tree in this case) and later uses it to classify a domain based on the sole knowledge of features.

The engineering of the decision tree requires ingenuity. First, I need to define the set of features to use. I opt for an extensive list guided by domain knowledge, and then let the classifier to choose which are those that better allow to separate core and support domains. Given domain $d = www.acme.com$, the system visits the main page at `http://www.acme.com/` by using the Selenium automatic browser, and automatically extract features [93]. Table 5.1 summarizes the list of features, giving a brief description and the expected behavior. For instance, for core domains, I expect the length of the main HTML response to be quite long, and to include large number of objects, possibly hosted in different domains, with the overall page resulting quite sizable. I expect the domain to start with www, and eventually accept a redirect to the same domain, i.e., HTTP response code can be 2xx or 3xx, but not 4xx. I expect the server to be a well-known solution, and serve a HTML page. Finally, core domain flows should appear separated in time from previous flows due to user think time. In practice, however, I expect the separation between core and support domains to be much blurred.

Given the list of core and support domains, I build a labeled dataset that I use for training and testing. I opt for the J48 implementation of the C4.5 algorithm offered by Weka [47]. Interestingly, the final decision tree results in a very simple, efficient, and intuitive model which I report below:

```
if (HTML resp. length <= 3375B) then Support
else
       if (redirect to == other) then Support
       else Core
```

Despite its simplicity, performance shows that overall accuracy is higher than 96% when tested against 1000 domains. Details are provided in Section 5.5, along with parameter sensitivity. Notice the high number of classes, making this result particularly remarkable, where traffic classifiers typically chose among few classes representing L7 protocols (e.g., HTTP, Peer-To-Peer, etc.).

Aiming at a self-configuration, during bootstrapping phase *WHAT* automatically builds the list of the most popular domains from the vantage point. Then it runs an active crawling phase to extract features and uses the classifier to select the set of core domains $C$.

### 5.3.4  BoD Learner

**Evaluation Window and BoDs:**

Given the set of core domains $C$, *WHAT* needs to learn the BoD $B_c$ for each $c \in C$.[3]

---

[3]I process domains to keep only *stems*: I replace numbers by a constant (e.g., `123-edge.acme.com` becomes `D-edge.acme.com`) and isolated characters by constants (e.g., `acme-a.cdn.net` becomes `acme-C.cdn.net`).

Table 5.1: Features extracted for a domain and used for classifying core and support domains.

| Feature | Core | Support |
|---|---|---|
| HTML resp. length | long | short |
| Object number | large | small |
| Domains in page | large | small |
| page size | large | small |
| www. in $d$ | likely | unlikely |
| Redirect to | same domain | other domain |
| HTTP resp. code | 2xx, 3xx | 1xx, 4xx, 5xx |
| Server | Apache, ISS | ngix, custom |
| Content-Type | text/html | application/xml, other |
| $\Delta T$ from previous | long | short |

*WHAT* considers the flow trace generated by each client, e.g., all flows generated by the same client IP address. Intuitively, after a client requests a web page, support domains are contacted, and I see flows annotated with domains in the traces (see Figure 5.1). Thus, the BoD can be learned by inspecting domain names of flows initiated immediately after the core domain flow. For this, I could rely on active experiments, e.g., randomly visiting pages in $c$, and then extracting names. While active experiments would generate a clean and reliable dataset, they suffer from limited (i) temporal and (ii) spatial scope, which questions the representativeness of the data. For instance, different pages in $c$ may present different content, with possibly user authentication complicating the access to internal pages. Similarly, the browser, device, or application being used may change the content being served.

An intuitive solution is to let *WHAT* extract the BoD from the passive traces directly at the vantage point, i.e., learning (and updating) the BoDs from the data the system is exposed to. Passive traces naturally factor all the above issues. *WHAT* considers valid triggers those flows directed to $c$ that appear after a idle period $\Delta T_{idle}$, i.e., likely due to a new user visit. When a trigger is observed, *WHAT* extracts all domains found in the time window following it. I call this the *Observation Window*, *OW*, of duration $\Delta T_{OW}$. A domain $d$ appearing in the OW becomes part of the BoD of $c$ ($B_c$) as support domain. In Figure 5.3, this is represented by coloring flows with the same color of the core domain. The longer $\Delta T_{OW}$, the more information is collected, with the chance to pollute the $BoD_c$ with *false support domains*. Algorithm 1 shows a pseudocode for the BoD update function. Traces from each client contribute to learn $B_c$.

There are however two aspects to be taken into account when learning the BoDs. First, support domains may appear immediately after visits to core domains, but also quite separate in time (e.g., a page reload, or a video being displayed after an ads, etc.). The time period $\Delta T_{OW}$ *WHAT* searches for support domains needs to be carefully chosen. I will discuss parameter settings in Section 5.5.

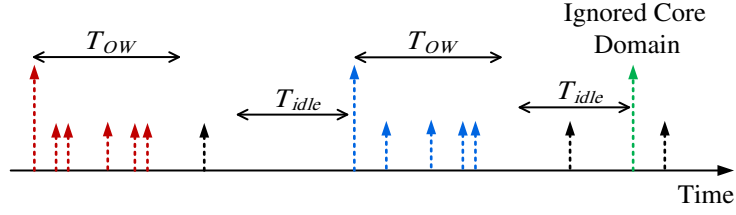

Figure 5.3: BoD Learning: a flow to a core domain triggers a new observation window if client was idle for more than $\Delta T_{idle}$.

---

**Algorithm 1** $BoD\_update(f, C, BoDs)$

---

**Input:**

    $f$                                        ▷ *The current flow*

    $C = \{c_1, ..., c_k\}$                 ▷ *Core Domains*

    $BoDs = \{BoD_{c_1}, ..., BoD_{c_k}\}$     ▷ *BoDs of core domains in C*

1:  $t = GetTime()$                           ▷ *Get current time*

2:  $d_f \leftarrow parse(f)$                         ▷ *Get the domain of f*

3:  $(t_c, c) \leftarrow \Delta OW$              ▷ *Retrieve current OW if any*

4:  **if** $OW \neq \varnothing \wedge t - t_c \geq T_{OW}$ **then**

5:     $OW \leftarrow \varnothing$                 ▷ *Remove the OW if expired*

6:    *// Put domains in the Bag if OW exists*

7:  **if** $OW \neq \varnothing$ **then**

8:     $BoD_c(d_f) += 1$

9:  **else**

10:    **if** $d_f \in C \wedge t - t_{last} > \Delta T_{idle}$ **then**

11:      $OW \leftarrow (t, d_f)$             ▷ *Open a new OW*

12:      $freq_{d_f} += 1$              ▷ *Update CD frequency*

13: $t_{last} \leftarrow t$                      ▷ *Update last flow time*

---

Second, not all support domains appear after every request to a website. More dangerous, background traffic and support domains triggered by other core domains may appear nearby $c$ by chance, poisoning $B_c$ with *false support domains*. *WHAT* needs then to observe a large number of OWs to accumulate support domains, and select those that are actual support domains. The assumption is that support domains emerge, whereas the irrelevant ones (including background and false domains) can be filtered out by means of thresholds and domains scores.

**Domains Score:**

The key idea is that domains that are triggered by a core domain should appear more frequently in observation windows than other domains. I leverage text processing methodology to implement a filtering process based on this idea. Even if text processing and traffic classification lie on different domains, in both cases we face the problem of information retrieval. I will show that text processing can be successfully used (with some modifications) for traffic classification too. Here, I rely on the $tf - idf$ (term frequency – inverse document frequency) of domains in bags to represent the scores. The $tf - idf$ is used in information retrieval to evaluate the importance of a word to a document in a collection. A word is more important when it appears often in a document (captured by the $tf$), but its importance is reduced by a factor representing how frequent the word appears in other documents in the collection (captured by the $idf$).

In this problem, a document is a BoD $B_c$ for the core domain $c$, a word is a domain name $d \in D$ and the collection of documents is the set of all bag of domains $BoDs$. Domains triggered by a single core domain should have high $tf$ and high $idf$, domains that are triggered by many core domains (e.g., advertisements) should have high $tf$ but low $idf$, while domains related to background traffic should have low $tf$ and low $idf$:

$$tf(d, B_c) = \frac{\sum_{W \in OW_c} |d \in W|}{|OW_c|} \tag{5.1}$$

$$idf(d, BoDs) = \log \frac{|BoDs|}{|B_c \in BoDs : d \in B_c|} \tag{5.2}$$

$$S(d, B_c) = tf(d, B_c) \times idf(d, BoDs) \tag{5.3}$$

where $tf(d, B_c)$ is the number of times $d$ appears in any observation window $W$ for the core domain $c$, normalized by the number of observation windows. Note that $tf(d, B_c)$ can be greater than 1, since a support domain $d$ can appear multiple times in the same observation window. $idf(d, BoDs)$ is the logarithm of the ratio between the number of BoDs in the collection and the number of BoDs containing $d$. Thus, the more BoDs a domain appears into, the closer to zero $idf(d, BoDs)$ is, and thus the smaller $S(d, B_c)$ is.

The outcome of the training phase is the creation of a BoDs for each core domain $c \in C$. Each domain $d \in B_c$ is associated two scores, namely

$$B_c = \{(d, tf(d, B_c), S(d, B_c)) | d \in D\}. \tag{5.4}$$

If $d$ appears in all BoDs, then $idf(d, BoDs) = 0$ and $S(d, B_c) = 0$, suggesting its presence is insignificant to characterize the document. Similarly, if $d$ does not appear in any observation window in $OW_c$, $tf(d, B_c) = 0$ and $S(d, B_c) = 0$. *WHAT* uses the $tf(d, B_c)$ score to filter out those core domains which appear too infrequently, i.e., $tf(d, B_c) < MinFreq$, since those are likely to be background or false support domains, i.e., domains appearing in the BoD by chance. Trade-offs are explored in Section 5.5. The score $S_{d, B_c}$ allows *WHAT* to assign ambiguous domains that appear into two BoDs (cfr. Figure 5.1c) during classification. In the following section I give details.

### 5.3.5 Traffic Classifier

Once core domains are identified, and their respective BoDs are built, *WHAT* processes traces to assign flows to the most likely core domain. It gets as input the set of flows $F$ generated by a single client, and processes them according to their generation time. Being designed for accounting, real time processing is not a main constraint, even if scalability is easy to obtain given *WHAT* operates on a per-flow and per-client basis.

*WHAT* uses Algorithm 2 to classify each flow $f$. It receives the core domains $C$, the BoDs and set of flows $F$ generated by a client. It then outputs flows annotated with the core domain that generated them, or as *unknown* in case no association is found.

---

**Algorithm 2** *classify* ($C$, $BoDs$, $F$)

---

**Input:**
    $C = \{c_1, ..., c_k\}$              ▷ *core domains*
    $BoDs = \{B_{c_1}, ..., B_{c_k}\}$      ▷ *BoDs of core domains in C*
    $F = \{f_1, ..., f_n\}$          ▷ *list of flows of a client to be classified*
**Output:**
    $O = \{(f_1, l_1), ..., (f_n, l_n)\}$     ▷ *labeled flows*
**Parameters:**
    $\Delta T_{EV}$                   ▷ *timeout without flows to expire BoDs*

1:  $W \leftarrow \varnothing$                   ▷ *set of currently active EVs*
2:  $O \leftarrow \varnothing$
3:  **for** $f \in F$ **do**
4:     // *retrieve start/end times and domain name of f*
5:     $ts_f, te_f, d_f \leftarrow parse(f)$       ▷ *$ts_f$ is also current time*
6:     // *remove expired EVs*
7:     $W \leftarrow \{(ts, te, c_i, B_{c_i}) \in W \mid ts_f - te \leq \Delta T_{EV}\}$
8:     // *obtain the best neighbor BoD among the active ones*
9:     $w_{best} \leftarrow \{(ts, te, d_f, B)\} \leftarrow BestBoD(ts_f, d_f, W)$
10:    $c = d_f$              ▷ *$d_f$ is the core domain of the best window*
11:    **if** $c \in C \wedge valid\_core(c, ts_f, w_{best}, F)$ **then**
12:       // *start an evaluation window for core domain c*
13:       $W \leftarrow W + \{(ts_f, te_f, c, B_c)\}$
14:       $O \leftarrow O + \{(f, c)\}$
15:    **else**
16:       **if** $w_{best} \neq \varnothing$ **then**
17:         $O \leftarrow O + \{(f, c)\}$
18:         // *enlarge time boundary of best EV*
19:         $te(w_{best}) \leftarrow \max(te_f, te(w_{best}))$
20:       **else**
21:         $O \leftarrow O + \{(f, \text{``}unknown\text{''})\}$

---

The algorithm is based on the concept of *Evaluation Window* (EV), i.e., the time window during which a support flow can appear after the observation of a core domain. For this, the algorithm maintains a list of active EVs, $W$. The list grows as new core domains are observed (lines 11-14), and entries are aged out based on a timeout $\Delta T_{EV}$, i.e., window ending time $te = \max_{f \in W} te_f$ is elapsed by at least $\Delta T_{EV}$, (line 7).

Differently from the training phase, the evaluation window duration is extended during classification. This happens when new support domains are found (lines 18-19). The rationale is that flows to support domains may be observed long time after the core domain since the terminal keeps downloading objects due to a user action (e.g., scrolling a webpage that triggers the download of new elements), or to the application fetching further blocks of data (e.g., a video player using adaptive streaming and downloading new portions of the video).

In case multiple active windows are alive, *WHAT* checks which is the most likely one using the function *BestBoD*() (line 9). This is detailed in Algorithm 3. I checked different options, and opted for a "closest in time" criteria: *WHAT* looks for the closest active window among $W$, and for which the domain $d_f$ of $f$ has a frequency above the $MinFreq$ threshold (lines 2-5).

---

**Algorithm 3** $BestBoD(ts_f, d_f, W)$

---

**Input:**
 $ts_f, d_f$              ▷ *start time and domain of flow to classify*
 $W = \{(ts, e, c, B)\}$            ▷ *set of active BoDs*
**Output:**
 $O = (ts_o, te_o, c_o, B_o)$          ▷ *best BoD*
**Parameters:**
 $MinFreq$          ▷ *Minimum $tf$ score for valid support domains*

 1:  $w_{cand} \leftarrow \emptyset$; Min$=\infty$
 2:  **for all** $\{w_i = (ts, te, c, B) \in W | tf(d_f, B) > MinFreq\}$ **do**
 3:   **if** $((ts_f - ts) \leq$ Min$)$ **then**
 4:    Min$\leftarrow (ts_f - ts)$
 5:    $w_{cand} \leftarrow w_i$
 6:  $O \leftarrow w_{cand}$            ▷ *take most recent EV*

---

At last, the most challenging problem is to resolve the ambiguity for a domain that may be both support and core domain. Recall this is the case of www.facebook.com, cfr. Figure 5.1c. To disambiguate, *WHAT* relies on the scores $S(d, B)$. Algorithm 4 provides details. It first considers $d_f$ a possible core domain. It computes the *core* score as the sum of $S(d, B)$ for those flows in the EV *after* the current flow time $ts_f$. As usual, only terms above the $MinFreq$ threshold are considered (lines 2-4). Next, it consider $d_f$ a possible support domain for the best candidate core ($w_{best}$) and computes the *support* score as the sum of $S(d, B)$ for flows against $w_{best}$ (lines 6-8). It consider both those *past* and *future* flows. The rationale is that EV would be extended in case $f$ becomes a support flow. At last, *WHAT* compares *core* and *support* to classify $f$ (lines 9-12).

---

**Algorithm 4** $valid\_core(d_f, ts_f, w_{best}, F)$

---

**Input:**

    $d_f, ts_f$                                           $\triangleright$ *candidate core domain and current time*

    $w_{best} = (ts, te, c, B)$                                $\triangleright$ *the best BoD*

    $F = \{f_1, ..., f_n\}$                         $\triangleright$ *list of flows of a client to be classified*

**Output:**

    $Valid$                                     $\triangleright$ *TRUE if the domain is a valid core*

**Parameters:**

    $\Delta T_{EV}$                               $\triangleright$ *timeout without flows to expire BoDs*

    $MinFreq$                      $\triangleright$ *Minimum $tf$ score for valid support domains*

1:    // *Possible core: compute score against $B_{d_f}$ for future flows*
2:    **for all** $\{f_i | ts_f \leq ts_{f_i} \leq ts_f + \Delta T_{EV}\}$ **do**
3:        **if** $(tf(d(f_i), B_{d_f}) > MinFreq)$ **then**
4:            $core \leftarrow core + S(d(f_i), B_{d_f})$
5:    // *Possible support: compute the score against $w_{best}$*
6:    **for all** $\{f_i | ts \leq ts_{f_i} \leq ts_f + \Delta T_{EV}\}$ **do**
7:        **if** $(tf(d(f_i), B) > MinFreq)$ **then**
8:            $support \leftarrow support + S(d(f_i), B)$
9:    **if** $(core > support)$ **then**
10:       $Valid \leftarrow TRUE$
11:    **else**
12:       $Valid \leftarrow FALSE$

---

## 5.4    Datasets

For training and testing I build upon two datasets. The first one is a passive trace collected from a large ISP network. It represents a realistic scenario of possible *WHAT* deployment. Almost no ground truth is available. To then thoroughly assess classification performance, I build a second dataset made of synthetic traces where I have the full ground truth knowledge. I use it to build benchmarking datasets and challenge *WHAT* classification capabilities.

### 5.4.1    ISP Flow Traces

This dataset includes flow summaries exported by Tstat in a real deployment. Measurement and collection methodologies are described in Chapter 2. In this chapter, I consider data from January 2015 to April 2016 for the ADSL dataset, containing flow summaries of 10,000 ADSL subscribers. Considering only HTTP and HTTPS TCP flows, I obtain 13.25 billion flows related to around 18 million domains. Data is stored in a Hadoop cluster for scalable processing.

To give the picture of how extended the web is nowadays, Figure 5.4 provides basic statistics about the trace. Figure 5.4a shows the growth of the number of unique domains over time. More than 18 million domains have been contacted by users via TCP.

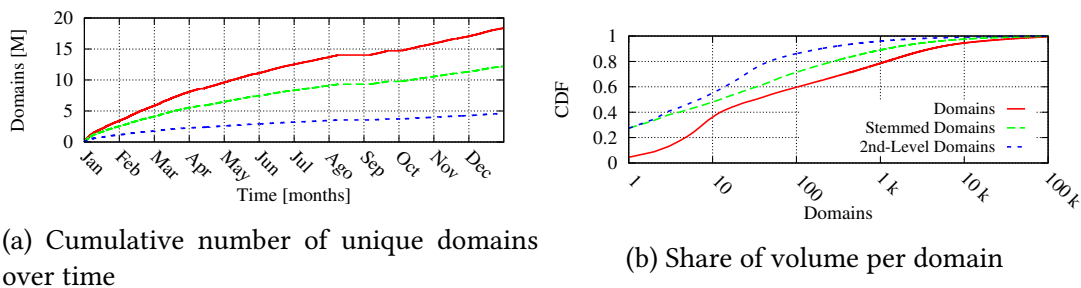

(a) Cumulative number of unique domains over time

(b) Share of volume per domain

Figure 5.4: Characteristics of the flow traces for the ADSL dataset. Most traffic is due to 10,000 domains. The probe suffered two outages during August and October 2015.

Some services rely on dynamic names (e.g., in CDNs) and inflate figures. For instance CloudFront and Gstatic, Amazon's and Google's CDN, account for 10% (1.8M) and 26% (4.7M) of the unique domains, respectively. My strategy to stem names partly reduces the number of unique domains. Yet, more than 11 million stemmed names are in the dataset. Similarly, the figure grows even limiting the domains to the 2nd-level domain name, with about 5 million names seen in 2015. These numbers show how complicated the picture is, and discourage the usage of synthetic traces for training and support the need to automatically train *WHAT* on the field.

Looking at the traffic share per domain,[4] Figure 5.4b shows the classic Pareto rule: most traffic is associated to few domains, with ≈ 30% coming from a single 2nd-level domain – i.e., `googlevideo.com`, a support domain serving YouTube videos. The top-10,000 (0.05%) domains serve more than 95% of the traffic, with the remaining ones in the tail contributing negligible volume each. Most of the negligible domains appear in the trace just once, and are typically machine generated domains, where the client embeds some information in the third/fourth-level domain. This plays in favor of *WHAT* which targets primarily accounting.

### 5.4.2 Synthetic Traces

To assess classification performance, I create a labeled dataset using data from volunteers, a mix of students, colleagues, and friends. I collect browsing histories of 30 users, extracting all visited URLs directly from `SQLite` databases used by Safari, Chrome and Firefox. Browsers log visited URLs and the time of visits. These are core domains, since users did actually visit the URLs. Some browsing histories spawned years and included more than 50 000 pages. I extract (up to) the most recent 5 000 URLs visited by each user, obtaining a dataset with more than 100 000 visits to 3 759 distinct domains. These form the ground truth of core domains.

---

[4]I compute the volume as the bytewise total amount of data exchanged. Any statistic can be computed once flows are annotated with the core domain.

Browsing histories allow me to characterize also users' browsing habits. I use those to obtain Cumulative Distribution Function (CDF) of the idle time between consecutive visits, and then use this to generate synthetic but realistic traces for benchmarking. I observe that 90% of the visits are separated by less than 1 min, with the remaining visits spread in a tail representing long pauses in volunteers' navigation.

To obtain a set of support domains, I revisit each URL by instrumenting a Firefox browser with Selenium. I let Selenium visit the URL, and then wait until the page is fully loaded (i.e., the `On Load` event is fired). The next URL in the list is then loaded after the browser is inactive for 1 second. Note that this creates artifacts, e.g., eventual video playout are stopped after 1 s from start. In parallel, Tstat records flows seen in the network, saving the same information available in the passive flow traces. This is a labeled trace, where 100,000 URLs are visited, referring to 3,759 core domains, and 9,764 support (possibly ambiguous) domains. Note how negligible those are compared to numbers in Figure 5.4. Crawling was done in April 2016 and lasted 5 days.

### Web Browsing Benchmark

In this first synthetic trace, each user is identified by an different IP address, and she visits URLs following the original sequence found in browsing histories. Inter-visit time is extracted according to the CDF distribution extracted from browsing histories, mimicking then user's idle time.

After the core URL visit, I populate the trace with support flows as from recorded by Tstat during the Selenium navigation. Timestamps are shifted to maintain the inter-arrival times of support flows, and IP addresses adjusted per user.

The traffic share per domain of the resulting trace is depicted in Figure 5.5a. Contrast this figure to Figure 5.4b. I see that this synthetic trace follows some patterns also seen in the actual flow traces, with around 10 000 domains making the majority of traffic, but missing the long tail. Most notably, it misses also the Pareto rule, i.e., the concentration of traffic around few domains serving videos (see left-most points in the figures). This is due to Selenium artificially stopping all video playout after 1 s. This trace is thus mostly representative of simple web browsing. Considering the tail, it is considerbly shorter than the real case due to the reduced size of the data.

### Web and YouTube Benchmark

To increase the effect of video streaming I form a second dataset. I revisit all YouTube URLs found in browsing histories of volunteers and let Selenium to play videos for longer time, i.e, to 5 min of each video, or until the video is over.

I then augment the web browsing trace by injecting YouTube video playing events. I form this trace with a worst case scenario in mind, in which all simulated users have YouTube videos going on at random moments.[5] No restrictions on starting time of videos are imposed. In fact, I determine starting time of videos independently of interactive visits to other URLs, thus mimicking videos played in background.

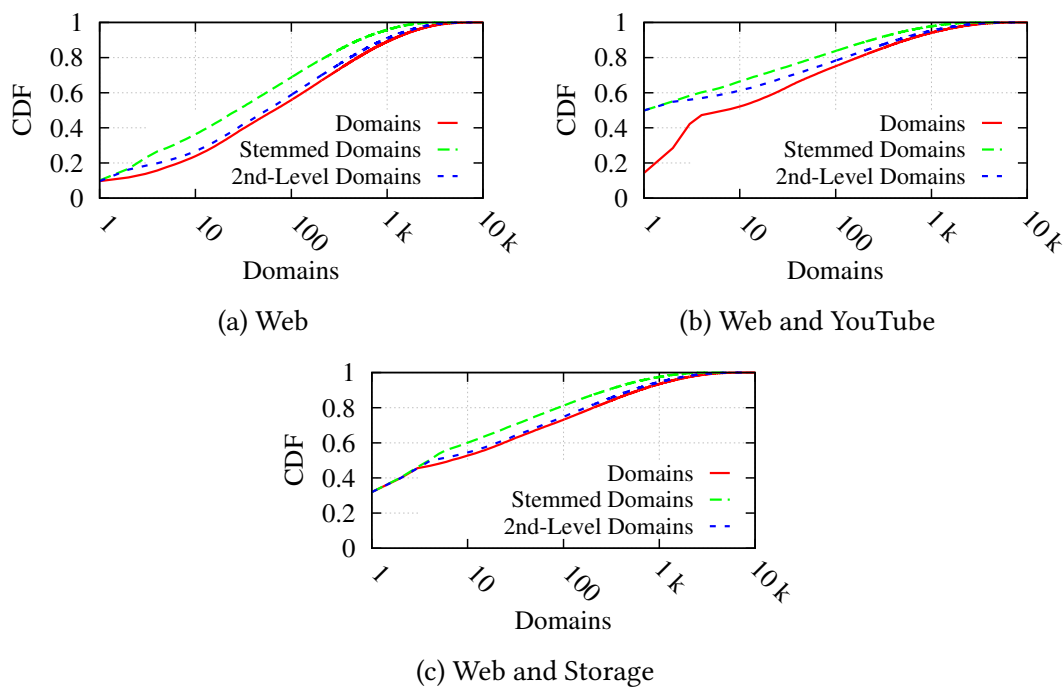

(a) Web

(b) Web and YouTube

(c) Web and Storage

Figure 5.5: Share of volume per domains in synthetic traces.

To determine the time between user video sessions, I rely on information coming from volunteers' browsing histories as previously done. Flows triggered by videos during the tests with Selenium are then added to the synthetic trace. Again, timestamps and IP addresses are adjusted accordingly.

The trace combining both interactive web and videos has the characteristics summarized in Figure 5.5b. Here I see that 50% of the traffic is related to `googlevideo.com` domain. It overestimates the share of video when compared to real traces (Figure 5.4b). This trace is a challenging scenarios, since all users are having web and YouTube sessions in parallel.

**Web and Storage Benchmark**

Since Browsing histories record only URLs visited within browsers, the trace misses any possible flows generated by other applications. To then simulate the presence of background flows, I characterize the traffic of cloud storage services. I use traces presented in [11]. In short, 11 services, including Dropbox, Microsoft OneDrive and Google Drive, have been installed in a testbed. Active experiments have been performed for around 3 months to characterize traffic of each service while idle and while handling workloads of different types (e.g., upload and download of single file, batches of files etc). I reprocess all traces presented in [11] with Tstat, thus creating a dataset with typical flows of background cloud services.

---

[5]In a real case, the chance of video playing during browsing would be much lower.

The third synthetic trace combines interactive web with cloud storage traffic. I create an extreme scenario again, in which all users have one cloud storage application constantly open in background, running idle.

Besides that, I add flows generated by cloud storage applications when actively exchanging content with servers. The arrival time of such flows is determined according to the model proposed in [44]. As before, flows are added to the web browsing trace. Figure 5.5b shows the resulting share of traffic per domain. I can see a high percentage of traffic to some few domains – i.e., cloud storage ones. This trace challenges *WHAT* to discriminate background traffic.

**NAT Scenarios**

The last synthetic traces simulate several users behind a NAT, so that the client IP address is not anymore a reliable user identifier. In a nutshell, I take the previous synthetic benchmark traces, and create a number of user replicas. Each replica starts navigation at a random position in the browsing history, looping back to the beginning when the list is over. *k* replicas are assigned the same IP address to simulate the presence of a NAT, so that flows from independent sessions are now multiplexed over the same timeline. This simulates multiple users browsing the web, but with the same identifier, e.g., behind a NAT, using multiple tabs, browsers, or devices.

## 5.5  WHAT Validation

### 5.5.1  Core Domain Discovery

I start by evaluating the performance of the decision tree used for Core Domain Discovery.

I need to build a labeled dataset to train and test the classifier. I consider the first 10 days of March, 2016 from the ISP flow trace and manually inspect the domain names. Following *WHAT* bootstrapping phase, I start from the ones with the highest traffic share, cfr. Figure 5.4b. I visit the front web page of each of them (if any), and manually mark domains as core or support. The procedure is followed until a dataset of 500 core and 500 support domains are found. As such, samples for the two classes are balanced, allowing a fair use of classification algorithms. I use Selenium and Tstat to automatically extract features listed in Table 5.1 for all 1,000 domains.

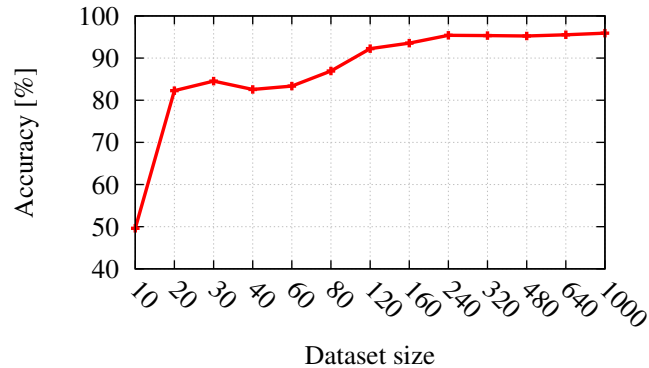

Figure 5.6: Core domain discovery accuracy per BoD learning period.

I use this labeled dataset to train and test the J48 decision tree with Weka, following the 10-fold cross validation to assess accuracy, i.e., the percentage of domains correctly classified. Results are in Figure 5.6. Points in the figure mark 10-fold cross-validation experiments, which are performed with an increasing training and testing dataset size (*x*-axis). For instance, for $x = 100$, I have randomly selected 100 domains, and used them for training a new tree.[6] The remaining $1000 - x$ domains are used for testing. Average accuracy is then computed over ten rounds.

Results let me conclude that even a moderate number of domains is sufficient to train the Core Domain Discovery module. Indeed, I see that 120 domains allow *WHAT* to reach accuracy above 90%. When 1000 domains are used, accuracy tops to 96%.

### 5.5.2 Classification Performance

I evaluate the performance of *WHAT* when classifying new flows – i.e., Algorithm 2. For next experiments, *WHAT* learns BoDs from ISP traces, and performance is assessed on synthetic benchmarks. I consider the set *C* of 500 most popular core domains as previously described and let *WHAT* learn the BoDs using the ISP trace starting from 9 a.m. of March $1^{st}$, 2016.

Figure 5.7 shows results for Web Browsing benchmarks. YouTube and Storage benchmarks are omitted since they lead to similar conclusions. The figure depicts the accuracy of *WHAT* when letting it learn BoDs from an increasing time. The *"optimal learning"* line marks the accuracy when *WHAT* learns BoDs from the same benchmark trace used for testing – i.e., a biased result that gives hints on the best possible performance of the algorithm in this benchmark.

---

[6]To avoid over-fitting, I use J48 decision tree with binary split and 10 minimum number of instances per leaf.

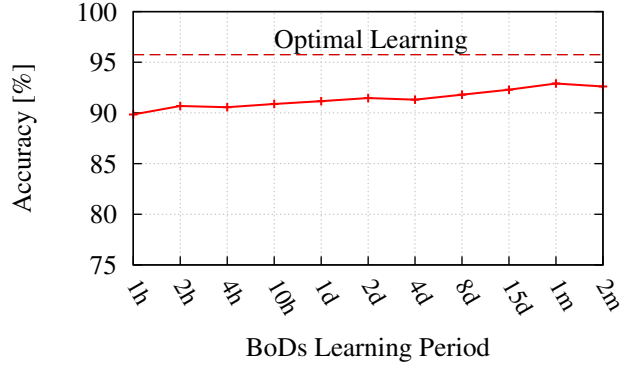

Figure 5.7: Accuracy vs. learning dataset size. BoDs learning starts on March $1^{st}$, 2016.

Accuracy in these experiments is computed as *the percentage of volume* in bytes against the correctly labeled benchmark trace. Therefore, errors occur either because flows have been labeled with wrong core domains, or mislabeled as "unknown". The same figures are obtained when considering the *percentage of flows*, i.e., the errors are equally distributed across heavy hitters (e.g., video or storage), and less frequent (and voluminous) domains.

Focusing on the left-most point in Figure 5.7, note that *WHAT* correctly classifies 90% of the traffic volume with a learning set of 1 day only. That is, most of the popular BoDs are learned by observing a single day of traffic in such medium-sized PoP/ISP. Increasing the learning set marginally improves results, with the best accuracy at around 93% with a 1-month long learning set. Classification errors are due to (i) *WHAT* heuristics to disambiguate domains; and (ii) domains seen in the benchmark that are not in the ISP flow dataset. Yet, overall *WHAT* delivers very promising results.

Figure 5.8 presents the accuracy of the three benchmarks in NAT scenarios. Results are obtained by increasing the number of users in the synthetic NAT trace. A similar trend emerges for the three benchmarks. More user aggregation reduces the performance of *WHAT*. This is not a surprise, since users navigating in parallel increase the probability of support domains to become ambiguous. Overall, *WHAT* performs very close to its best accuracy when up to five users are active at the same time. The accuracy drops to less than $\approx 70\%$ when more than 20 users are aggregated. Finally, note that the presence of background traffic such as Youtube or Storage marginally reduces accounting capabilities despite injecting sizable amount of traffic.

### 5.5.3 Parameter Tuning

*WHAT* relies on a number of parameters, which are evaluated next to understand their effects on classification accuracy. I discuss Evaluation Window ($\Delta T_{EV}$) and the minimum $tf$ score to include domains in BoDs (*MinFreq*), since they have the highest impact on the system. Best choices for all parameters, including those omitted for the sake of brevity, are listed in Table 5.2.

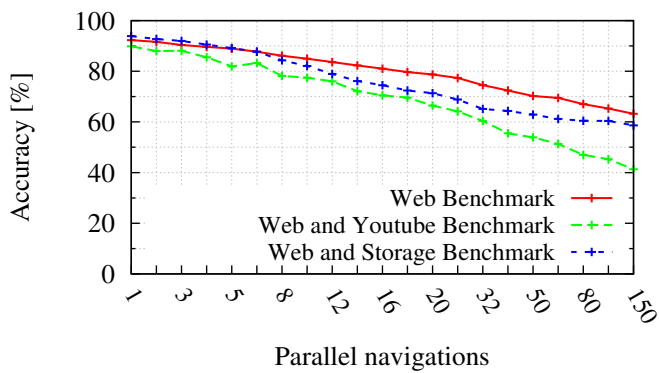

Figure 5.8: Accuracy of benchmarks in NAT scenarios.

Table 5.2: Best choices of parameters.

| Parameter | Best Value |
|---|---|
| Training set size | 1 Month |
| $\Delta T_{OW}$ | 10 Seconds |
| $\Delta T_{idle}$ | 1 Minute |
| $\Delta T_{EV}$ | 2 Seconds |
| $MinFreq$ | 2% |

For experiments in this section, *WHAT* is trained with one month of traffic from the ISP flow traces, and tested with the various benchmarks. Accuracy is calculated as in the previous section − i.e., the percentage of bytes in benchmarks that is labeled with the correct core domains.

**Evaluation Window ($\Delta T_{EV}$):**

Figure 5.9 depicts how accuracy varies according to $\Delta T_{EV}$. Lines represent results for four benchmarks. Results for the NAT scenario are calculated considering five users generating web traffic behind a NAT.

Focusing on the Web Browsing benchmark (red line), notice how the accuracy starts at $\approx 80\%$ when $\Delta T_{EV} = 0.1$ s, grows at the best figures (e.g., $\approx 90\%$) when $\Delta T_{EV} = 5$ s, and consistently decreases for larger values. Very small values of $\Delta T_{EV}$ cause *WHAT* to miss support domains, whereas large $\Delta T_{EV}$ values increase the chance to account for background or unrelated flows.

Similar pattern is observed for the YouTube benchmark (green line). However, accuracy slightly decreases in this case, and the system performance degenerates faster when $\Delta T_{EV}$ is too large − compare the red and green lines in the right part of the figure.

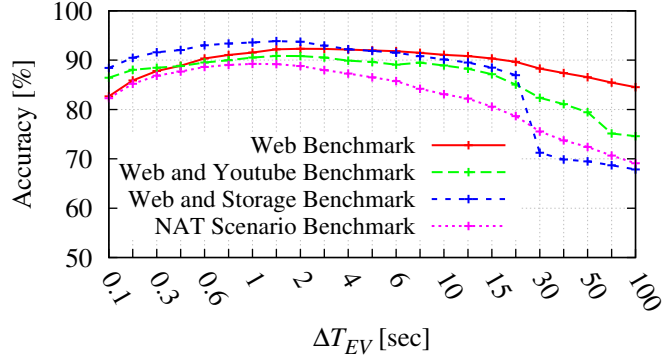

Figure 5.9: Accuracy vs. evaluation window.

It happens because YouTube BoD shares support domains with other Google services' BoDs. Large $\Delta T_{EV}$ increases the probability that a YouTube support domain becomes ambiguous because other BoDs related to Google are active. Similar conclusion arises for the Storage benchmark (blue line), where a jump is seen when $\Delta T_{EV} = 30$ s. This jump seems to be caused by periodic traffic of Google Drive, which would be misclassified with large $\Delta T_{EV}$. Again, this happens because support domains serving Google Drive are in BoDs of other Google services.

Finally, $\Delta T_{EV}$ becomes even more important in scenarios where traffic of multiple users are mixed in NATs. $\Delta T_{EV}$ needs to be carefully set, since accuracy decreases faster for large number of simultaneous users.

Overall, $\Delta T_{EV} \in [1,4]$ s provides the best trade-off.

**$MinFreq$ Threshold:**

The impact of $MinFreq$ threshold is illustrated in Figure 5.10. Curves for four benchmarks are depicted, and similar methodology as in previous section is used to calculate the accuracy. The $x$-axis marks the value of the threshold – e.g., $x = 2\%$ depicts results for which any domain with $tf$ lower than 2% in a BoD is not considered.

The importance of the $MinFreq$ to filter out noise from BoDs becomes clear. As an example, when $MinFreq$ is too large (e.g., 20%), domains that are very popular in BoDs may be ignored, resulting in a sharp decrease on accuracy. That is, I observe a reduction on accuracy in particular for YouTube benchmark where `googlevideo.com` domains are ignored.

On the other extreme, when $MinFreq$ is too low, unrelated support domains pollute BoDs. Focusing on results for $MinFreq = 0.05\%$, notice how accuracy is around 90% in the Web Browsing benchmark, but it is reduced to around 80% in the YouTube benchmark. This happens because background videos cause YouTube domains to randomly appear in many BoDs during learning. A similar pattern is not seen for Storage benchmarks because the tested services are less chatty than YouTube and, thus, their domains are filtered out even by a low *MinFreq*.

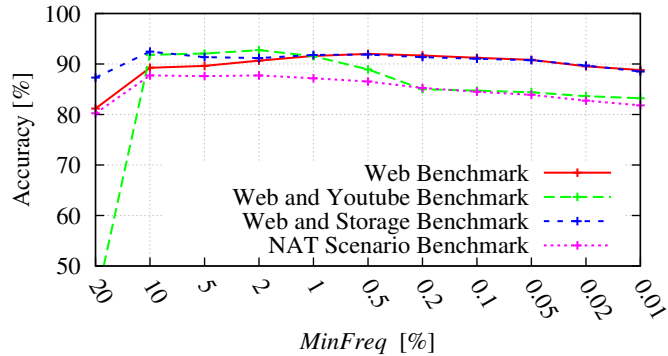

Figure 5.10: Accuracy vs. *MinFreq* threshold.

Overall, $MinFreq \in [1,5]\%$ provides best trade-off.

### 5.5.4   Stability of Learning

At last, I check how BoDs evolve over time, possibly giving insights about how frequently learning phase should be run. Figure 5.11 illustrates the stability of some example BoDs. The figure is produced by building BoDs for selected core domains using consecutive ten days long periods from the ISP flow traces. Let $B_c(i)$ the BoD of core $c$ in the i-th period. I compute the number of support domains in $B_c(i + 1)$ which were not present in $B_c(i)$.

Interesting, the BoD of very popular core domains, such as `www.facebook.com` and `www.google.it` are stable. This reflects the fact that the infrastructure of these giants slowly changes over time.

Instead, BoDs of smaller but still popular core domains show variations. For instance, `www.ilmeteo.it` (a popular weather service) and `www.libero.it` (a popular portal) show 5 to 40 new domains appearing in each BoD every ten days. Not shown here due to lack of space, I observe also 5 to 40 domains that disappear from the BoD. This is due to the dynamic domains used by trackers, e.g., by different advertisement campaigns served by different platforms over time.

Interestingly, two large spikes are seen. Checking the BoDs which include more than 500 support domains, I observe the sudden emerging of new support services ranging from tracking and ads platforms, CDNs and difficult to identify domains. *WHAT* helps to uncover the changes in the remote services.

These results show the need of updating the BoDs, which can be easily solved by periodically running the learning phase.

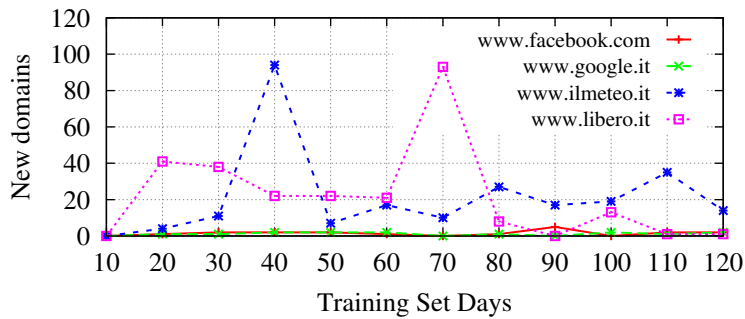

Figure 5.11: Delta of domains in time.

## 5.6   Case Study

In this section, I present a case study to illustrate the applicability of *WHAT*. The system is applied to the ISP flow trace considering 10 days at the beginning of March, 2016. All experiments focus on *C* containing the 500 most popular core domains, while BoDs are learned using the whole month of February, 2016.

### 5.6.1   Ranking Domains and Services

In this first case, I show how *WHAT* helps to account traffic to services. I present three traffic rankings in Figure 5.12: (i) *Traffic per domain* (Figure 5.12a) illustrates the volume grouped by mapping flows to domains; (ii) *Traffic to core domains* (Figure 5.12b) provides the above figure, but only considering core domains; (iii) WHAT *output* (Figure 5.12c) provides figures which account the volume of support domains to core domains.

Figure 5.12a illustrates how support domains dominate traffic share. Notice that domains are stemmed in the labels. Sub-domains of CDNs, such as `googlevideo.com`, `fbcdn.net` and `nflxvideo.net`, emerge among the main traffic sources.[7] Numbers inside plots mark the traffic observed for each domain.

While a network administrator could correlate traffic with services in the mentioned examples (i.e., YouTube, Facebook and Netflix), enumerating all support domains of a service requires a time-consuming manual process. Moreover, it misses the fact support domains may be triggered by other services – e.g., YouTube videos embedded in Blogs or News web sites.

Figure 5.12b shows the result of a possible (naive) approach, of focusing only on core

---

[7]`isDcdn.se.skyvod.cdn.xxxxxx.it` is a CDN serving Sky IPTV: 2nd-level domain is anonymized for privacy reasons.

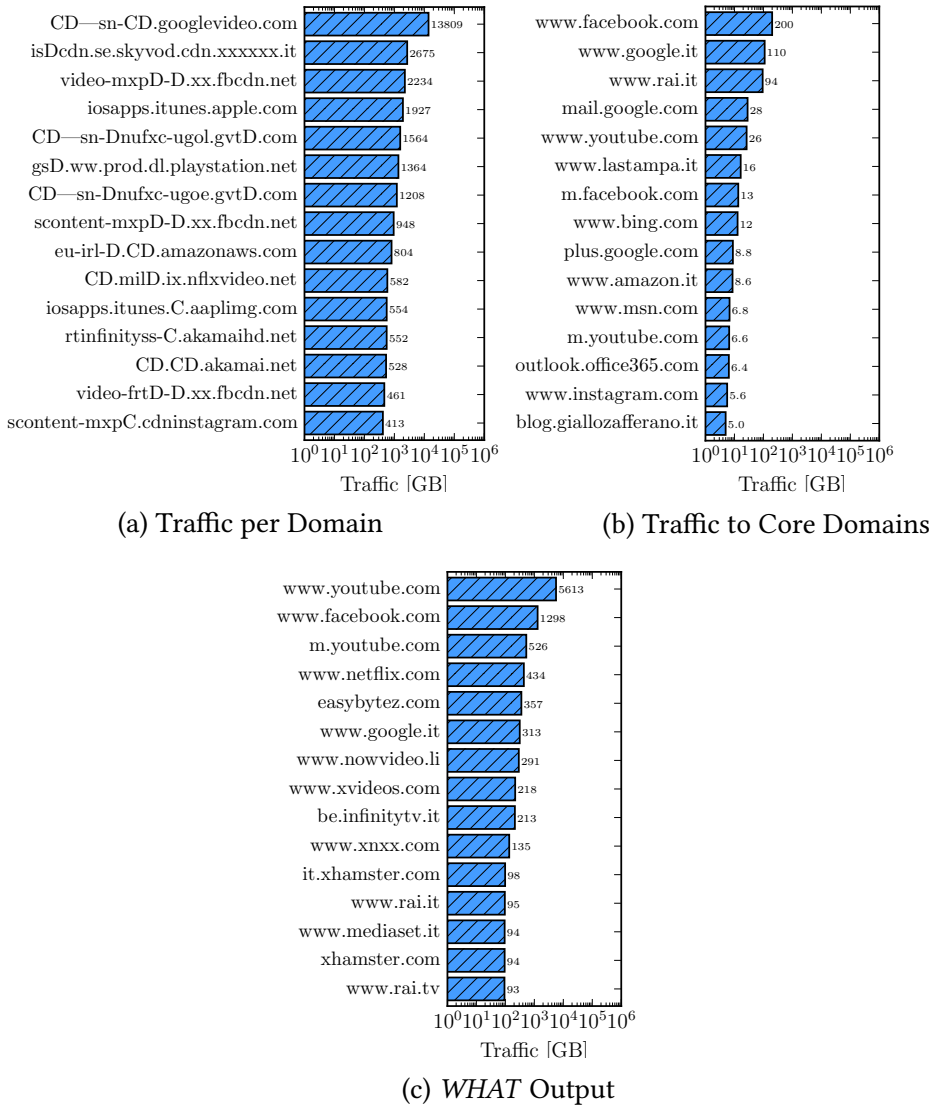

(a) Traffic per Domain

(b) Traffic to Core Domains

(c) *WHAT* Output

Figure 5.12: Ranking traffic: Domain rankings highlight support sites. *WHAT* highlights important services.

domains. Here I list the most popular core domains, which are naturally strongly biased towards popular websites in the country where the data has been collected. While popular core domains are meaningful, their traffic shares by no means represent network resource consumption by users accessing those services. Notice, for example, that YouTube is only fifth in this list accounting for a mere 26 GB over ten days by 10,000 users.

Finally, Figure 5.12c shows *WHAT* output. Here I see a more realistic picture of the actual network usage. Video streaming, i.e., YouTube, NetFlix, adult video providers, and Rai (a broadcaster) emerge among the most popular services in term of traffic volume. Some support domains seen in Figure 5.12a (`playstation.net` and `skyvod.cdn.xxx.it`) are not reflected in any service on Figure 5.12c because they host background services (e.g., gaming consoles and IPTV), which are out of *WHAT* classification scope.

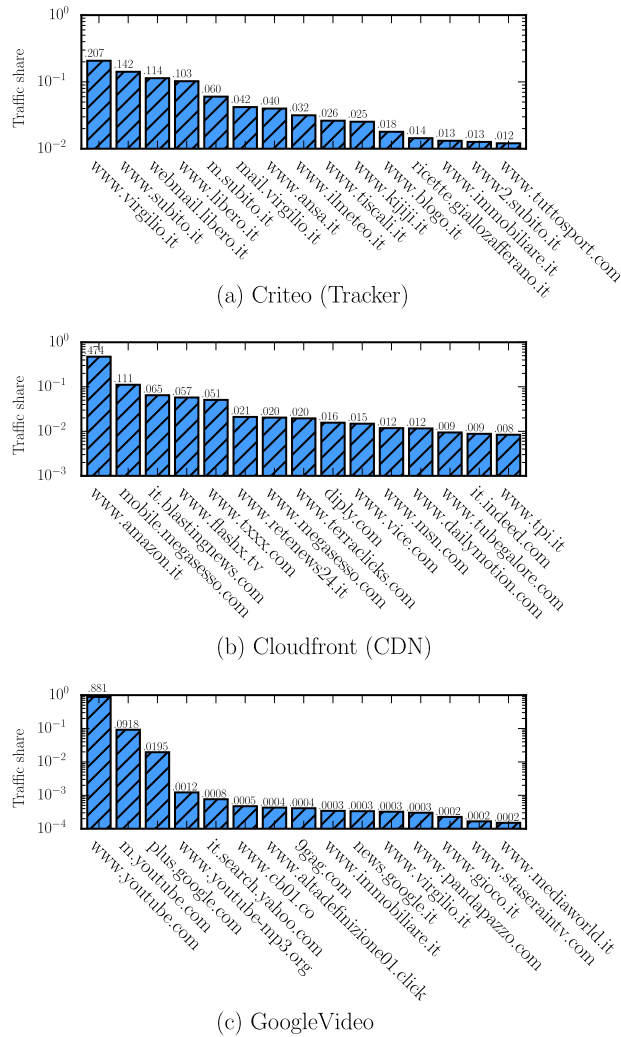

Figure 5.13: Traffic of popular support domains according to core domains.

In sum, the figure illustrates how *WHAT* can help administrators to correctly account traffic to services going beyond per flow metering.

### 5.6.2 Support Domains Pervasiveness

Support domains can be triggered by many core domains. *WHAT* methodology provides an interesting opportunity for understanding how flows of different parties contribute to the traffic volume of core domains. Moreover, it allows to identify which core domains trigger pervasive support domains, such as trackers.

An example is provided in Figure 5.13a. It depicts the list of core domains that trigger a popular third-party tracking domain, i.e., `criteo.com`. This service is included in many websites to collect information about users. *WHAT* unveils the breakdown of Criteo traffic accounting it to core domains. The figure that emerges illustrates well how this tracker is pervasive, covering an extensive list of popular sites (e.g., news, blogs etc). *WHAT* helps administrators to spot tracker partners.

Figure 5.13b uncovers services that rely on a popular CDN (Cloudfront). It shows the contribution of each core domain for the CDN traffic volume in the ISP network. Cloudfront belongs to Amazon, and it is not a surprise that `amazon.it` emerges as the main core domain relying on that CDN. However, *WHAT* automatically exposes other services hosted in Cloudfront, such as adult video sites, news outlets, etc.

Finally, Figure 5.13c uncovers which services trigger `googlevideo.com`, i.e., which are the services where users consume YouTube videos. As expected, YouTube website is responsible for the vast majority of the traffic. However, since third-party sites embed YouTube videos too, *WHAT* learns those relations. This example decisively illustrates the applicability of *WHAT* for accounting applications: Whereas YouTube serves the content, completely different services and websites can be responsible for the video download. *WHAT* exposes such relations to network administrators and in general enables any analytic that entails the accounting of traffic to services.

## 5.7 Related Work

With recent trend of increasing web service traffic, a number of traffic classification works focus on the traffic contents [65, 64, 82, 25]. Kim et al. [65] surveys behavioral techniques that leverages machine-learning to achieve traffic classification comparable to far more privacy invasive DPI techniques. A behavioral approach Karagiannis et al. [64] discovers traffic signatures unique to Peer-to-Peer (P2P) network. While the behavioral traffic classification successfully identifies some services, they are far from being comprehensive [82, 25]. Moreover, since their methods requires laborious manual analysis to extract behavioral signatures, they were unscalable. *WHAT* on the other hand, leverages machine learning to minimize human intervention to be self-adaptive to different deployment scenarios.

DNS traffic based classification has been a popular area of research as an alternative to the behavioral classification. Agar et al. [2] is one of the first research proposing to leverage DNS traffic for classification. The authors built a map of the whole World

Wide Web using DNS information classified. But because the method requires DNS traffic to be actively generated, the method cannot be widely used in operational networks. Plonka et al. [85] proposed passive DNS analyses by leveraging available DNS information on the wire. The authors classify IP traffic in a tree structure comprised of three classes. In contrast to [85], I neglect well-known protocols (e.g., FTP or P2P) in *WHAT*. Instead, I focus on typical services that make the majority of web traffic nowadays, and develop a system to classify flows according to services generating the flows.

In addition to DNS queries, Tongaonkar et al. [101] and Foremski et al. [37] propose to use Server Name Indication (SNI) strings found in TLS handshakes for classification. While authors discover better coverage thanks to SNI, their primary target for classification were on the protocols (e.g., SIP, HTTP, etc). *WHAT* conducts finer-grained level of classification on web services rather than their protocols.

## 5.8   Conclusions

This chapter presented *WHAT* (Web Helper Accounting Tool) describing how it leverages machine learning to autonomously identify the *core domain* a user is accessing and the set of associated *support domains* automatically contacted as a consequence. This dramatically recudes the manual effort needed to understand the mixture of web traffic, and help ISPs obtain visibility on the activity of users. *WHAT* uses this to create a model of web service access and uses it to categorize future traffic flows. The extensive evaluation presented in the chapter show how *WHAT* offers accurate metering of individual web activities, which is not enabled by traditional traffic classification and metering solutions.

# Chapter 6

# AWESoME: Big Data for Automatic Web Service Management in SDN

The work I present in this chapter is mostly taken from my paper "*AWESoME: Big Data for Automatic Web Service Management in SDN*" published in the IEEE Transactions on Network and Service Management journal in March 2018 [104].

## 6.1 Introduction

The Software Defined Network (SDN) paradigm has changed the way networks are managed [66]. Thanks to a logical centralized controller and well-defined interfaces to program forwarding devices, SDN controls the traffic in a consistent manner and dramatically eases interoperability across different vendors. Yet, network managers face complex traffic engineering and policing requirements when operating the network to meet quality levels, prioritize traffic and enforce polices. Traditionally, such requirements might translate into complex matching on packets or flows, e.g., to drop P2P packets or regulate flows related to specific services.

The complexity of the web has introduced more challenges in the overall picture. On the one hand, the widespread adoption of cloud services and CDNs puts into question the identification of the services behind the traffic flows because a single server supports multiple services, e.g., providing content for several sites. On the other hand, the convergence towards encrypted protocols – i.e., HTTP(S) [80] – has rendered Deep Packet Inspection (DPI) based matching ineffective. Nowadays, the access to a single service might result in the generation of several traffic flows to multiple servers, e.g., CDN nodes, advertising platforms, video servers, etc., that are shared by different services and, as such, cannot be easily associated to the specific web service originating the traffic.

Figure 6.1 illustrates this problem by showing the diverse servers contacted by a user after visiting two simple web sites, i.e., an e-learning platform and a news website.

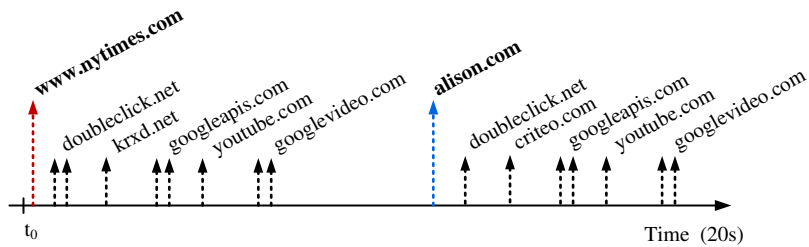

Figure 6.1: Flows opened when visiting two websites. I search flexible mechanisms to independently manage all traffic triggered by each site – e.g., for traffic engineering and policing.

Arrows mark flows to the contacted domains of first- and third-party platforms involved in the services. Both sites rely on the same third parties for video services (i.e., YouTube), analytics and web tracking. This poses unique challenges to a network manager wanting to give higher priority to the e-learning platform (on the right), while segregating the news site traffic (on the left). Prioritizing only the first-party servers would fail to give the intended treatment for the video content of the e-learning platform that is hosted on YouTube, whereas prioritizing *all* YouTube traffic would give high-priority also to leisure videos triggered by the news site.

My goal is to allow administrators to manage *all* traffic of a service comprehensively, i.e., steering all traffic generated by the user accessing a given service, and not just the traffic related to first-party servers. A novel approach to traffic management is required where policies are based on the *services* that users are contacting, which in turn must be translated into rules that can be imposed on *packets* and *flows*.

I solve this problem by proposing AWESoME. It defines a novel paradigm in which the network administrator imposes policies based on the *service* being accessed, e.g., giving priority to `alison.com` in Figure 6.1, while segregating `nytimes.com`, and treating third-party traffic according to the accessed first-party service. Using big data approaches, AWESoME automatically learns *groups* of flows related to the services and steers them despite being served by the same CDNs, servers, clouds, and with the same (encrypted) protocols.

AWESoME is a SDN application that leverages standard SDN functionalities to steer traffic in the network. At the core of the SDN application is a novel annotation-module operating at edge elements, which is able to associate each flow to the originating service in real-time and with high accuracy. It leverages DNS information and big data to automatically learn from the traffic. It achieves an overall accuracy higher than 90%, that, despite not suitable for security purposes, is well-suited for traffic engineering and management goals.

In contrast to previous works that also aimed at bringing service-awareness to SDN, but focused on per-flow management [8, 59, 81, 88], AWESoME addresses the challenge in the more comprehensive and seamless way based on the following three premises:

- **Comprehensive policing of services**: AWESoME creates forwarding rules that cover complex relations among flows (e.g., as in Figure 6.1). It achieves that by learning which domains are typically contacted when accessing each service. Models to translate high-level descriptions of services into low-level rules are learned automatically from traffic with unsupervised algorithms, minimizing human intervention. Flow dependencies have already been studied and exploited for data-center management [21, 22], but I extend those methodologies to operate at the edge of the network.

- **Early classification with low overhead**: AWESoME takes final forwarding decisions since the very first packet of each flow. This limits the load on the controller and application, making it compatible with actual technology. This is achieved by extending methodologies that rely on the DNS for traffic annotation [7, 37, 78, 85].

- **Compliance with SDN specifications**: AWESoME has been designed to be fully compliant with the basic SDN architecture and the latest version of Open-Flow [83], although it could also be deployed with other communication protocols between controller and forwarding elements. It requires no changes to existing APIs and SDN controllers, hence allowing adoption of AWESoME to existing SDN platforms to be simple.

I thoroughly evaluate accuracy and scalability of AWESoME in the classification and steering of web service traffic using traces collected from both volunteers (which offer ground truth) and operational networks (which challenge AWESoME in both ISP and corporate environments). Results show that AWESoME (i) identifies traffic per service with accuracy greater than 90%, more than adequate for traffic management; (ii) limits decision time to less than a hundred microseconds, with negligible load overhead to SDN controllers; (iii) adds a compatible number of rules to forwarding devices and, therefore, it is feasible for real deployments.

To allow other researchers to reproduce and validate my results, I release to the public ground truth traces and Python scripts implementing the core components of AWESoME.[1]

Next I introduce terminology, deployment scenarios, and AWESoME architecture (Section 6.2). I then detail the core annotation algorithms (Section 6.3), before introducing the dataset (Section 6.4) that I use to validate performance (Section 6.5). I conclude by discussing related work (Section 6.6) and summarizing my findings (Section 6.7).

---

[1]Available at: `https://bigdata.polito.it/content/open-datasets`

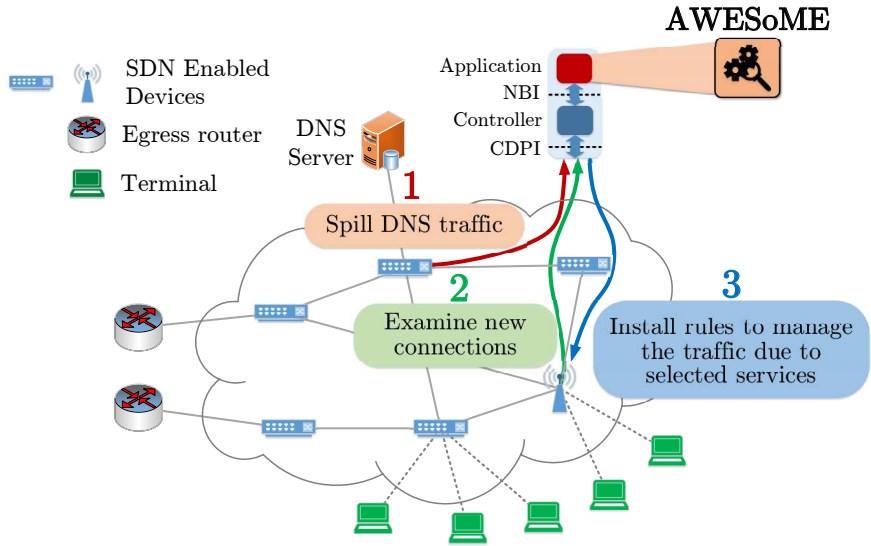

Figure 6.2: Typical corporate SDN deployment.

## 6.2 Definitions and Architecture

### 6.2.1 Per service management approach

I aim at enabling management operations that target the control of all traffic involved in the access to web services, i.e, all objects a browser or a terminal downloads when users access the given web service. I call this *per service* management. I envision several scenarios where the *per service* approach will help administrators to manage the network. To name some examples, AWESoME allows network managers (i) to block non-authorized services in the network, (ii) to route traffic of given services on specific paths with performance guarantees, (iii) to route suspicious traffic of unknown services to specific devices (e.g., through a security firewall), (iv) to regulate the traffic of pre-selected services. Even if I do not provide a specific evaluation of AWESoME performance in all such tasks, I argue that the introduced *per service* management is an enabling building block for all these operations.

I will use corporate networks as a running example in the chapter (see Figure 6.2) although AWESoME is applicable to other scenarios too. In the scenario depicted in Figure 6.2 the corporate network has two links to external networks that deliver different performance, potentially at different costs. In this example, the network administrator may want to forward priority services (e.g., the e-learning platform illustrated in Figure 6.1) to the best performing link, whereas traffic from non-priority services is forwarded to the best-effort link. AWESoME must guarantee that all traffic of the selected services flow to the desired path. Therefore, all network elements in the corporate network must be programmed to forward traffic according to the per service management approach.

## 6.2.2   Core and support domains

Servers being contacted by clients are identified by their IP addresses, but they are typically reached using their Fully Qualified Domain Names, or *domains* for short. Like in the the previous chapter, I divide *domains* in *Core* and *Support* (see Section 5.2). Services that people (or applications) intentionally access are identified by their *Core Domain*: www.nytimes.com, alison.com are core domains (tall arrows in Figure 6.1). Unfortunately, only a minor fraction of traffic related to a service is served by the core domain, with *Support Domains* (short arrows in Figure 6.1) being contacted for analytics, ads, video and image download.

Table 6.1 quantifies the traffic related to core and support domains for popular sites. It details the breakdown of flows served by the core domain, by support domains whose name is trivially linked to the core domain (e.g., `www.nytimes.com` and `css.ny-times.com`), and by generic support domains (e.g., `ads.com`). Notice how a large fraction of flows is exchanged with support domains, and that a simple approach taking into account only traffic to the core domain would fail in identifying most of the flows.

More than that, generic support domains are often shared across different websites, and some core domains also appear as support domains for other services (e.g., online social networks). Figure 6.3 quantifies these cases again for a set of popular sites. It shows 9 websites, grouped into 3 categories. By visiting each site I have collected all contacted support domains. Over 275 total domains, 43 are shared by websites of different categories and 6 domains are present in all categories.

Per service management, therefore, is required to identify *core* and *support* domains. In continuity with terminology introduced in Section 5.2, I call *Bag of Domains (BoD)* the set of all support domains contacted when accessing the given core domain. For each core domain, its Bag of Domains must be automatically built from traffic using big data approaches.

## 6.2.3   SDN as enabling technology

I consider an SDN, where users access the Internet via their devices connected to SDN enabled switches or wireless access points, as sketched by Figure 6.2. The SDN controller manages the network, translating the requirements from the SDN applications to SDN datapath commands. AWESoME interacts with the SDN controller via the NorthBound Interface (NBI), as a standard SDN application. AWESoME operates by installing three types of rules in the network elements: (i) default rules, (ii) per flow rules, and (iii) policing rules.

*Default rules* are installed on edge switches to forward selected packets to the AWESoME application running on the controller. These rules are summarized in Table 6.2a: (1) all DNS response packets are normally forwarded, and mirrored to the controller, (2, 3) the

Table 6.1: Traffic generated by visiting 10 popular services.

| Service | Percentage of flows to | | | Total domains |
|---|---|---|---|---|
| | Core domain | Related support domains | Generic support domains | |
| www.bbc.com | 19.4 | 35.3 | 45.3 | 90 |
| www.nytimes.com | 17.4 | 43.7 | 38.9 | 63 |
| washingtonpost.com | 34.8 | 21.2 | 44.0 | 90 |
| www.ieee.org | 37.8 | 24.3 | 37.9 | 17 |
| www.acm.org | 43.5 | 0.0 | 56.5 | 8 |
| researchgate.net | 5.2 | 75.5 | 19.3 | 29 |
| www.facebook.com | 21.9 | 63.0 | 15.1 | 12 |
| www.google.com | 8.9 | 77.8 | 13.4 | 141 |
| twitter.com | 6.8 | 86.8 | 6.4 | 6 |
| www.youtube.com | 5.8 | 76.9 | 17.2 | 30 |

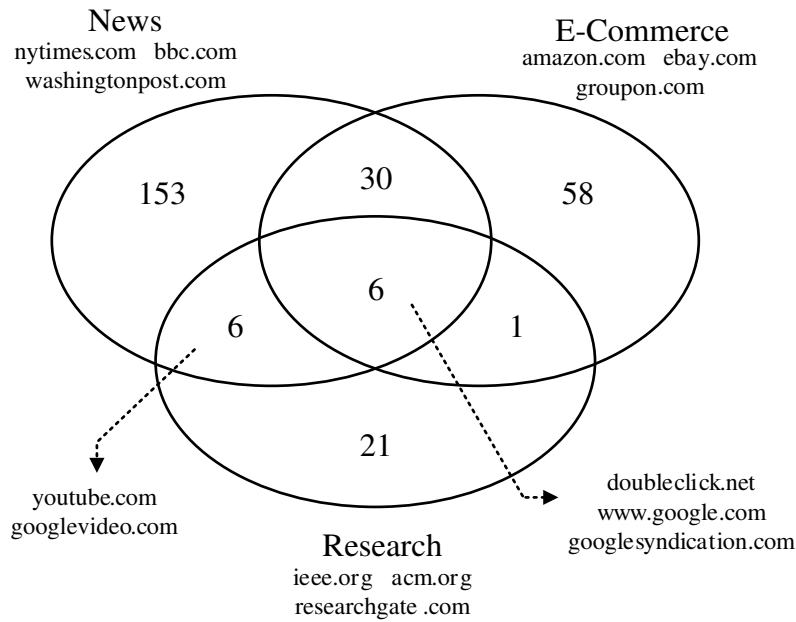

Figure 6.3: Support domains shared across different categories of sites. Analytics and advertisement domains are always present.

first packet of all TCP and UDP flows are forwarded to the controller.[2] The first rule is

---

[2]The flow must match the TCP flags using the `OFPXMT_OFB_TCP_FLAGS` field available since Open-Flow 1.5.0. These rules are given low priority to avoid overriding more specific rules.

Table 6.2: Rules to be installed on the SDN switches across the network.

(a) Default rules installed at edge switches control the traffic that needs to pass the controller for taking decisions.

| N | Match | Action | Description |
|---|---|---|---|
| 1 | IP_PROTO=UDP and UDP_SRC=53 | Forward, Forward to Controller | Spill DNS responses |
| 2 | IP_PROTO=TCP and TCP_FLAGS=PURE_SYN | Forward to Controller | Intercept new TCP connections |
| 3 | IP_PROTO=UDP and UDP_DST!=53 | Forward to Controller | Intercept all UDP non-DNS traffic |

(b) Transient per-flow rules are installed at edge switches to tag each flow (e.g., $f_1$ and $f_2$) with the respective service label.

| N | Match | Action | Description |
|---|---|---|---|
| 1 | IP_PROTO=TCP and IPV4_SRC=$IP_{SRC}^{f_1}$ and TCP_SRC=$TCP_{SRC}^{f_1}$ … | Push VLAN tag, VID=0x001 | Tag as *Gold* |
| 2 | IP_PROTO=TCP and IPV4_SRC=$IP_{SRC}^{f_2}$ and TCP_SRC=$TCP_{SRC}^{f_2}$ … | Push VLAN tag, VID=0x002 | Tag as *Silver* |

(c) Stable policing rules are installed in core switches to steer packets according to the application scenario faced by AWESoME. In this example, traffic of each class is forwarded to particular network path (see also Figure 6.2).

| N | Match | Action | Description |
|---|---|---|---|
| 1 | VLAN_VID=0x001 | Output on $P_1$ | Forward *Gold* traffic towards the reliable link |
| 2 | VLAN_VID=0x002 | Output on $P_2$ | Forward *Silver* traffic towards the best-effort link |

used to maintain a database that allows AWESoME to associate a flow with a domain name via previously issued DNS requests [7, 37, 78, 85]. The latter rules let AWESoME handle each new flow to subsequently impose the most appropriate actions.

Like any SDN solutions based on such reactive paradigm, the default rules may force the controller to examine a large number of packets. In Section 6.5.5 I will show that the load is still limited for a network with moderate number of users. For very large deployments, controller load-balancing solutions should be considered [31, 52]. Reactive SDNs are also exposed to Denial-of-Service attacks – e.g., malicious nodes that exploit rules to overwhelm controllers with lots of packets. Different solutions have been proposed to tackle the issue [62, 112], and they could be employed in my scenario.

Once AWESoME has taken the decision about a new flow, it installs a *per flow rule* on the edge switch to handle the packets of the new flow. Per flow rules aim at guaranteeing that different flows associated to a single service are treated equally in the network. They are transient and thus maintained only while the given flow is active. Table 6.2b lists rules installed to handle the example presented in Figure 6.2. Flows that are identified as belonging to selected applications are *tagged* as priority (i.e., Gold class, implemented as VLAN tag `0x001`), whereas the remaining flows are tagged as best-effort class (i.e., Silver, implemented as VLAN tag `0x002`).

Notice that only the first packet of each flow needs to be inspected by AWESoME. Per flow rules guarantee that subsequent packets of the flow do not transit through the controller, but are directly forwarded by edge switches. Since the system adopts a reactive SDN paradigm, packets transiting through the controller are retained by the switch until the controller takes a decision. More in detail, when such a packet arrives to the switch, a copy is sent to the controller using a *PacketIn* message, and holds it in a local buffer. When (eventually) the controller answers with a *PacketOut* message, it is actually forwarded in the network. As a result, clients can only establish connections *after* AWESoME has programmed the edge switch.

Finally, AWESoME programs core switches with pre-defined policing rules. These rules are stable and installed when the manager deploys an application based on AWESoME. In Figure 6.2, traffic of each category needs to be forwarded to the particular reserved path. As such, rules to handle and forward the classes are installed in core switches (see Table 6.2c). For this example on traffic engineering, policing rules are built based on the VLAN tags determined at edge switches. Other mechanisms can be exploited too, such as MPLS labels or Provider Backbone Bridges (PBB) tags.

## 6.2.4   AWESoME architecture

Figure 6.4 provides a schematic diagram of the AWESoME SDN application. Four elements are identified, each in charge of a logically independent operation, which together enable per service management:

1. *BoD-Training* automatically learns and updates the BoDs in background;

2. *Flow-to-Domain* tags flows with domains;

3. *Domain-to-Service* links named-flows to services;

4. *Service-to-Rule* translates the service into the appropriate actions (i.e., the rules described in previous section).

Below, I describe each of them, while performance and parameter tuning are discussed in Section 6.5.

**BoD-Training — automatically building BoDs**

The BoD-Training block is responsible to automatically build the BoD associated to each core domain. This is the key module in the AWESoME approach, and, it runs the same algorithm described in the previous chapter in Section 5.3.4.

**Flow-to-Domain — flow labeling using DNS**

This step associates a server domain to each flow, i.e., to create *named-flows*. This helps the association of a flow to a given service, since the information offered by IP addresses is much coarser than the one carried by the domain of the server being contacted [106]. This is because a single cloud (and CDN) server may host many services.

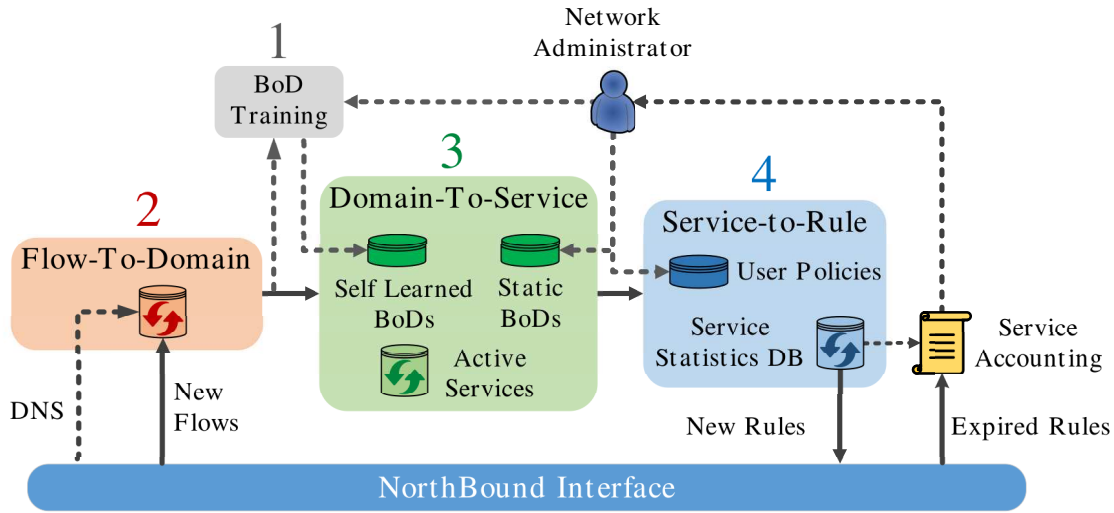

Figure 6.4: AWESoME architecture. Databases with arrows are maintained in real-time.

Intuitively, the same server IP address hosts a multitude of web services which are better identified by their domains. It has been shown that this operation can be solved by leveraging DNS traffic [7, 37, 78, 85]. The Flow-to-Domain block builds a local cache of domains that terminals have resolved in the past, maintained as a key-value store. Below, I describe the two actions of building and using the key-value pairs as *Insert and update* and *Lookup*.

*Insert and update:* For each DNS response forwarded by the controller, AWESoME extracts the $Client IP$ address, the domain being queried (*Queried Domain*), and, from each Answer record, the list of resolved $\{Server IP_i\}$ addresses. For each key $\{Client IP,$ $Server IP_i\}$, it inserts (or rewrites) an entry with value *Queried Domain*. The time such entries must be preserved in the store (and expired when old) is discussed in Section 6.5.

*Lookup:* Whenever a TCP or UDP packet is forwarded to the SDN controller, AWESoME parses the IP and layer 4 headers and accesses the name store with the key $\{Client IP,$ $Server IP\}$ to fetch the original *Queried Domain* the client previously resolved. In case there is not such key, the store returns the $Server IP$. The packet is then forwarded to the Domain-to-Service block, along with the *Queried Domain* or, if not available, the $Server IP$.

Using DNS information has several advantages with respect to more intrusive flow classification methods. First, it does not require to use costly DPI technology to extract hostname or SNI (Server Name Indication) from HTTP or HTTPS requests. Second, DNS information is not protected by encryption, and even DNSSEC does not provide confidentiality. Most importantly, the lookup is done on the very first packet of each flow, eliminating the need of keeping per flow state and waiting for more packets to take a final decision at the controller. As such, when clients finally open a connection, the network is already programmed to handle the traffic accordingly. On the downside, erroneous domain associations can happen due to collisions (i.e., a rewrite operation) – the same *server I P* being contacted by the same *client I P* for two different *Queried Domain*. As I will show later, AWESoME is robust to such events.

### Domain-to-Service — associating services to flows

Once the flows are labeled with DNS names, AWESoME associates the named-flows to services. This is the core of AWESoME engine and its details are provided in Section 6.3. Leveraging a well-known text mining technique, Bag-of-Words, I model semantics of the domain names, namely, *Bag-of-Domains (BoD)*. A BoD is created for each core domain and includes all support domains that are contacted when the service identified by the core domain is accessed.

I further group the domains into two types: *Self Learned BoDs*, and *Static BoDs* based on the characteristics of the domains. Automatically built by AWESoME while analyzing the traffic, Self Learned BoDs are BoDs comprised of interactive web services, which users *explicitly* access from their browsers, e.g., interactive web applications. On the other hand, manually built by network operators, static BoDs are comprised of *background* services that are periodically accessed by terminals without user intervention (e.g., software updates, file sync with cloud storage services, calendar or mail services, etc).[3] The traffic generated by such services is quite different from the interactive ones where core domains and support domains are expected to appear close in time (see Figure 6.1). Background services challenge the assumption of temporal correlation between flows, and extending AWESoME to learn Static BoDs automatically is left for future work.

The list of recently accessed core domains by each *Client I P* is stored in the *Active Service* database. Keeping a cache of active services is important since the same domain normally appears in multiple BoDs – cf. Figure 6.1 – and thus it must be associated to a core domain that has been actually visited. Given a flow, the Domain-to-Service block checks if its domain appears in the BoDs of *Client I P* Active Services, so to associate to the most likely service the user has recently accessed. In case a domain is not in the active services, it falls back to match against Static BoDs.

The packet is then forwarded to the Service-to-Rule block, along with the service corresponding to the BoD the packet was associated with.

### Service-to-Rule — policy enforcement

Once a flow has been associated to a service, the Service-to-Rule block is a classic policing module which enforces actions by requesting the SDN controller to install rules on the switches. Policies are stored in a *User Policies* database, which is accessed with the service name as key, and returns the corresponding rules.

---

[3]In the current implementation, regexp and wildcards are supported in the specification of static BoDs to simplify the administrator's task.

Policing rules are installed (e.g., in core switches) when the AWESoME application is started, whereas the per flow rules are pushed whenever a flow must be steered. Table 6.2 has already exemplified the rules created for the particular traffic engineering case used as illustration, but other rules can be defined too – e.g., to block services, route traffic to security devices or to regulate the traffic per service. Rules expire using the `Idle Timeout` standard OpenFlow feature.

In case of "default" action, no extra rule has to be added for TCP flows since only the SYN-TCP packet will be forwarded to the controller. The lack of explicit connection indication in UDP forces AWESoME to insert a rule for each UDP flow. However, only the first packet of UDP flows transits through the controller, while the others are directly forwarded by the switch, as a transient per-flow rule is inserted.

Scalability is evaluated in Section 6.5.5. Again, *per flow* rules have to be installed on the edge switches only – i.e., those switches that are directly connected to clients or work as ingress point to the SDN. Upstream devices instead operate on a *per service* basis, e.g., using IP Type of Service, MPLS labels or PBB tags, which are all supported by SDN.

The Service-to-Rule block additionally maintains the *Service Statistics DB*, with flow identifier (e.g., the classic 5-tuple) as key, and service information as value. When a rule expires at switches, its flow identifier is passed along with statistics (byte and packet amount) to the SDN controller (standard in `FlowRemoved` messages) that, in turn, exposes them to the AWESoME application. Such statistics are collected in the Service Statistics DB, later used for BoD training, and exposed to the network administrator. This enables for instance per service accounting, anomaly detection, billing, etc.

## 6.3 How Service Association Works

The core of the service management is the ability to associate each flow to the originating core domain, i.e., the service the user originally intended to access. AWESoME solves this by leveraging the bag-of-words model which is commonly used to succinctly representing complex textual data in natural language processing [50]. In the context of AWESoME, I extract bag-of-words features from each domain and "classify" it into a service. Hence I call the process Bag-of-*Domain* (*BoD*) training. Due to the complex composition of web pages and the intertwined nature of the Internet, it is not trivial to design the BoD training with minimal human intervention.

### 6.3.1 Automatic BoD training

Let $C$ be the set of core domains of interest provided by the network administrator. AWESoME training consists of building a $BoD_c$, for each core domain $c \in C$. To this end, it employs the algorithm already proposed in the last chapter, in Section 5.3.4.

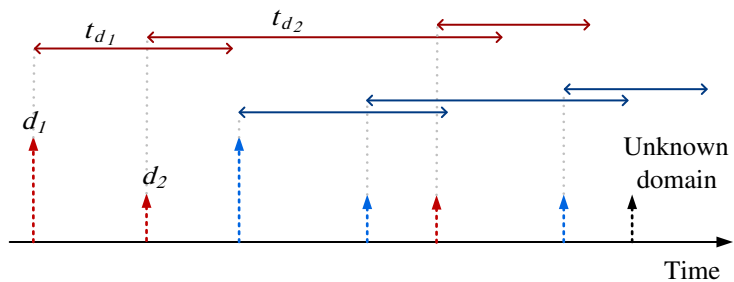

Figure 6.5: Domain-to-Service: Blue and red services are active at the same time; the EWs are extended as new flows are associated to the core domain.

Domains that constitute the BoDs are stored in a LRU cache of limited size (e.g., 5 000 entries). This is more than adequate – cf. Table 6.1 – and limits memory usage.

Then, AWESoME needs to compute the average flow duration per each domain in $BoD_c$. This is done using the flow duration information as exposed by the Service-to-Rule block. For each domain $d$, AWESoME maintains the average flow duration $t_d$. To cope with possible changes in service behaviors, a standard exponential moving average estimator is used (parameter $\alpha = 0.1$). AWESoME is however almost insensitive to the parameter as site changes occur in much longer time periods than the re-training of BoDs.

Flow duration is fundamental to AWESoME as I will show next, and OpenFlow `Idle Timeout` mechanism lets the controller to derive it. After no packet has matched a rule for a configurable period, a `FlowRemoved` message is sent by the switch to the controller. Flow duration is obtained by computing the time between the rule install action and the `FlowRemoved` message, subtracting the `Idle Timeout` set in switches.

AWESoME takes advantage of the last days of traffic to build the BoDs used by the Domain-to-Service module. A discussion about the time needed to build BoDs is provided in Section 6.5. The training dataset potentially becomes large in real scenarios, and thus the BoD-Training module is implemented in a state-of-art big data platform, namely Apache Spark. The statistics to build BoDs are continuously collected in the BoD-Training module. Periodically, e.g., once per hour, BoDs are computed and given to the Domain-to-Service for on-line annotation of traffic.

### 6.3.2 Domain-To-Service classification module

Armed with core domains and their respective BoDs, AWESoME has to associate named-flows with the service identified by the core domain. It first tries to associate the flow to any BoD in the Self Learned BoDs. In case of no match, it then tries with Static BoDs. For the sake of simplicity, I describe only the first stage as the second is

---

**Algorithm 5** $annotate(f, C, BoDs, T)$

---

**Input:**

 $f$                        ▷ *The current flow to annotate*

 $C = \{c_1, ..., c_k\}$                   ▷ *Core Domains*

 $BoDs = \{BoD_{c_1}, ..., BoD_{c_k}\}$          ▷ *BoDs of core domains in C*

 $T = \{t_{d_1}, ..., t_{d_t}\}$              ▷ *Domain average flow duration*

**Output:**

 $O = (f, CoreDomain)$                ▷ *Annotated flow*

 1:  *// Retrieve start time and domain of f*

 2:  $t = GetTime()$                 ▷ *Get current time*

 3:  $d_f \leftarrow parse(f)$               ▷ *Get the domain of f*

 4:  *// Remove expired Services*

 5:  $AS \leftarrow \{(ts, te, c_i, BoD_{c_i}) \in AS \,|\, t \le te\}$

 6:  *// Obtain the best BoD among the AS*

 7:  $as_{best} \leftarrow \{(ts, te_{best}, c_{best}, BoD)\} \leftarrow BestBoD(d_f, AS)$

 8:  **if** $d_f \in C \wedge as_{best} == \varnothing$ **then**

 9:   *// $d_f$ is a core domain – Start a new AS for $d_f$*

10:   $c = d_f$

11:   $AS \leftarrow AS + \{(t, t + t_c, c, BoD_c)\}$

12:   $O \leftarrow (f, c)$

13:  **else**

14:   **if** $as_{best} \neq \varnothing$ **then**

15:    $O \leftarrow (f, c_{best})$           ▷ *The flow is assigned to $c_{best}$*

16:    *// Update the AS validity time*

17:    $te_{best} \leftarrow \max(t + t_{d_f}, te_{best})$

18:   **else**

19:    $O \leftarrow (f, \text{``} unknown\text{''})$        ▷ *Flow not classified*

---

identical.

Figure 6.5 gives an example of a possible timeline where a *ClientIP* accesses first to the red service, and then to the blue services.

AWESoME uses Algorithm 5 to annotate each flow $f$. It is a modified version of the algorithm proposed in the previous chapter, in Section 5.3.5, modified to operate in real time. It receives: (i) the current named-flow $f$, (ii) the set of core domains $C$, (iii) the BoDs, (iv) average duration $t$ for each domain. It processes each *ClientIP* separately and keeps separate data structures. It outputs a flow annotated with the core domain, or *unknown* in case no association is found.

The algorithm is based on the concept of *Evaluation Window* (*EW*), i.e., a time during which a support flow can still appear after the observation of the core domain $c$. The algorithm maintains a list of Active Services, *AS*, i.e., those core domains previously seen, and for which it is still possible to associate some flows. The list grows as new core domains are observed (lines 8–12), and entries are aged out, i.e., window ending time $t_e$ is passed (line 5).

First, AWESoME checks if there exists a Active Service $as_{best}$ whose BoD contains the domain of $f$. In case more than one AS matches, I consider the $as_{best} = BestBoD(d_f, AS)$ as the one whose evaluation window start time is the closest in time (line 7). Intuitively, I consider the most recent visited core domain as the most likely one to associate the current support domain. I tried other choices, e.g., considering random choice, weighted choice by the frequency of occurrence in BoDs, etc., with worse results.

Next, AWESoME has to resolve the ambiguity for domains that can appear as both support and core. If $d_f$ is a possible core domain, and there exists no AS in which it appears as support domain (line 8), then it is considered a new core domain, and a new evaluation window is opened (lines 9-12). The rationale is that the domain has been contacted because of an intentional visit from the user.

On the contrary, $d_f$ is considered a support domain if there exist an active service $as_{best}$ (line 14). The flow is associated to the core domain $c_{best}$ (line 15), and the evaluation window ending time $te_{best}$ is extended (line 17) to consider the average duration of the current flow $t_{d_f}$. The rationale is that flows to support domains may be observed long time after the core domain, since the terminal keeps downloading objects due to a user action, e.g., scrolling a web page that triggers the download of new elements, or the download of a new video chunk in a streaming service. This is sketched in Figure 6.5 where the evaluation windows are represented by horizontal arrows, which extend the AS ending time.

Finally, in case of no match with any AS, the flow is associated to the "unknown" class (line 19), and AWESoME looks for a matching in the Static BoDs.

It is important to notice that the Domain-To-Service module operates on a per-flow and per-*ClientIP* basis and, thus, the processing is amenable for per-client parallelization.

## 6.4 Datasets

I validate AWESoME and evaluate its performance using trace-driven analysis. First, I thoroughly assess classification performance using traces where ground-truth is available – i.e., where I have information about the core domain responsible for the visit to support domains. Then, I use passive traces collected at operational networks to study realistic AWESoME deployments.

### 6.4.1 Ground-truth traces

I rely on ground-truth data from volunteers, following an approach similar to what I have done previously in Section 5.4.2. I collect browsing histories of 30 users, directly extracting the URLs they intentionally visited in the past months, which are stored in a local database by their browsers. I automatically revisit each URL by instrumenting a Chrome browser. I let Chrome visit the URL and wait until the page is fully loaded (i.e., the On Load event is fired).

In parallel, I record all network activity in the environment to have a complete picture of the traffic that would be managed by AWESoME. The outcome of these steps is a dataset of named-flows, where each entry is annotated as a core domain, if it was a URL given as input to the instrumented browser; or as a support domain, if it was triggered by a core domain visit.

In total, I collected 973 000 flows, referring to 3 760 and 97 640 unique core and support domains, respectively. Crawling was done in December 2016 and lasted 5 days. I build three traces from this raw dataset:

- **Simple-browsing:** It mimics the original behavior of volunteers. Each volunteer is given a unique *Client I P* address, and I simulate page visits in the same sequence and with the same visit time of the original browsing history. The arrival time of support flows after a core domain visit respects what is seen during crawling.

- **Tab-browsing:** I create this scenario by repeating the previous steps, but starting 5 independent navigation threads per *Client I P* in parallel. To avoid any kind of synchronization among threads, each navigation starts following the browsing histories at a random position. This scenario emulates (i) an extreme case of multi-tabbed browsing where the same user has 5 tabs concurrently and continuously browsing the web; or (ii) 5 users concurrently and continuously browsing the web behind a NAT (i.e., identified by the same client IP address). The latter is a typical setup in ISP environments where a single home gateway acts as a NAT, and a handful of household devices access the Internet contemporarily with the same identifier. Core and support domains of many visits may appear simultaneously in the trace. This challenges the disambiguation of core and support domains.

- **Simple-browsing + video streaming:** My crawling based on volunteers' histories notably miss video streaming sessions, since videos may continue playing after the On Load event is fired. Traffic generated by video servers might be quite different from interactive browsing because flows to retrieve video chunks have low temporal correlation with the core domain request [63]. Using the instrumented browser, I record all traffic generated when accessing 250 arbitrary URLs from 15 sites with embedded videos. I let the video play for 5 min before moving to another page. I finally mixed the Simple-browsing trace by simulating a second parallel thread for each volunteer. This thread continuously watches videos, with the user changing page every 5 min, without any pause in between. This is again an extreme case to test.

### 6.4.2 Operational network traces

I capture flow-level datasets from operational networks using passive meters. My captures include four measurement locations: two ISP networks (see Section 2) and two corporate networks. The datasets are summarized in Table 6.3. To preserve users' privacy, IP addresses have been anonymized, and I kept only the information required for the study. Trace collections has been approved by ISP and corporate security board.

Table 6.3: Traces collected from operational networks. A detailed description of ADSL and FTTH can be found in Section 2.

| Dataset | Duration | Flows | Unique Domains | Client IP addresses |
|---------|----------|-------|----------------|---------------------|
| *ADSL* | 12 months | 13 billion | 18 million | $\approx 10\,000$ |
| *FTTH* | 12 months | 4 billion | 6 million | $\approx 1\,000$ |
| *Corp* 1 | 1 day | 6 million | $\approx 38\,000$ | $\approx 1\,600$ |
| *Corp* 2 | 3 days | 32 million | $\approx 64\,000$ | $\approx 6\,000$ |

**ISP traces**

ADSL and FTTH traces include data exported by flow exporters deployed at different Points of Presence (PoP) of a large ISP. A detailed description of the measurement methodology is provided in Chapter 2. The flow exporters provide the Flow-to-Domain mapping performed by AWESoME by processing the DNS traffic on-the-fly. I employ data from January to December 2016. In total, I have observed more than 17 billion flows, 18 million domains.

I additionally dumped DNS traffic in the ADSL and FTTH PoPs for 6 hours in December 2016, simultaneously to the flow exporting, for some specific analysis that will follow.

**Corporate traces**

I rely on proxy logs from enterprise networks to assess AWESoME performance in corporate scenarios. They come from two different enterprises in different states of the USA. The proxies provide web connectivity to thousands of employees of two companies. They save logs for (i) each HTTP request and (ii) each CONNECT command for HTTPS tunnels. Clients are consistently identified by IP addresses. No UDP traffic is allowed.

I directly construct a named-flow log from each of the raw proxy logs, creating the *Corp* 1 and *Corp* 2 datasets. I proceeded as follows: for each CONNECT and for each HTTP request entry I create a flow record for the involved client and server. The domain is extracted directly from the hostname in HTTP request and from the CONNECT command. Naturally, this approach will over-estimate the number of flows in the network, since TCP flows are reused by clients when communicating with a HTTP server.

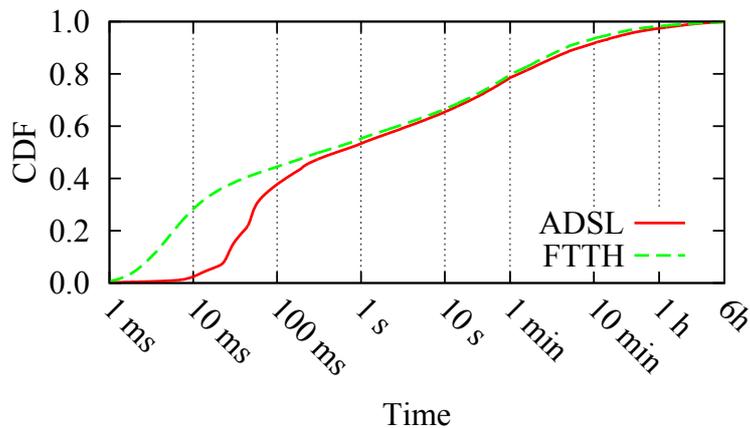

Figure 6.6: Time between TCP flows and their associated DNS query. AWESoME needs to cache information about 1-hour of DNS traffic to annotate flows.

## 6.5 AWESoME Performance

### 6.5.1 Flow-to-Domain evaluation

I evaluate the Flow-to-Domain block aiming to answer two questions: (i) What is the percentage of flows that can be annotated with DNS information? (ii) How long should the DNS information be cached to perform flow annotation?

I start focusing on the first question to check how many flows would remain unnamed due to lack of DNS data. To answer this question I use the 6-hour-long dataset in which I have both named-flows and raw DNS traffic from the ADSL probe. I look for web flows, i.e., to port 80 and 443, which have a domain associated. In particular I simulate the Flow-to-Domain module with infinite memory. To avoid boundary effects, I warm up the domain key-value store loading the initial 5 hours of the DNS trace and then I use the last hour of the flow trace to perform a lookup while still processing the DNS trace at the same time.

I found that 93% of the web flows are annotated. Manual inspection reveals two main causes for the missing domains: (i) ≈ 1% of services contact support servers using directly IP addresses; (ii) possible loss of DNS packets during the passive captures. AWESoME handles the first case by adding IP addresses to BoDs, whereas the second case is a measurement artifact that should not happen in real SDN deployments.[4]

Figure 6.6 presents the Cumulative Distribution Function (CDF) of the delay between the flow and the previously issued DNS query. For both traces, more than 90% of the flows starts within 10 minutes since the corresponding DNS query. The percentage grows to 97% considering a 1-hour interval.[5] These large time gaps between DNS queries and the flows are mostly due to large TTLs of DNS responses and client-side caches – i.e., clients can open flows to servers long after resolving their names thanks to the local DNS cache. Nevertheless, the figure shows that the Flow-to-Domain block must be sized to hold in its key-value store the information extracted from about 1-hour of DNS traffic in order to output high-quality named-flows. In the largest of the traces this corresponds to manage about 1 000 000 entries.

---

[4]In SDN, packets sent to the controller are always received thanks to the usage of reliable transport protocols between switches and the controller.

### 6.5.2   Domain-to-Service accuracy

I next evaluate the core part of AWESoME – i.e., the association of services to named-flows. I use the ground-truth traces for this validation. I only check the accuracy for self learned BoDs, since static BoDs are manually provided by network administrators.

I estimate the accuracy of AWESoME by checking whether the service determined for each flow matches with the ground-truth. In this experiment, learning of BoDs is performed using the ground-truth trace. All 3 760 core domains are considered with 3 760 BoDs built from the trace itself. I here test such a case in which every service would be managed independently to evaluate AWESoME performance in extreme cases. In more practical scenarios (e.g., Figure 6.2), one would expect only a limited number of key services to be classified and managed. I also repeat experiments with different settings to study the effects of AWESoME parameters. The obtained figures are similar to those reported before in Section 5.5, in which I evaluated the performance of *WHAT* for classification. Despite the similar results, the annotation algorithms are different, provided the need of AWESoME of running in *real time*, without any knowledge of the future behavior of the client. Indeed, AWESoME performance is slightly lower if compared to *WHAT* (see Section 5.5).

Figure 6.7 depicts the performance of AWESoME when varying the $MinFreq$ threshold used for learning. Recall that a support domain is discarded from the BoD if it appears less frequently than $MinFreq$. Curves for three scenarios are depicted.

Focusing in $MinFreq = 6.25\%$, notice how the accuracy of AWESoME is high, reaching close to 93% in the Simple-Browsing trace. Errors are related to flows annotated with ambiguous domains (i.e., belonging to more than one BoD) or left as "unknown" (e.g., no active window during classification). Even for the extreme traces, AWESoME delivers accuracy close to 85%. That is, AWESoME can identify flows per service with high accuracy, even in challenging situations that should be uncommon in real deployments. In particular, for the 5-Tab-Browsing trace the performance penalty is very small. This means that AWESoME can successfully operate when users perform tab-browsing, or in typical ISP scenarios where devices of a household access the network with the same client identifier.

---

[5]Differences for small *x*-values occur due to variations in the RTT between clients and the flow exporters.

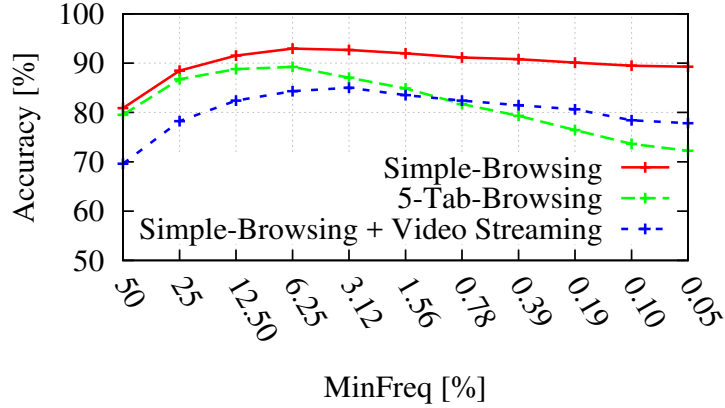

Figure 6.7: Accuracy when varying the *MinFreq* threshold. AWESoME accuracy surpasses 93% in the Simple-Browsing trace, and 85% in extreme scenarios.

I however remark that the tested 5-Tab-Browsing scenario does not guarantee that AWESoME would work with *any* level of parallelism. Carrier Grade NAT, in which hundreds or thousands of users are aggregated, is an example where the deployment of AWESoME requires planning. Switches inside the NAT-ed network need to be part of the SDN as well, thus aggregating a moderate number of users, which will ensure AWESoME delivers performance as in Figure 6.7.

Notice also the importance of *MinFreq* to filter out noise from BoDs. When *MinFreq* is large (e.g., 50%), domains that are popular in BoDs are ignored, resulting in a sharp decrease on accuracy. On the other extreme, when *MinFreq* is low, false support domains pollute the BoDs. Focusing on results for *MinFreq* = 0.1%, notice how accuracy drops to 90% in the Simple-Browsing trace, and to less than 80% in extreme scenarios. This happens because BoDs get very large with lots of false support domains that hinder the annotation process.

I omit for brevity analyses with other parameters. Overall, the best parameter choices are $T_{OW} = 10$ s, $T_{idle} = 5$ s and *MinFreq* = 6%, resulting the best figures shown in Figure 6.7.

### 6.5.3 Training set size and location

AWESoME learns BoDs by observing traffic. I now answer two practical questions regarding training in real deployments: (i) What is the amount of traffic that needs to be observed for learning representative BoDs? (ii) Should training be performed with traffic of the managed network or generic BoDs can be distributed to different deployments?

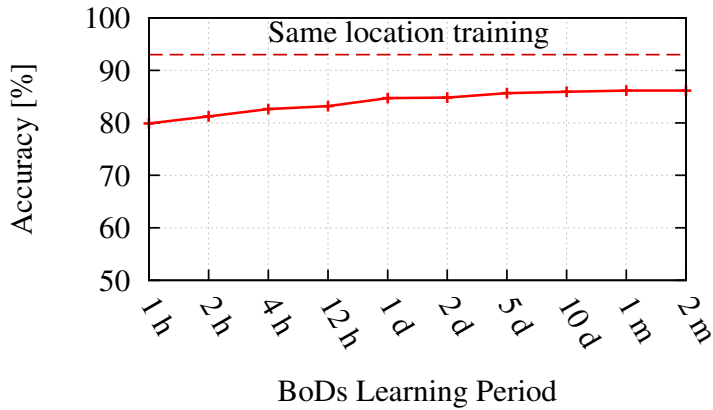

Figure 6.8: Accuracy vs. training dataset size. BoDs learned with ADSL, accuracy calculated with Simple-Browsing trace − 1-month training window is sufficient.

Figure 6.8 shows the effects of the training dataset size. For this experiment, AWESoME learns BoDs using ADSL trace, and performance is assessed with the Simple-Browsing trace. I extend the training dataset duration in each experiment round. The *"same location training"* line marks the best result obtained with training performed with Simple-Browsing trace. Again, AWESoME has to learn 3 760 BoDs. Here, I want to study the effect of different learning periods, and, thus, the study is limited to the ADSL trace. *Corp* 1 and *Corp* 2 are captured very far in space, and this would lead to worse results. This effect is evaluated later in this section.

Focusing on the left-most point in Figure 6.8, note that AWESoME correctly identifies 80% of the flows when the training set contains 1 hour of traffic only. That is, most of the popular BoDs are learned by observing a single hour of traffic. Increasing the training set improves results, with the best accuracy at around 87%. Thus, AWESoME needs to be trained for around 10 days to reach its best performance in this scenario. Further results, omitted for brevity, show that BoDs change slowly and are well-captured by the continuous training.

Since AWESoME requires historical data for training, the size of the training dataset may become large. For ADSL and FTTH, this corresponds to millions of flow records, which result in several GBs of traces. This calls for the use of scalable data processing approaches, and AWESoME training is thus built on Apache Spark to scale with the size of training dataset.

Figure 6.8 points to a decrease in performance when training is done with data from a different network. I explore this effect in Table 6.4. It reports the fraction of flows identified by AWESoME in a trace when training is done on another dataset. Columns indicate the training dataset, and rows indicate the testing dataset. I consider as core

Table 6.4: Fraction of flows classified by AWESoME when varying training and testing locations. The Alexa top-100 websites are core domains in this analysis.

|  |  | Training | | | |
|---|---|---|---|---|---|
|  |  | ADSL | *Corp* 1 | *Corp* 2 | *Crawling* |
| Classif. | ADSL | 1 | 0.45 | 0.48 | 0.32 |
|  | *Corp* 1 | 0.72 | 1 | 0.81 | 0.40 |
|  | *Corp* 2 | 0.42 | 0.47 | 1 | 0.34 |

domains the top-100 Alexa sites, since most of them are common across traces.

Cells report fractions taking as reference the flows which are annotated when training and testing are done with the same network. For instance, the first row shows that when training is performed with *Corp* 1, AWESoME annotates only 45% of the flows in ADSL that would be identified if both training and testing are done in ADSL. The remaining 55% of flows are marked as "unknown". This happens because the BoDs learned from different vantage points differ because of variations in the domains used by CDN servers or different content (ads) per location. Additionally, some BoDs are completely empty in a trace because of regional browsing preferences.

Interesting, last column of Table 6.4 shows that active crawling is not sufficient for generating comprehensive BoDs. I learned BoDs by active crawling the homepage of top-100 Alexa sites. Those classify as little as 32% of the flows for the ADSL trace. Therefore, AWESoME deployments must include mechanisms for in-place training.

### 6.5.4   Per service performance

I investigate further AWESoME performance by breaking down results for popular services in my datasets. Figure 6.9 shows precision and recall obtained when learning BoDs using 10 days training on *ISP*1 and applying them to the Simple-Browsing trace.[6]

Figure 6.9 shows that precision is typically higher than 97% excluding Facebook and Linkedin. That is, false positives are generally very rare unless for those service that are (i) extremely popular and (ii) both core and support domains. AWESoME may consider a support domain as a new core in these cases. Recall is typically higher than 80% – i.e., some support domains are not associated to the right service, typically becoming unknown. For management purposes, this translates into a marginal probability of wrongly treating few flows of a service of interest, i.e., AWESoME most common errors mark as *unknown* traffic that should not be considered unknown, but rarely assign to a wrong class (the precision is generally higher than recall). For instance, in some case, some support domains are not identified and handled as all other flows in default classes. Finally, I argue that this a pessimistic scenario, as AWESoME has been instrumented to discern all 3 760 Core Domains in the trace i.e., it must classify the traffic into the same number of classes. In a real deployment, I expect lower misclassification probability.

---

[6]Precision is calculated as the percentage of flows correctly identified as belonging to a service, whereas recall indicates the percentage of flows of the service that is identified.

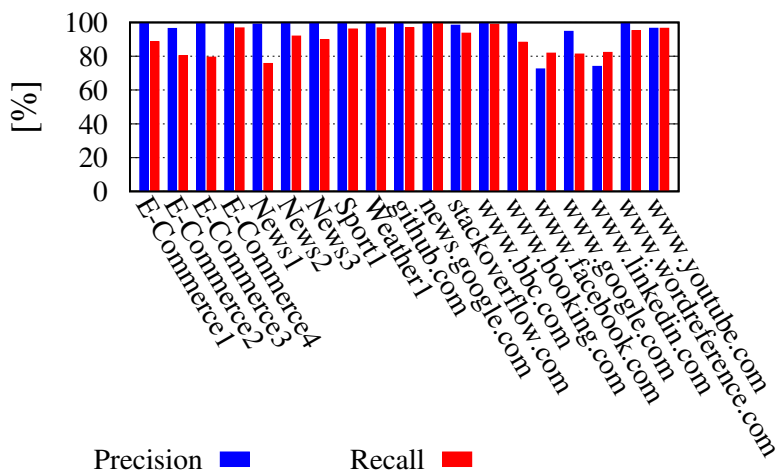

Figure 6.9: Precision and recall for popular services.

### 6.5.5 Is AWESoME scalable?

Finally, I evaluate three key aspects for a practical AWESoME deployment: (i) its overall run-time to take a decision when a new packet is received by the controller; (ii) the number of packets that need to be handled by the SDN controller; (iii) the number of rules that are installed in forwarding devices.

AWESoME has been prototyped in Python. Figure 6.10 shows the CDF of the execution time of my prototype for each packet that arrives at the controller. I found that AWESoME running on a commodity server takes less than 100 $\mu s$ to take a decision for more than 99% of the packets reaching the SDN controller. That is, AWESoME internals add only negligible delays per flow.

I now focus on the number of packets the controller has to handle. I use the operational network traces for this. Figure 6.11 depicts the number of packets per second forwarded to the controller. Different experiment rounds are executed, including the top-$n$ most active *ClientIP* addresses in each round. Remind that client IP addresses are equivalent to home gateways in ISP traces and to unique users in corporate traces. Box plots depict the distributions of packets per 1-second time bins, with boxes ranging from the 1st to 3rd quartiles, and whiskers marking 5ft and 95ft percentiles. Only ADSL, *Corp* 1 and *Corp* 2 are shown to improve visualization.

In summary, the packet arrival rate at the controller is very low. Focusing on the right-most point in Figure 6.11, notice that 4 096 terminals generate less than 1 000 packets/s for more than 95% of the time bins. For the sake of comparison, my AWESoME Python implementation can handle more than 40 000 packets per second. That is, even for large numbers of clients, AWESoME deployment is scalable thanks to its ability to take decisions using only DNS traffic and a single packet per flow.

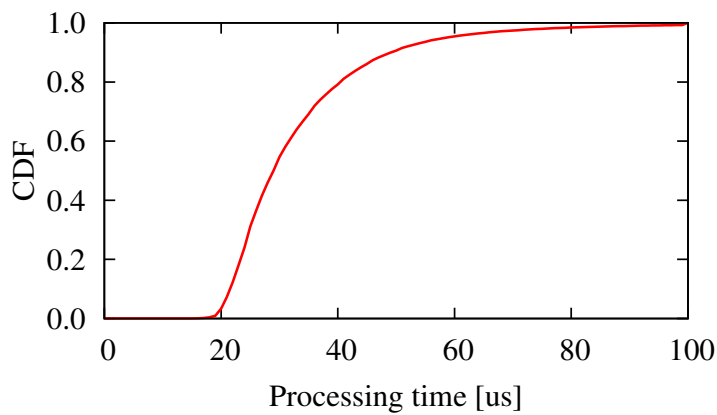

Figure 6.10: Processing speed of AWESoME for each packet arriving at the controller in a single-core of a commodity server.

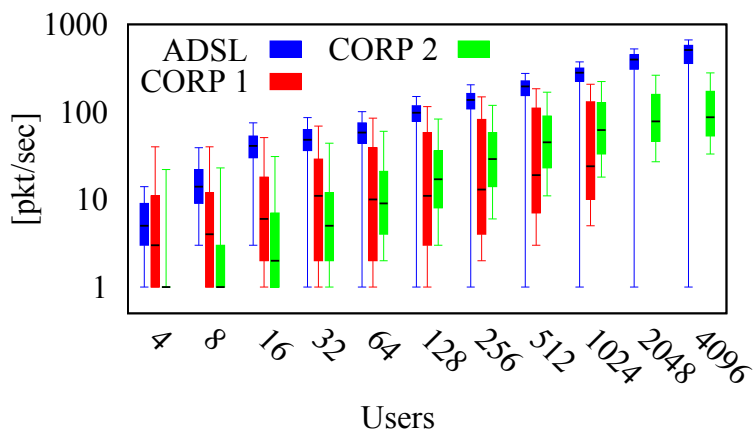

Figure 6.11: Packet arrival rate at the controller. Even for large numbers of clients, the number of packets handled by the controller is limited.

Finally, I investigate the number of rules that are installed on the SDN switches. This aspect must be necessarily taken into account, since switches typically can host a limited number of rules (i.e., < 10 000). Notice that AWESoME imposes the largest load in edge switches, where packet policing and tagging are performed on a *per flow* basis (see transient *per flow* rules in Table 6.2b). Other upstream elements (i.e., SDN switches in the core of the network) instead operate on a *per service* basis, and are programmed using *stable rules* as illustrated in Table 6.2c. As such, the loading on switches that are upstream in the network should be lower than on edge switches. In the example of Figure 6.2, stable rules are based on VLAN tags and impose only one rule per traffic class in the core switches.

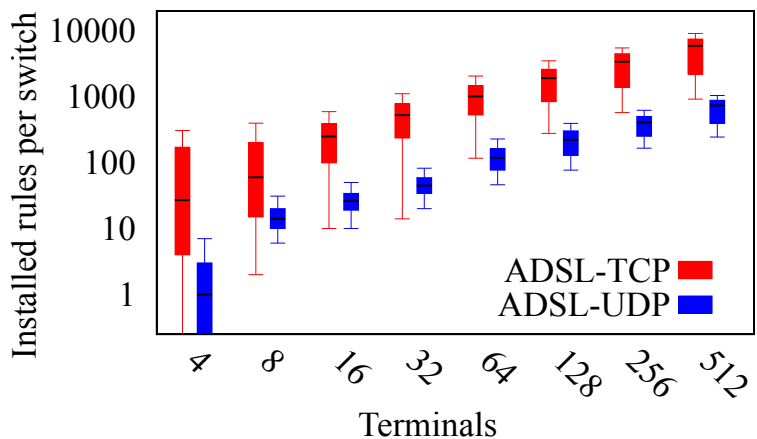

Figure 6.12: Active rules assuming that the top-*n* clients are connected to a bottleneck-switch and *all* services are managed. The number of installed rules is limited.

I estimate the number of rules that would need to be installed in edge switches creating a scenario where I assume: (i) the top-*n* most active clients are connected to a single edge switch; (ii) the network administrators policy *all* services in the network, thus requiring rules to manage every TCP/UDP flow individually; (iii) rules stay active for `Idle Timeout = 120 s` after the last flow packet.

Figure 6.12 illustrates the distribution of the number of rules installed in the edge switch. The distribution is calculated monitoring the flow table while replaying `ADSL` trace. The box plots follow the same characteristics as in Figure 6.11. This study is limited to `ADSL` for the sake of brevity. Similar results are obtained using the available traces. Focusing on the right-most point of Figure 6.12, notice that the flow table occupancy is low even when more than 500 terminals are connected on a single switch. The switch would rarely observe more than 5 000 active rules. The figure also breaks down numbers per TCP and UDP flows, showing that most rules would be related to (long-lived) TCP flows. In real deployments, where only few services of interest are managed, I expect AWESoME to put a negligible number of rules per switch.

## 6.5.6 Limitations and future work

Previous sections have shown that AWESoME is able to steer traffic per service with an overall accuracy of about 90%. Whether or not this accuracy is sufficient depends on the target application. For traffic engineering in a corporate scenario (see Figure 6.2)

AWESoME accuracy is appropriate. For example, it would allow to steer/segregate unwanted video traffic to slower paths, prioritize corporate cloud services, etc. AWESoME would steer 90% of the flows on the paths selected by the network administrator, with the wrongly routed flows imposing minor loads to the remaining paths. Comparing this error rate to today's alternatives (e.g., routing based on IP addresses of core domains – see Table 6.1), I believe AWESoME is a step forward for traffic engineering. I also argue that 10% of errors is enough for most traffic engineering tasks, where a mistake would result only in a longer/slower path without compromising the use of the service.

However, some scenarios may not tolerate *any* false positives, which is the case for some security applications. Devising per service tagging with zero false positive rates for security purposes is left for future work.

AWESoME has some limitations originating from assumptions and design decisions. Those decisions are justified by my goal of keeping AWESoME as simple as possible. For instance, Algorithms I and II assume that services are interactive and, as such, core and support domains appear close in time. BoDs for background services cannot be learned by these Algorithms since the assumption does not hold for background services. My experience with the traces suggests that background services are, in general, easier to identify thanks to the machine-to-machine nature of the traffic and the low number of domains supporting the services. While AWESoME allows administrators to specify Static BoDs, extending the system to automatically learn BoDs of background services is a promising direction for future work.

Finally, AWESoME assumes edge switches are part of the SDN and aggregate a moderate number of users – e.g., users in home NAT or in a corporate LAN. AWESoME cannot be deployed if large numbers of users are aggregated behind a single address, such as in Carrier-grade NAT, unless edge switches inside the Carrier-grade NAT are part of the SDN.

## 6.6 Related work

### 6.6.1 Web service traffic identification

Many approaches for traffic identification have been proposed [15, 65], and different alternatives could be coupled with SDN to implement per service traffic management. DPI has been employed not only to classify traffic of web services [101, 118, 117], but also to bring service visibility into SDN [59]. The DPI-based approach however suffers from weaknesses when applied to SDN: (i) the number of packets to be forwarded to controllers or SDN applications can be high for common protocols; (ii) as encryption gains momentum, essential information cannot be observed, thus reducing its applicability.

AWESoME adopts a behavioral identification approach – i.e., traffic behavior is used to infer the services generating packets [64]. The AWESoME approach is innovative in that it builds models based on server hostnames as they are resolved by clients. As such, AWESoME can differentiate services even if they use exactly the same protocols (e.g., HTTPS) and are hosted in the same infra-structure (e.g., in the case of CDNs and cloud hosting).

The idea of annotate traffic on-the-fly using DNS information has appeared in [7, 37, 78, 85]. However, AWESoME not only annotates flows and classifies traffic on a per-flow basis, but also automatically clusters third-party flows triggered by services. Thus, AWESoME is able to manage traffic even if flows are annotated with uninformative or ambiguous hostnames.

AWESoME relies on the fact, exploited in other works [63, 87], that flows triggered by a service present temporal correlations. AWESoME extends the approach to named-flows, tunes it to operate in real-time scenarios, and integrates the algorithms into SDN, so to control the network based on complex traffic relationships.[7]

Finally, authors of [21, 22] exploit relationships between flows for traffic management. They leverage groups of flows, or *coflows*, to boost performance of MapReduce applications. Their solutions are designed for data centers and require application-level modifications, while AWESoME uses a completely in-network approach. Moreover, AWESoME aims at managing services at the edge of the network. Thus, AWESoME needs to manage a vast number of services that may behave differently from each other. Automatic identification of coflows is proposed in [120], but the solution is also limited to data centers, facing limitations if deployed at edge networks. AWESoME instead builds models for the services automatically, identifying service traffic based on the DNS.

### 6.6.2 Service-awareness in SDNs

SDN has become very popular from academic environments [72] to large-scale data-centers [58], sparking a host of applications, such as SDN-based routing [111] and Internet exchanges [46]. Most of the SDN applications proposed to date (see [66] for a comprehensive survey) however are a good fit to forwarding rules expressed using information from L2-L4 headers, as it is typical of popular SDN implementations.

Authors of [89] mention the lack of support for L7 applications in SDN. They make a first step towards it by solving in SDN traffic steering functions traditionally performed by middle-boxes – e.g., firewalls, proxies, intrusion detection systems etc. Like us, they advocate a solution that requires no changes to SDN standards. AWESoME is a next step into bringing L7 support to SDN. It builds upon the traffic exchanged with the DNS to perform advanced traffic steering, enabling flexible managing of complex web services.

Few works have proposed low-level (stateless) forwarding rules to comprehensively manage complex services. Some authors have focused on specific services [114] or scenarios where communication patterns are well-known [121]. Authors of [86] propose a collaboration between CDNs and ISPs to allow content-based traffic engineering and an efficient server selection. In contrast, AWESoME learns how generic services communicate only based on network traffic and then relies on traditional L2-L4-based forwarding rules to handle the corresponding traffic.

---

[7]An off-line version of the algorithm used by AWESoME to learn bags of domains has aleady been used in Chapter 5.

Other works propose extensions to the SDN architecture either to delegate to switches (i.e., the data plane) management tasks that are based on L7 information, or to customize controllers and the data plane for stateful management [74, 79]. AWESoME instead is a SDN application that requires no particular changes in the lower layers of the SDN architecture.

I am aware of only few works that propose SDN applications to manage general web services [8, 59, 81, 88]. They use different methodologies to classify flows – e.g., forwarding the first $N$ packets of each flow to controllers or implementing machine learning algorithms. These works however miss dependencies among flows, as depicted in Figure 6.1, and, as such, fail in handling CDN/cloud traffic. Moreover, AWESoME requires to analyze only the first packet of each flow, which reduces the load at controllers.

## 6.7 Conclusions

I introduced the concept of "per service" management with SDN. This allows the network administrators to define policies to handle all traffic exchanged by terminals when accessing complicated web services today served by multiple domains and servers.

I presented and evaluated AWESoME to enable the per service management with SDN. It leverages DNS and the concept of Bag of Domains to associate the first packet of each flow to the originating service. I showed that AWESoME is accurate and poses a marginal load on the SDN controllers and switches, thus enabling fine grained control in practice.

I believe the concept of per service management can foster new studies, e.g., to improve the classification up to make it compatible with security applications, where high accuracy is mandatory, or to develop anomaly detection based on BoDs and per service accounting.

# Chapter 7

# PAIN: A Passive Web Performance Indicator for ISPs

The work I present in this chapter is mostly taken from my paper "*PAIN: A Passive Web Speed Indicator for ISPs*" presented in the 2nd Workshop on QoE-based Analysis and Management of Data Communication Networks (Internet-QoE 2017) [103].

## 7.1   Introduction

Metrics related to Quality of Experience (QoE) are key to understand how users enjoy the web. Such metrics are of prime importance to all actors involved in the service delivery. From Content Providers, which need to monitor users' satisfaction to maintain or increase their user base, to Internet Service Providers (ISPs), which need to be aware of performance offered by the network and factors affecting web browsing experience [92]. The idea that unsatisfied users are more prone to switch providers is widely disseminated. More than that, there are many anecdotal evidences that a small deterioration of quality levels could result in losses of revenues to providers.[1]

Given the importance of QoE monitoring, Content Providers have developed a number of alternatives to estimate users' QoE. On the contrary, there are hardly any methods to estimate users' QoE at ISPs [10, 4, 92], even if they are equally blamed for poor users' experience. Bad performance in the network and, in particular, in the last-mile is historically the first suspect when users' quality degrades. This has motivated major Content Providers to publicize rankings of ISP performance.[2] It is no exaggeration to say that ISPs are evaluated based on the experience of end-users while interacting with third-party services, with video and web browsing being the most important. In addition, ISPs need to measure the impact of possible network configuration changes on performance – e.g., to decide whether the deployment of web caches or new content delivery nodes is advantageous, or to tune configuration parameters of their networks.

---

[1]https://www.fastcompany.com/1825005/how-one-second-could-cost-amazon-16-billion-sales

[2]For an example, see `https://ispspeedindex.netflix.com/`

Users' QoE is intrinsically subjective, thus hard to be assessed and quantified. Previous works [4, 10, 18] have proposed objective metrics that have been shown to be correlated with users' Mean Opinion Score (MOS), even if a model to predict MOS is still hard to get [12]. These metrics however either are computed at the server-side (i.e., available to Content Providers only) or require ground truth from in-browser instrumentation (i.e., not scalable for the monitoring of a large number of sites at ISPs). Passive solutions that provide visibility into web performance are rare, and generally complicated by the need to analyze payload to reconstruct web pages [92].

Here, I introduce PAIN (PAssive INdicator), a completely unsupervised system to monitor web page performance using passive traffic logs. The adoption of encryption (e.g., HTTPS) makes solutions that reconstruct web sessions from payload [10, 4, 92] no longer effective. PAIN instead relies only on L4-level statistics (e.g., Netflow), annotated with the original server domain[3] information [7] to compute a synthetic indicator of the web page rendering time. Despite the passive approach, it combines machine learning approaches and techniques guided by the domain knowledge to reach the challenging goal of measuring browsing speed for encrypted websites. The design of PAIN is complex, but then its deployment is straight-forward, as no manual intervention or tuning is needed.

I validate PAIN in a testbed, which browses web pages while collecting also classic client-side objective metrics. I show that PAIN is able to spot changes in network conditions, reporting quality degradation when the page load time increases. PAIN metrics are strongly correlated with objective metrics based on client instrumentation, that are, in turn, shown to be correlated to users' QoE by literature works [10]. Finally, PAIN outperforms alternatives, either by avoiding expensive training or by effectively working with encrypted traffic.

I demonstrate the practical application of PAIN in a case study. I run PAIN on ISP network traces for one full year. First, I show how PAIN can help the ISP understand its users' experience, e.g., highlighting web browsing performance of users connected with different Internet access capacity. In particular, it allows to study the penalty in performance among users with good/poor access link conditions, topic of particular interest for the ISP, and explicitly requested to our research team. Then, I show how PAIN lets the ISP quantify variations in web browsing performance, e.g., pinpointing sudden performance variations of websites.

PAIN is open-source, and it is released as a module of the NetLytics Big Data platform [102]. It can be fed using Tstat [107], Squid [99] and Bro [84], to extract performance metrics directly from raw log files.

---

[3]I use the term *domain* informally throughout the chapter, meaning Fully Qualified Domain Name (FQDN).

In the following, Section 7.2 details the problem and envisioned deployment scenario, while Section 7.3 summarizes related work. Section 7.4 describes PAIN design and algorithms. Section 7.5 introduces the employed datasets. Section 7.6 validates PAIN, while Section 7.7 describes my experience of running PAIN on operational network traces. Finally, Section 7.8 concludes the chapter.

## 7.2 The complexity of QoE estimation

### 7.2.1 Objective QoE-related metrics

Given the intrinsic subjectiveness of QoE, measuring it is hardly possible without involving the users directly. Therefore, large-scale measurement campaigns are usually infeasible. Not a surprise, several approaches exist to estimate QoE with objective metrics calculated without human intervention.

In this chapter I focus on users' experience while browsing the web. Two of the most popular objective metrics to estimate users' QoE in this scenario are:

(i) **OnLoad** time: The time browsers fire the `onLoad` event – i.e., when all elements of the page, including images, style sheets and scripts have been downloaded and processed. This metric is widely used in literature to quantify web performance despite some well-known pitfalls: (i) a single slow page elements could negatively affect **On-Load** time, (ii) asynchronous scripts might be programmed to load after the **OnLoad** event is fired by the browser.

(ii) **SpeedIndex**: Proposed by Google[4], it represents the delay to render the visible portions of a page. It is computed by capturing the video of the page loading in the browser and tracking its visual progress.

These metrics are computed by the web browser at client-side. Collecting them requires the access of users' devices. Content providers and websites usually instrument services to collect such metrics from web browsers and upload results to servers as pages are loaded.

### 7.2.2 Challenges for estimating QoE from network traffic

QoE estimation based on Deep Packet Inspection [92] can no longer work, due to the deployment of encrypted protocols. New methods to estimate QoE must therefore be compatible with the data visible in the network.

ISPs can still rely on flow-level monitoring [51], which provides coarse-grained data about the activity collected at the network and transport layers. Moreover, ISPs usually control key Internet services, e.g., the DNS. PAIN exploits flow-level measurements and DNS information to build models for the traffic of given websites. In the remainder of the chapter I assume that both flow level and DNS measurements are available at the ISP. Proposed protocols designed to encrypt DNS traffic (like DNSCrypt and DNS-over-TLS), despite being only at an early phase, would however complicate the design of PAIN.

---

[4]https://developers.google.com/speed/docs/insights/about

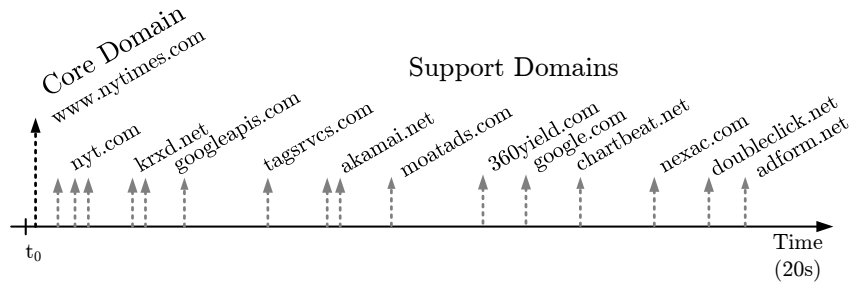

Figure 7.1: Sample of flows in a visit to *www.nytimes.com*. I use the time to contact support domains to monitor performance.

Estimating QoE metrics from such coarse-grained data is not trivial. The complexity of websites has dramatically increased over the years [54], and loading a web page requires reaching dozens of servers and fetching hundreds of objects. I provide some real examples of flow-level measurements obtained during visits to arbitrary websites to illustrate the challenges for extracting meaningful performance metrics. Once users reach a website, her browser opens multiple flows to different servers to fetch HTML objects, scripts and media content. I call the domain associated with the first contacted server the *Core Domain* and the remaining contacted domains *Support Domains*.

Figure 7.1 provides a simplified example: arrows represent the time in which flows to support domains start while the user is visiting the core domain *www.nytimes.com*. In this example, loading the web page requires the browser to issue 16 flows to 12 different servers. PAIN has to infer a performance indicator from this kind of traces, which are influenced by browser configurations, website designs, network configuration, etc.

Figure 7.2 depicts a complete example, where I report *all* flows to support domains opened during a visit to *www.bbc.co.uk*. This visit has taken around 6 seconds to load all objects. The browser has contacted 94 (unique) support domains. Black lines in the picture represent notable browser events. The browser starts rendering the page at 0.7*s* and finishes parsing the HTML document at time 1.6*s*, when the browser has downloaded mainly HTML objects and JavaScripts. Then, it starts to download other page objects (e.g., images and style sheets), firing the `onLoad` event only at 5.4*s*. After this, the browser continues to download elements from other servers (and opening new flows). In this example, the page triggers 27 additional connections to domains hosting analytics, advertisements, etc.

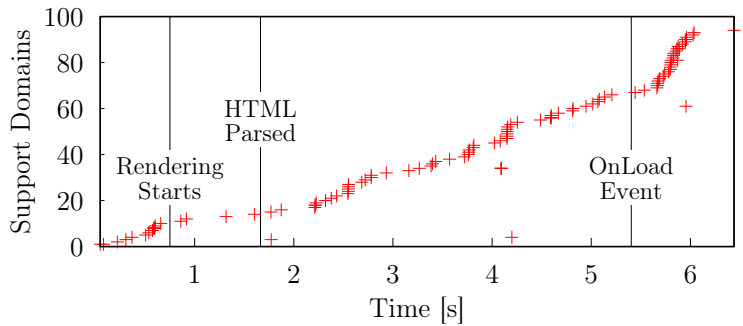

Figure 7.2: Support domain flows for a visit to *www.bbc.co.uk*. The browser contacted 94 support domains (*y*-axis) during 6 seconds (*x*-axis). Notable browser events are reported as vertical lines.

PAIN profits from support domains to estimate page load speed from flow timings. It is an unsupervised system that automatically learns typically contacted support domains after a core domain visit, and creates models describing the typical order in which support flows appear after the core domain visit. PAIN then considers the delay to observe support flows as performance indicators. PAIN assumes that clients' location does not vary, and the monitored set of clients is continuously connected to the Internet under a well-known access technology.

## 7.3 Related Work

Several previous works focus on estimating QoE from passive network measurements. Authors of [18] show that indirect metrics serve as indicators of users' MOS. According to [30], packet losses are strongly correlated with users' session abandonment events. Considering web browsing QoE, past works focus on the difficulty of its estimation, and propose objective metrics to this end. Egger *et al.* [32] show that user perceived web page load times may deviate from technical page load times, while Wang *et al.* [113] claim that in-browser computation and blocking Javascript are significant factors affecting perceived QoE. Metrics such as the `onLoad` time or `SpeedIndex` have been shown to be correlated with QoE metrics [10]. Authors of [10] also propose `ByteIndex` and `ObjectIndex` – metrics based on the bytes delivered to the client to render a page. Authors of [14] propose the `Above-The-Fold` metric to overcome the limitations of the naive `onLoad` approach. It is used in combination with classical metrics to predict users' MOS [27].

Past works targeting the ISP scenario either require DPI, or rely on ground truth from client browsers to train machine learning classifiers. Ibarrola *et al.* [53] build a network emulation system that, based on data collected thanks to volunteers, estimates

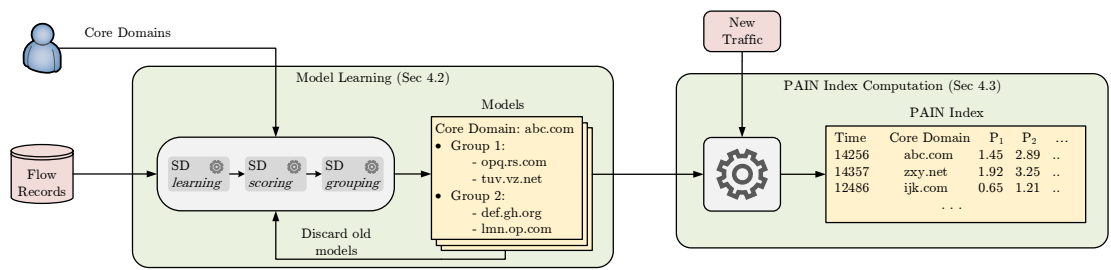

Figure 7.3: Architecture of PAIN. It learns and clusters support domains using flow records and a list of target core domains. The resulting groups are used to estimate performance.

QoE when varying network conditions. Shaikh *et al.* [95] study the correlation between physical layer metric with QoE of a single object page building on a lab testbed. A similar approach is used by Aggarwal *et al.* [3], where mobile devices carefully instrumented provide the ground truth to train models predicting QoE. Other works rely on deep packet inspection (DPI) of HTTP transactions to gather useful knowledge, but fail in a world where encryption is the norm [45]. Balachandran *et al.* [4] create models to predict web QoE from passive measurements on cellular networks examining the sequence of HTTP requests. Similarly, Sandvine industry products [92] build dependency graph of web pages to assess PLT, but are limited to not-encrypted traffic. Differently from past works, I have decided to follow the unsupervised approach, avoiding the need of a resource-consuming testbed to gather client-side metrics. PAIN automatically builds the models from flow-level traces, with no need to access to payload, and seamlessly operates with encrypted data carried over TLS/QUIC.

## 7.4   The PAIN system

PAIN is an unsupervised system composed by two blocks (see Figure 7.3). The *Model Learning* module analyzes flow records exported by monitoring devices and creates a model for each core domain of interest, i.e., it discovers and clusters support domains associated to specific websites. It must be continuously updated to cope with changes in web-page structure. The *PAIN Index Computation* module extracts the actual performance index using the previously built models. All algorithms scale linearly with respect to the input size (i.e., each flow record must be inspected just once), and support scalable processing using big data approaches offered by Apache Spark.

### 7.4.1   Input data

PAIN expects two inputs: (i) Flow records from traffic, and (ii) the list of Core Domains of interest.

Flow records are annotated with time and domain information: Given a flow $f$, identified by client and server IP addresses, client and server port numbers and the transport-layer protocol, $ts_f, te_f$ are the start and end timestamps, i.e., the time of the first and last packet of the flow. Each flow record must be enriched with information about the server domain $d$ requested by the client.

Flow meters typically export information from the network and transport layers, missing the association between server IP addresses and domain names. To get the server domain, different methods can be used. For example, DNS logs can be employed to extract queries/responses and annotate records in a post-processing phase [7]. Equally, some flow meters export such information on-the-fly directly from the measurement point for popular protocols [51]. For instance, Deep Packet Inspection allows one to extract the Server Name Identification (SNI) from encrypted TLS flows, or the server `Host:` header from plain-text HTTP flows.

The list of Core Domains is a user-defined list containing the set of core domains the ISP is interested in monitoring, such as popular websites accessed by users of the network. Since PAIN operates with L4-level measurements and domains names, the analyst must specify only the domain names to be monitored, and not full URLs. This allows PAIN to deal with encrypted traffic.

### 7.4.2   Model learning

The Learning Module observes the timings of flows as seen for the network traffic after a Core Domain. The first task is to learn which support domains are due to the core domain visit. PAIN learns that by focusing on the flows *commonly* occurring after core domains appearance. Section 7.6.5 shows that support domains are rather stable, and change slower than monthly.

Given that downloaded HTTP objects while rendering a page vary from visit to visit (e.g., because of caching, persistent connections, modification in the content, personalized content etc.), PAIN analyzes the order in which *groups* of support domains *typically* appear. The rationale is that some support domains may be missing in a visit, while others may not be relevant for indicating page rendering events (see Figure 7.2). PAIN uses groups of support domains to build models that are robust to such variations, i.e., tolerate missing or out-of-order support domains.

The combination of these building blocks lets PAIN model the typical behavior of the websites hosted in a core domain.

**Support domains learning**

PAIN learns support domains based on the methodology explained in Section 5.3.4. Let $C$ be the set of *core domains* of interest provided as input. PAIN training consists of learning the set of *support domains* $S_c$, for each core domain $c \in C$. The intuition is simple: When a client is observed opening a flow to the core domain $c$, the domains of flows that follow shall be considered within $S_c$. As illustrated in Section 5.3.4, PAIN evaluates flows after the core domain in a window $\Delta T$ long.[5] The impact of $\Delta T$ on the final results is discussed later in Section 7.6.1. Traffic from all clients contributes to $S_c$, so that information is accumulated over time and in different conditions, i.e., identities, browsers, devices, configurations, etc.

**Support domain scores**

Intuitively, the timeline of support flows reflects the speed at which a web page is rendered (recall Figure 7.2). Page elements hosted by third-party sites (e.g., images and advertisements) are requested after other components of the page (e.g., scripts) are processed. PAIN leverages this behavior to calculate a score for $d \in S_c$. The score is higher for support domains appearing further away in time from the core domain $c$ (e.g., right-most points in Figure 7.2).

However, the set of support domains varies from visit to visit. $S_c$ is constructed from many observation windows and not all support domains appear in every observation window due to caching and persistent connections. Equally, nothing prevents browsers or mobile apps from opening flows to third-parties in a different order while rendering pages.

To determine the score for each $d_i \in S_c$, PAIN computes a *dependency matrix* $\mathscr{M}_c$ of order $|S_c|$ for each core domain $c$. Each cell $\mathscr{M}_{c_{i,j}}$ represents the number of observations windows $OW_c$ in which the support domain $d_i$ has appeared *after* the support domain $d_j$ in time. Note that $\mathscr{M}_{c_{i,i}} = 0$. Similarly, $\mathscr{M}_{c_{i,j}} = |OW_c|$ only if $d_i$ appears always after $d_j$, and both $d_i$ and $d_j$ are in all observation windows for the core domain $c$. The score of $d_i$ is calculated as:

$$score(c, d_i) = \sum_j \mathscr{M}_{c_{i,j}} \tag{7.1}$$

Note that $score(c, d_i)$ is high if $d_i$ appears often later than other domains in the observations windows of $c$. Similarly, it is lower if $d_i$ usually appear close to the core domain. Algorithm 6 reports a pseudocode for the score calculation function. It processes one core domain at a time. PAIN computes the dependency matrix $\mathscr{M}$ (lines 1-6), and, then, uses it to provide the scores (lines 7-8).[6]

---

[5] The $\Delta T$ parameter has the same role of both $\Delta T_{OW}$ and $\Delta T_{EV}$ in Chapter 5.

[6] In PAIN implementation, Algorithms are executed on-the-fly as new traffic comes into the system.

---

**Algorithm 6** Compute the scores of support domains for the core domain $c$.

---

**Input:**

    $c$                                                   ▷ *Core domain to be processed*

    $S_c$                                                  ▷ *Support domains of c*

    $OW_c$                                           ▷ *Observation windows for c*

1:  $\mathcal{M} \in \mathbb{R}^{|S_c| \times |S_c|}$                     ▷ *Define the dependency matrix*

2:  **for** $ow$ in $OW_c$ **do**             ▷ *For each observation window*

3:    **for** $d_i$ in $ow$ **do**           ▷ *For each support domain in ow*

4:      **for** $d_j$ before $d_i$ in $ow$ **do**      ▷ *Supports before $d_i$ in time*

5:        **if** $d_i \in S_c \wedge d_j \in S_c$ **then**

6:          $\mathcal{M}_{i,j} \mathrel{+}= 1$      ▷ *Increment $\mathcal{M}_{i,j}$ if supports are in $S_c$*

7:  **for** $d_i$ in $S_c$ **do**       ▷ *Compute score for each support domain*

8:    $score(c, d_i) = \sum_j \mathcal{M}_{i,j}$         ▷ *Sum the row of $\mathcal{M}$*

---

**Support domain grouping**

After scoring, PAIN identifies groups of support domains. I propose a simple rule that considers *groups* of support domains, which gives robust outcomes based on my tests. By clustering the support domains in some *few* groups, I filter out the noise caused by missing support domains, besides creating groups of domains that are strongly correlated to web performance.

More precisely, I sort $d_i \in S_c$ in increasing order of $score(c, d_i)$ and split the domains in $n$ groups in $G_c$, where groups have at least $|G_{c_k}| = \left\lceil \frac{|S_c|}{n} \right\rceil$ support domains. $G_{c_1}$ will contain those support domains that often appear the closest to the core domain flow, wheres $G_{c_n}$ will have the support domains that often appear the furthest to the core domain. $n$ is a parameter to be investigated.

The set $G$ – i.e., groups of support domains for core domains $C$ – is the output of the Model Learning module.

### 7.4.3 PAIN index computation

The index computation module analyzes live traffic to provide a performance index. Like in the training phase, PAIN analyzes the traffic flows on a per-client basis, chronologically sorted by time. When it encounters a flow to a core domain $c$, it opens an *observation window* $\Delta T$ long. PAIN considers all support domain flows generated by the client within the OW, and accounts them to the corresponding group.

I measure the time at which flows in each group are observed. A visit to a group is considered completed at the time when the *last* flow in the group is observed. For each group $G_{c_i}$ with $i \in 1, \dots, n$, PAIN calculates the index $P_i$, equals to the time difference between the starting of the last flow in the group $i$ and the starting time of the core domains $c$. Note that groups can be absent if none of its support domains is in $OW$. This can be typically caused by two phenomena: (i) the browser cache contains all the objects that are hosted on a particular domain and (ii) the browser already opened a persistent connection toward the target domain. In this case, I do not consider the sample.

I tested different criteria in place of *last* per group (e.g., average and median) and all lead to worse results. The intuition is that the web page performance is mainly driven by the ability of the browser to obtain objects to render the page, which correlates well with the time late flows are observed in the network. Using the last flow per group also highlights possible degradation of specific servers involved in serving the content.

The tuple $P = \{P_1, \ldots, P_n\}$ represents the performance index for a given visit to the core domain $c$. By considering all visits from all clients to $c$, PAIN builds statistics on the performance faced by clients. Due to the intrinsic noisiness of flow-level measurements, PAIN assumes relevance when multiple measures are aggregated to contrast different users, time periods or locations.

### 7.4.4   Design decisions, caveats and limitations

The decision of making PAIN a completely unsupervised system is motivated by my goal to monitor a vast range of websites. The system is expected to receive only the list of core domains of interest. It learns models directly from traffic, without requiring human intervention or any information collected at the client-side. Due to this design choice, PAIN does not directly provide QoE figures, as this would require expensive campaigns involving users directly. However, Section 7.6 shows that PAIN index is strongly correlated with objective metrics that have been identified as good proxies to users' satisfaction [27]. Even if I did not perform experiments in that direction, it is possible to calculate boundaries for the correlation between PAIN and QoE leveraging partial correlation properties.[7] Given my experiments, it is possible to demonstrate that correlation between PAIN index and QoE is always positive and in the interval $(0.3, 0.9)$.

Other designs would be possible too, such as by using supervised algorithms. The system could train the model from network traffic assuming client-side metrics are present. Such a supervised design would result in a system that requires ground truth data captured at client-side for each core domain of interest. The supervised approach would allow one to guess the actual value of objective metrics (e.g., `OnLoad` and `SpeedIndex`), but I argue that the absolute values of such indicators are far less useful than contrasting across different users and conditions. PAIN is fully able to pinpoint variations in objective metrics (see Section 7.6.2) despite not being able to estimate their absolute values.

The deployment of the supervised alternative requires a resource-consuming testbed, in which training should be performed periodically for each monitored websites. I have decided to follow the unsupervised approach, since it broaden the PAIN deployability and dramatically enhances training scalability. In Section 7.6.4 I consider a simple supervised approach and compare it to PAIN. I show that it brings limited benefits.

---

[7]Given $\rho_{AB}$ and $\rho_{AC}$, one can obtain bounds for $\rho_{BC}$ with $\rho_{BC} = \rho_{AB}\rho_{AC} \pm \sqrt{1 - \rho_{AB}^2}\sqrt{1 - \rho_{AC}^2}$.

Table 7.1: Description of datasets.

| Dataset | Size | Collected on | Collection Environment |
|---|---|---|---|
| `SynthTypical` | 11 GB | Testbed | 10 websites × 4 (emulated) devices × 8 emulated typical access links |
| `SynthDegraded` | 11.4 GB | Testbed | 2 websites × 4 (emulated) devices × manually degraded access link conditions |
| ADSL | 495 GB | ISP network | > 100 K websites × 10,000 ADSL installations × 1 year |

## 7.5  Datasets

In this section I describe my validation datasets. I employ both synthetic datasets generated using a testbed, and real world traces collected in an operational network. They are summarized in Table 7.1.

### 7.5.1  Synthetic traces

**Testbed**

Synthetic traces produced in a testbed allow to compare PAIN to objective metrics directly collected in the browser. I instrument a PC with *WebPageTest* [115], a tool for web performance assessment. WebPageTest emulates networks based on DummyNet [17], a network emulation tool. Given a list of URLs, it automatically navigates through each page while saving detailed statistics. Many options are available, including the choice of client browser (Chrome and Firefox), device (PCs, tablets and smartphones) and network emulation (e.g., 3G, DSL and Cable). It thus provides the means to emulate users' browsing considering realistic clients and network conditions.

WebPageTest exports the HTTP Archive (HAR) [49] for each page visit. It contains information about the visit as well as statistics for each object: from HTTP-headers, to network-level statistics that describe the TCP connections opened to download objects, including the time in which the TCP connection starts, and the domain associated with it.

Additionally, WebPageTest computes many objective metrics related to QoE. Here, I consider the `OnLoad` and the `SpeedIndex` (see Section 7.2).

**Synthetic datasets**

I build two datasets to validate PAIN, namely `SynthTypical` and `SynthDegraded`, with respectively *typical* and *degraded* network conditions.

Table 7.2: Browsers and emulated devices in the testbed.

| Browser | Device | Operating System |
|---|---|---|
| Mozilla Firefox | PC | Windows 10 |
| Google Chrome | PC | Windows 10 |
| Google Chrome | Nexus 7 | Android |
| Google Chrome | iPad Mini | iOS |

Table 7.3: Settings in the `SynthTypical` dataset. *Native* corresponds to a scenario with no traffic shaping.

| Name | Down Link | Up Link | RTT |
|---|---|---|---|
| Native | 1 Gbit/s | 1 Gbit/s | native |
| FIOS | 20 Mbit/s | 5 Mbit/s | 4 ms |
| Cable | 5 Mbit/s | 1 Mbit/s | 28 ms |
| DSL | 1.5 Mbit/s | 1 Mbit/s | 50 ms |
| LTE | 12 Mbit/s | 12 Mbit/s | 70 ms |
| 3G Fast | 1.6 Mbit/s | 768 Kbit/s | 150 ms |
| 3G | 1.6 Mbit/s | 768 Kbit/s | 200 ms |
| 3G Slow | 780 Kbit/s | 330 Kbit/s | 200 ms |

The `SynthTypical` dataset is built by letting WebPageTest visit 10 popular domains in Italy (listed in Table 7.4). For each domain, WebPageTest visits the homepage and 9 internal pages for a total of 100 pages.

Since PAIN must work seamlessly regardless of client configurations, I consider 4 different browser and device combinations, which I summarize in Table 7.2. I consider both Firefox and Chrome running on PCs and I leverage Chrome's features to emulate its use on a smartphone and on a tablet. I did not consider other browsers such as Explorer or Safari, as not handled by the Linux version of WebPageTest.

I further consider 8 access network technologies summarized in Table 7.3. These are emulated by WebPageTest by imposing traffic shaping policies that mimic actual parameters of the technologies. The *Native* case has no shaping – i.e., the 1 Gbps Ethernet network connecting the testbed is used without changes. For other cases, DummyNet enforces typical bandwidth and Round Trip Time (RTT) faced by users of a given technology.

I visit each page twice for each setup: (i) with empty browser cache; and (ii) few seconds later for profiting from caching. The traffic is expected to vary strongly, since many objects are cached in the second case, complicating the identification of support domains. In total, WebPageTest recorded 6 400 visits while building this first dataset (all visits have been completed in about 48 h).

The second dataset, `SynthDegraded`, represents artificial conditions, in which I enforce link delay or bandwidth limits. I simulate scenarios in page load time increases caused by worsening network conditions. I simulate 10 cases: (i) adding from 100 ms to

Table 7.4: Support domains for websites in `SynthTypical` dataset, together with the probability they appear after `onLoad`.

| Core domain | Support Domains | | | After OnLoad |
|---|---|---|---|---|
| | Min | Median | Max | |
| www.corriere.it | 30 | 57 | 137 | 2.2 % |
| www.ebay.it | 2 | 50 | 223 | 40.5 % |
| www.gazzetta.it | 25 | 58 | 138 | 6.5 % |
| www.ilmeteo.it | 17 | 56 | 185 | 18.5 % |
| www.lastampa.it | 14 | 34 | 81 | 8.7 % |
| www.meteo.it | 27 | 52 | 91 | 6.6 % |
| www.mymovies.it | 24 | 45 | 147 | 11.0 % |
| www.repubblica.it | 27 | 53 | 216 | 23.0 % |
| www.subito.it | 26 | 52 | 119 | 7.0 % |
| www.wordreference.com | 2 | 14 | 68 | 6.0 % |

500 ms extra per-packet delay and (ii) imposing a limit from 2.5 Mbit/s to 312.5 kbit/s on uplink and downlink access bandwidth. Again, I visit each page twice (cold and warm cache) and with 4 browsers. For the sake of brevity, I performed these experiments for 2 websites only, namely *www.repubblica.it* and *www.subito.it*. WebPageTest has performed 8 000 extra visits for building this second dataset (completed in about 60 h).

### 7.5.2 Support domains at a glance

I first provide high-level statistics about support domains in `SynthTypical` dataset (see Table 7.4). With these numbers, I aim at complementing Figure 7.2, illustrating the challenges to extract knowledge from support domains and their complex relations with the page loading process. Table 7.4 lists the websites in `SynthTypical` dataset. The 3rd column reports the median number of support domains across all visits: They vary from less than 20 to more than 50. The number and order at which support domain flows are opened significantly vary across visits (see 2nd and 4th columns of Table 7.4). This is no surprise as webpages of single website might be very different. However, considering $25^{th}$ and $75^{th}$ instead of minimum and maximum leads to considerably more consistent results, with a variations in the order 20-30%. More than that, support domains are often contacted after the `OnLoad` event has fired, e.g., due to browser pre-fetching or the presence of analytics scripts programmed to run after the page is loaded. I quantify the percentage of these cases in the 5th column of the table. Extreme is the case of *www.ebay.it*: More than 40% of connections are issued after the browser completed loading the page.

These results already hint for the importance of PAIN grouping step. For example, if one would naively take the delay of the last support flow as a performance indicator, the obtained metric would have very low correlation with objective QoE metrics observed at the client-side. Next, I validate several aspects of PAIN.

Table 7.5: Support domain persistence across different devices and subpages (`SynthTypical` dataset).

| Core domain | Support Domain Persistence | |
|---|---|---|
| | Devices | Subpages |
| www.corriere.it | 0.78 | 0.69 |
| www.ebay.it | 0.68 | 0.16 |
| www.gazzetta.it | 0.81 | 0.67 |
| www.ilmeteo.it | 0.68 | 0.90 |
| www.lastampa.it | 0.79 | 0.61 |
| www.meteo.it | 0.66 | 0.87 |
| www.mymovies.it | 0.59 | 0.69 |
| www.repubblica.it | 0.74 | 0.56 |
| www.subito.it | 0.58 | 0.82 |
| www.wordreference.com | 0.41 | 0.89 |

I then perform a second analysis aiming at quantifying the variability of support domains. Indeed, contacted support domains may change when varying the device used to load the page or when accessing different subpages of the same website. I thus compare the list of support domains obtained when considering each subpage separately. Then I compute the Jaccard index similarity coefficient [57] for all the pairs of subpages, and report the median values in Table 7.5, separately for each websites. On average, subpages share the majority of support domains, with *www.ilmeteo.it* reaching a 0.9 median similarity coefficient. A low value is observed only for *www.ebay.it*, where manual inspection reveals that few subpages have a simpler structure reflecting in a lower number of support domains. I repeat the operation for the 4 emulated devices in the `SynthTypical` dataset. Overall, varying the device used for accessing the webpage does not affect the contacted support domains. The lowest value is observed for *www.wordreference.com*, where the median similarity coefficient is 0.41. These results show that the set of contacted support domains is rather stable when varying subpages and employed device, allowing PAIN to create reliable and stable models, where device type and subpages cannot be observed by passive monitoring.

### 7.5.3   ISP flow traces

This dataset includes flow summaries exported by Tstat [107] in a real deployment. I use the `ADSL` dataset containing flow summaries for 10,000 ADSL subscribers. The measurement methodology and a further description can be found in Chapter 2. Important to this analysis, the ISP provides me the access link speed of each ADSL customer. Moreover, each customer is provided a fixed IP address and, thus, by inspecting the (anonymized) client IP addresses in the dataset, PAIN isolates flows per ADSL installation and use them as the per-client traces. Each trace includes information about traffic of all users' devices connected at home.

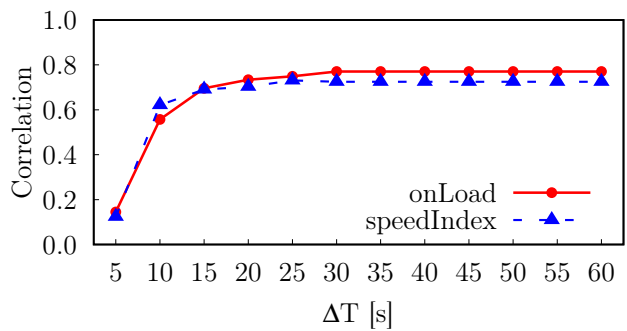

Figure 7.4: Spearman Correlation of $P_3$ with `onLoad` and `SpeedIndex` when varying $\Delta T$ (`SynthTypical` dataset).

I here consider data for the ADSL dataset for the whole year 2017. Considering only HTTP and HTTPS flows, I obtain 15 billion flows related to around 100,000 websites. This trace represents a realistic scenario of a possible PAIN deployment. No ground truth about associations of support and core domains is available in the dataset.

## 7.6 Validation

### 7.6.1 Tuning of parameters $\Delta T$ and $n$

I now tune the parameters $\Delta T$ and $n$. I rely on the `SynthTypical` dataset. I vary each parameter while comparing the PAIN index to the metrics exposed by my testbed, i.e., `onLoad` and `SpeedIndex`. Indeed, I want the PAIN index to be correlated with objective metrics. I quantify correlation using the Spearman's rank correlation coefficient between PAIN index and objective metrics [98]. A Spearman coefficient higher than 0.5 is usually considered a strong correlation indication.

I first observe the impact of the observation window choice ($\Delta T$) in Figure 7.4. Only the correlations between objective metrics and the 3rd group of support flows (i.e., $P_3$) are shown to improve visualization. Notice in the figure that PAIN achieves high correlation coefficients when $\Delta T$ increases. When $\Delta T$ value is larger than $30s$, results do not improve further. In a nutshell, PAIN is not very sensitive to $\Delta T$, as long as it is not too small. Provided that support domains are grouped, and each group is used to extract $P_i$, PAIN index remains mostly unaffected, even if some support domains are not associated to the respective core domain because $\Delta T$ is expired. In the following, I set $\Delta T = 30s$.

I next perform a similar analysis for $n$, the number of groups. I report results for the

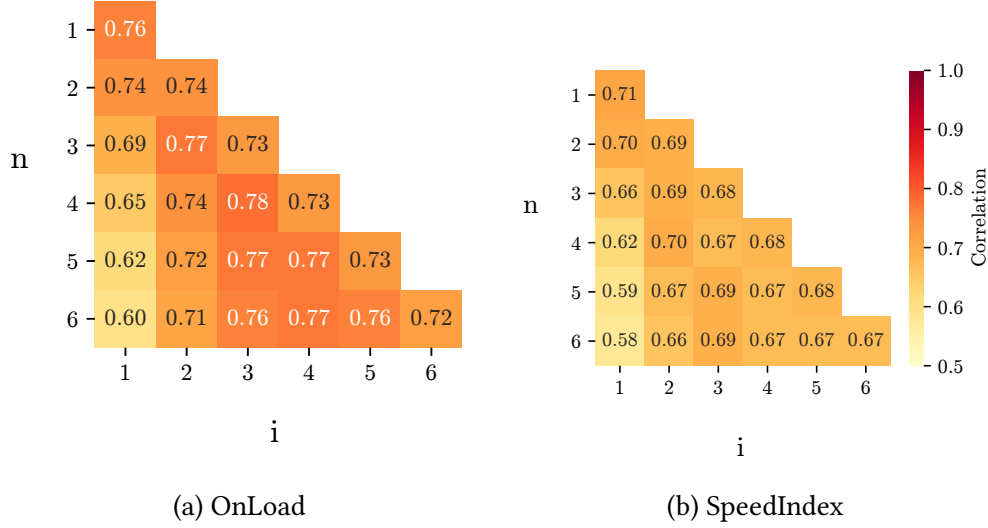

(a) OnLoad                           (b) SpeedIndex

Figure 7.5: Correlation of PAIN index with `onLoad` and `SpeedIndex` when varying the number of groups $n$ (`SynthTypical` dataset).

`SynthTypical` dataset for `onLoad` and `SpeedIndex` separately in Figure 7.5a and Figure 7.5b, respectively. Each row $j$ represents an experiment with a different $n \in [1,6]$. The column $i$ reports the correlation of $P_i$ when using $n = j$. For example, the left-most cell on the last row represents the correlation of $P_1$ with `onLoad` when using $n = 6$.

I notice that PAIN is not very sensitive to $n$ either, and the best values are observed for $n = 4$ and for $P_3$ and $P_4$ (they are the groups correlated the most with objective QoE metrics). Notice also that $P_3$ is slightly better than $P_4$, in particular when taking `onLoad` as reference. Using a small value of $n$ provides poor information. Considering multiple groups (i.e., more than one $P_i$), on the other hand, makes PAIN more robust to outliers.

I take $P_3$ with $n = 4$ for the remaining experiments. However, Figure 7.5a shows that small variations of $n$ and $P_i$ do not affect the results, and my experiments reinforce this claim.

## 7.6.2   Effects of network conditions

I check whether PAIN is able to reflect worsening on network conditions using the `SynthDegraded` dataset. Figure 7.6 illustrates PAIN index values when varying delay and bandwidth to reach the two websites in the dataset. Each point in the figure depicts the median value for the PAIN index over all tests with the given setup. Each point is the result of 80 runs, and includes experiments for different browsers, subpages, etc. Variability of such values is low and in no case higher than 15% from the median.

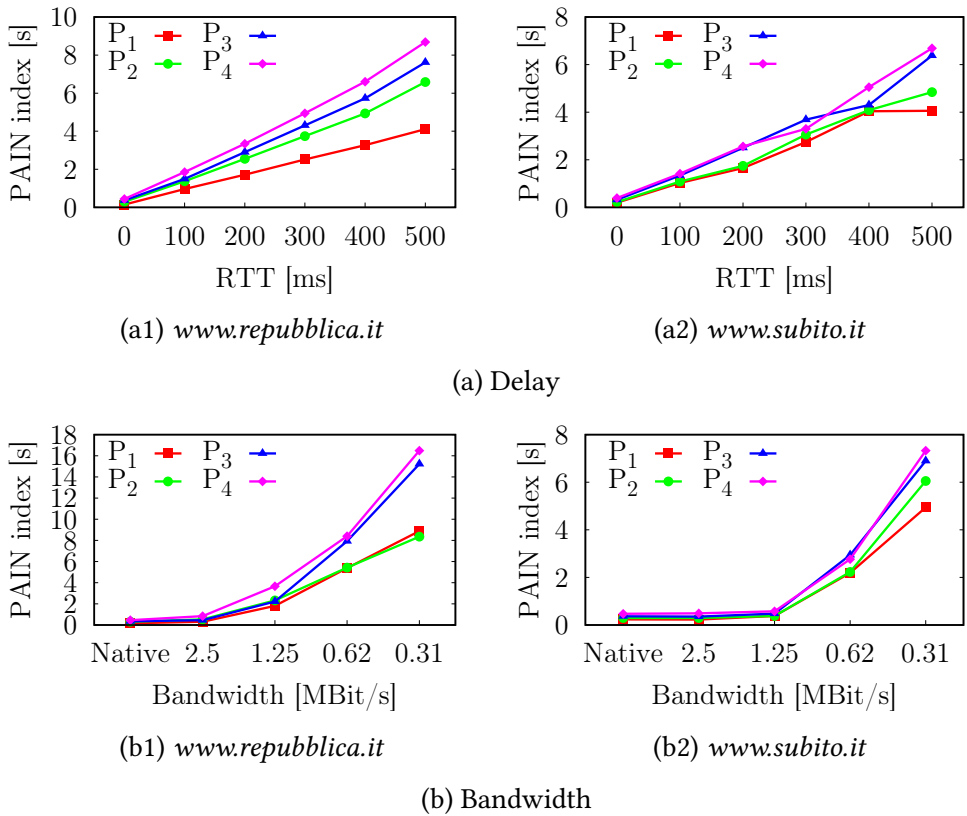

(a1) *www.repubblica.it*

(a2) *www.subito.it*

(a) Delay

(b1) *www.repubblica.it*

(b2) *www.subito.it*

(b) Bandwidth

Figure 7.6: Median time of PAIN index when varying delay and bandwidth (`SynthDegraded` dataset).

Consider Figure 7.6a, which refers to *www.repubblica.it* and *www.subito.it*, when RTT$\in$ [0,500]*ms*. PAIN index increases alongside the delay, starting from around 0.5*s* and up to almost 10*s* when RTT is 500 ms for *www.repubblica.it*. That is, $P_i$ reflects the network conditions and increases in case of degradation. Actual PAIN index values are sometimes inverted from their expected order for extreme values of RTT (e.g., $P_3$ larger than $P_4$). This is due to timers firing on the page that make not-deterministic the visiting order of support domains.[8] Similarly, in Figure 7.6a2, $P_1$ has a slightly lower value for 500 ms than for 400 ms. Indeed, this confirms that $P_1$ and $P_2$ are not good as sensible as $P_3$ and $P_4$ to network conditions, reinforcing results of Figure 7.5.

Similar considerations hold for Figure 7.6b, which shows the impact of download link capacity. When the available bandwidth is reduced, PAIN index increases. Observe that a bandwidth of 1.25 Mbit/s already implicates performance degradation for *www.repubblica.it*, while still no penalty is suffered by *www.subito.it*.

---

[8]I notice that the order at which a browser opens connections towards support domains is subject to variations, especially for those support domains appearing early in the page load process, and, thus, belonging to $P_1$ and $P_2$.

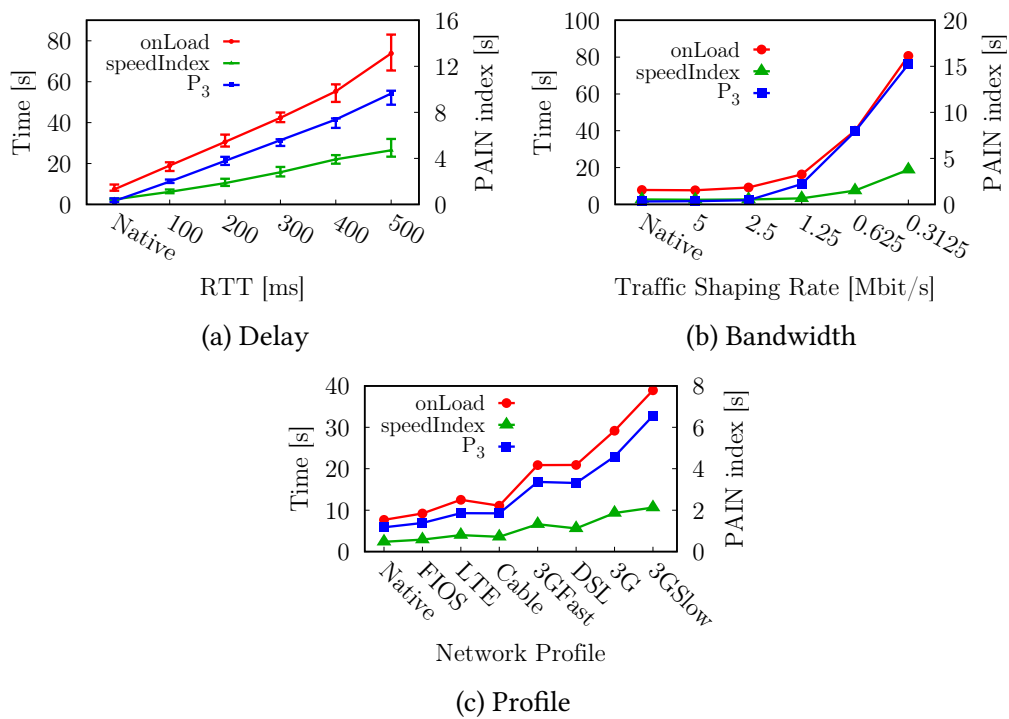

(a) Delay

(b) Bandwidth

(c) Profile

Figure 7.7: *www.repubblica.it* `onLoad`, `SpeedIndex` and $P_3$ for various setups (`SynthTypical` and `SynthDegraded` datasets).

In summary, results show that $P_i$ reflects the network conditions, allowing ISPs to track degradation on the network that impacts website performance. Very similar results are obtained considering $P_3$ and $P_4$.

### 7.6.3 Comparison to objective metrics

I have already seen in Figure 7.5 that PAIN index is correlated to objective QoE metrics. I now detail that analysis, by directly comparing the values of $P_3$ to the `SpeedIndex` and `onLoad`. I set $n = 4$ and $\Delta T = 30s$. Figure 7.7 reports results obtained for a single website in different scenarios. Similar figures are obtained for other cases. Again, the figure reports median values over 80 runs. Figure 7.7a also reports error bars that span over $25^{th}$ and $75^{th}$ percentiles. I use this figure to quantify the variability of my results that is always limited to less than 15% of the median value. Similar results are obtained for the other two figures, but they are not reported as error bars would overlap and compromise the readability.

Each point in Figure 7.7 represents the median value for all visits with the given network condition. Since the metrics have different absolute values, I use the *y*-axis in the left-hand side to report `SpeedIndex` and `onLoad` times, and the *y*-axis in the right-hand side to report values of the PAIN index. Thus, the figure shows whether the metrics present similar rate of variation given changes in the network conditions.

Focusing on Figure 7.7a notice how the three metrics grow almost linearly with the RTT. The rate of variation in PAIN (see blue line) is similar for `SpeedIndex` (green) and `onLoad` (red) ones. When varying the bandwidth in the degraded scenario (Fig. 7.7b), the values of PAIN index change similarly to the rate observed for `onLoad` time, but faster than `SpeedIndex`. PAIN is more sensitive to deterioration on the available capacity. Yet, results show that the PAIN index is directly related to page load times. Observe also that all three metrics are basically constant when the bandwidth is larger than 2.5 Mbit/s (see points in the left part of the figure). That is, the web page performance is not affected when a minimum bandwidth is available, and all three metrics reflect such behavior. Finally, Figure 7.7c reports the values for typical network scenarios. Again, I see similar patterns among the metrics, with the rate of variation of PAIN index in between the other metrics.

In summary, results show that the metrics are correlated, and they vary according to the network conditions similarly. Absolute values are in different ranges, but they all reflect degradation in quality.

### 7.6.4 Comparison to alternative approaches

I validate PAIN against two possible alternatives:

(i) **BestCheckpoint**: I use a *supervised* mechanism to extract a performance metric that tries to maximizes the correlation with objective metrics. Considering a training dataset and a core domain *c*, I extract the delay to observe each support domain $s \in S_c$ after all visits to *c*. Then, I compute the correlation coefficient between the delays for each $s \in S_c$ and the objective metrics (`SpeedIndex` and `onLoad`). I select the most correlated support domain to serve as landmark.

When evaluating new traffic, the delay to observe the landmark is considered as the performance metric for the given core domain. Note that this supervised approach requires the availability of per-site objective metrics at training time.

(ii) **BeaconCheckpoint**: This approach has been proposed by authors of [54]. It consists in leveraging the analytics objects typically present in web pages to identify when page loading is complete. The intuition comes from the fact that analytics services wait for the browser to finish rendering the page before sending back statistics to the server. Here, I consider the *Google Analytics* script that uploads statistics to Google servers after the `onLoad` event is fired by the browser. After finding a flow to the core domain of interest, I search the HTTP requests to *Google Analytics* URL. Note that such an approach requires non-encrypted traffic and works only for sites embedding analytics scripts (e.g., only present in 4 websites in `SynthTypical`).

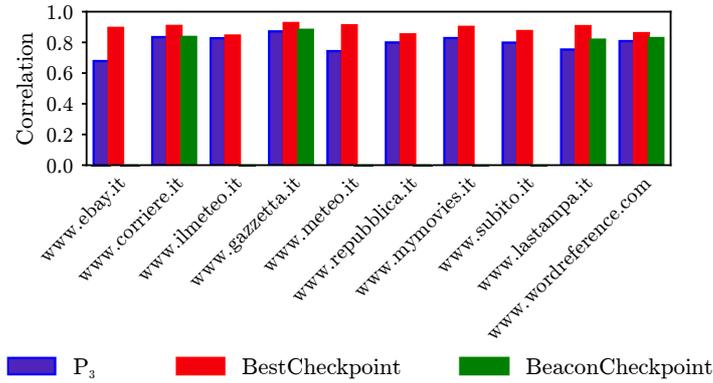

(a) OnLoad

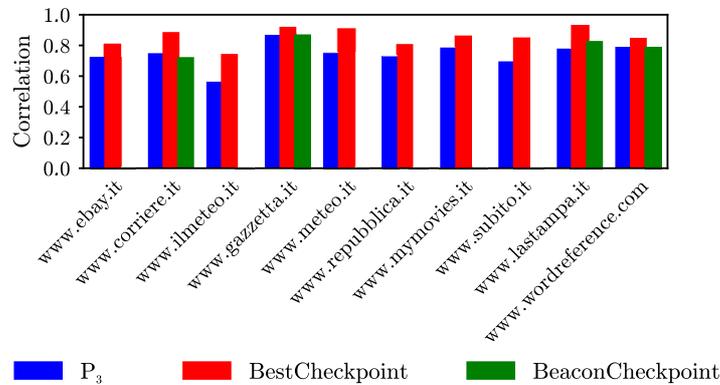

(b) SpeedIndex

Figure 7.8: Correlation of PAIN, **BestCheckpoint** and **BeaconCheckpoint** with objective metrics (`SynthTypical` dataset).

The delay between the core domain flow and *Google Analytics* HTTP request is reported as performance metric.

Figure 7.8 shows the correlation of PAIN, **BestCheckpoint** and **BeaconCheckpoint** with `SpeedIndex` and `OnLoad`. **BeaconCheckpoint** can be computed only for 4 websites. As I have seen before, PAIN correlation coefficients are positive and very high. Considering `onLoad` in Figure 7.8a, they range from 0.67 for *www.ebay.it* to 0.90 for *www.gazzetta.it*. Most values are close to 0.8 for both metrics. **BestCheckpoint** and **BeaconCheckpoint** are also positively correlated to the objective metrics. For example, for *www.gazzetta.it*, they achieve 0.92 and 0.88, respectively. **BestCheckpoint** is more strongly correlated to `onLoad` than PAIN. This is expected because of the supervised approach. Yet, absolute differences are small, showing that PAIN can achieve

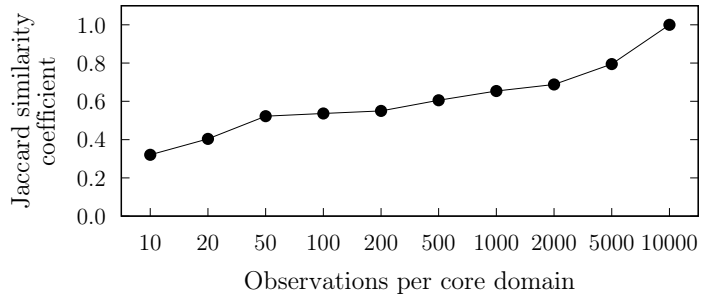

Figure 7.9: Support domains learned with increasing number of observations per core domain, compared to those learned with 10,000 observations (ADSL dataset).

similar performance without the burdens of building ground truth for training the models.

Similar conclusions hold for `SpeedIndex` in Figure 7.8b. PAIN correlations coefficient span from 0.55 for *www.ilmeteo.it* to 0.86 for *www.gazzetta.it*, with other metrics in similar ranges.

Summarizing, PAIN index is strongly correlated with both objective metrics for different sites. PAIN achieves similar performance than other approaches, which are however hardly feasible in real deployments.

### 7.6.5 Learning duration and periodicity

Next I investigate the number of observation needed to learn support domains, and for how long the models remain valid. This information defines the duration and periodicity of PAIN learning. Since PAIN is unsupervised, it learns models directly from live traffic. Large learning periods should help creating robust models. On the other hand, sites may change over time invalidating the models.

I first evaluate how the size of the learning sample impacts PAIN. I perform experiments with the ADSL dataset. Since I aim at checking how the models behave in large samples and long periods, I focus on the top-100 ranked pages in Italy by Alexa.

In Figure 7.9, I let PAIN learn support domains with an increasing number of observations per core domain. I then compare the selected support domains with the set obtained with the largest observation period – i.e., when all core domains have been observed at least 10,000 times. The *y*-axis reports how similar the two sets are using the Jaccard similarity coefficient [57]. Clearly, the right-most point has value 1 (perfect similarity). Other points confirm that the larger the observation period is, the more stable the sets become. This figure suggests that some thousands observations are required to learn stable sets of support domains. It also shows that shallow models can be learned with few tens of observations, but to get a complete one, much more are needed.

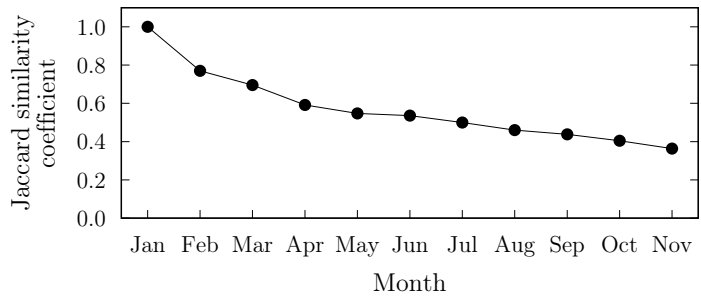

Figure 7.10: Persistence of support domains over the months of a year (ADSL dataset).

A question still remains: How often should PAIN learning be performed? Performing learning sporadically may let models get outdated and reduce the metrics precision. I quantify this phenomenon in Figure 7.10. I let PAIN run on the ADSL dataset using the previous subset of domains, learning support domains separately for each month. Then, I compare the set of support domains learned at each month with those learned during the January 2017. Again, I use the Jaccard coefficient as similarity metric.

The figure shows that support domains learned on February have a 0.77 similarity coefficient with those learned on January. The similarity decreases to 0.69 on March, and finally to 0.36 on November. It is clear that even in short periods, e.g., a couple of months, the learned support domains diverge significantly. While PAIN grouping approach partly compensates for such variations, these results suggest that continuously updating support domains is advisable to retain PAIN performance. In other words, an *updated* model is better than a model trained on *large* data.

In summary, PAIN requires a large number of observations to learn models of support domains for the websites. Few thousands of samples per core domain seem sufficient to bootstrap the system. On the other hand, learning must be continuous, with models being updated to avoid using outdated sets of support domains.

## 7.7   Case studies

I now report my experience when using PAIN in a real deployment. I exploit the ADSL dataset, containing flow-level measurements of around 10,000 ADSL customers over one year. More concretely, I run PAIN to understand (i) whether web browsing performance changes for different ADSL installations; (ii) the impact of large server-side events on users' experience.

PAIN learns the models on the ADSL dataset on a per-month basis. I focus on the top-100 Alexa rank for Italy. PAIN is set with best parameters ($n = 4$, $\Delta T = 30s$).

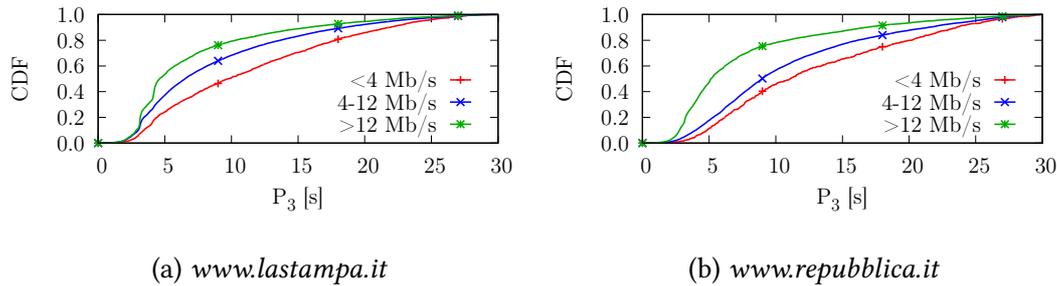

<div align="center">

(a) *www.lastampa.it*         (b) *www.repubblica.it*

</div>

Figure 7.11: Distribution of PAIN $P_3$ index according to the access-link capacity for all visits in ADSL dataset.

### 7.7.1 Performance per ADSL capacity

For ISPs, it is important to understand the impact of access link capacity on web browsing. For example, ISPs are interested in knowing whether users with poor connectivity are significantly impaired while surfing the Web, e.g., to propose upgrades to such users. PAIN allows them to estimate how objective metrics (i.e., OnLoad and SpeedIndex) vary across users, where these are not measurable with passive monitoring.

I know the download access link capacity of each ADSL installation in the ADSL dataset. I thus divide users in three categories: (i) slow ($< 4$ MBit/s), (ii) medium ($4-12$ Mbit/s) and (iii) fast ($> 12$ Mbit/s). I then compute PAIN $P_3$ for users of each group.

Results for two popular news websites in Italy are reported Figure 7.11. For *www.lastampa.it* (Figure 7.11a), distributions are clearly not overlapping. PAIN index decreases significantly when the access capacity increases. Indeed, the median value moves from $9.6s$ for slow users to $4.3s$ for fast users. For *www.repubblica.it* (Figure 7.11b), differences across users are even more pronounced. PAIN index median value is $12.3s$ for slow users and $5.1s$ for fast users.

These results allow quantifying the role of access capacity on page load time in the real world, where previous experiments relied only on testbeds [73, 70].

### 7.7.2 Impairments due to server-side events

ISPs can rely on PAIN index to monitor anomalies causing real impact on users' performance. To this end, I illustrate some noticeable episodes emerging from the ADSL dataset. I let PAIN run on the entire dataset for the top-100 Alexa services. I then manually went through the obtained time series to find episodes worth of attention, such as abrupt changes in PAIN index. Prominent cases have been further investigated, to uncover possible reasons behind the sudden changes.

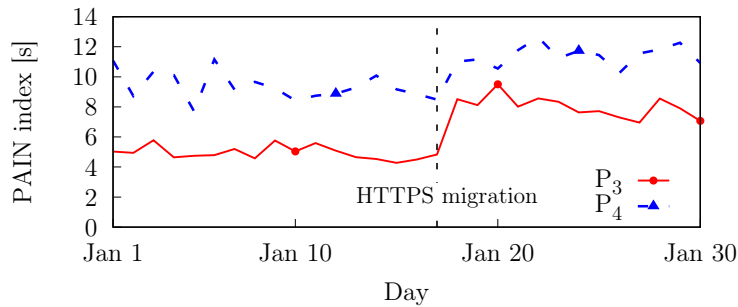

Figure 7.12: PAIN index for *www.poste.it* over 1 month (ADSL dataset).

Figure 7.12 reports an episode related to *www.poste.it*.[9] On January 18[th] 2017, the median PAIN index incurs sudden increase: The median value for $P_3$ grows from the [4,6]$s$ range to the [8,10]$s$ range, while median $P_4$ increases from [8,10]$s$ to [10,12]$s$ ranges (see *y*-axis the figure).

Investigating the root-cause for this change in behavior, I discovered that the website switched all services to HTTPS on that date. As such, the additional load imposed to both servers and clients is likely causing a performance impairment.

Figure 7.13 depicts a second prominent episode uncovered by PAIN, related to *www.re-pubblica.it*. Recall that this site hosts a major Italian news portal. The website passed a major reorganization of layout and content on 27[th] February 2017. The portal claimed at the time that the reorganization would lead to performance improvements for its users.

PAIN is able to measure the page performance before and after the restructuring. Figure 7.13 depicts $P_3$ and $P_4$ evolution in time. The median values computed per day are reported with thin lines, with thick lines marking the exponentially weighted moving average (EWMA) of the values. The performance of the website has improved after the migration day. $P_3$ decreases from $\approx 8s$ to $\approx 6s$, while $P_4$ from $\approx 11s$ to $\approx 9s$.[10]

In summary, these case studies illustrate how PAIN can be used to spot changes in page load performance, due to intrinsic characteristics of the network or external events (e.g., websites modifications). PAIN can be used to trigger alerts in case of sudden changes in performance, driving ISPs to further investigate the problems that are relevant to users' experience.

---

[9]*www.poste.it* is the website of the Italian national mail service.

[10]A one-tailed T-Test confirms that differences for values before and after the migration are statistically significant.

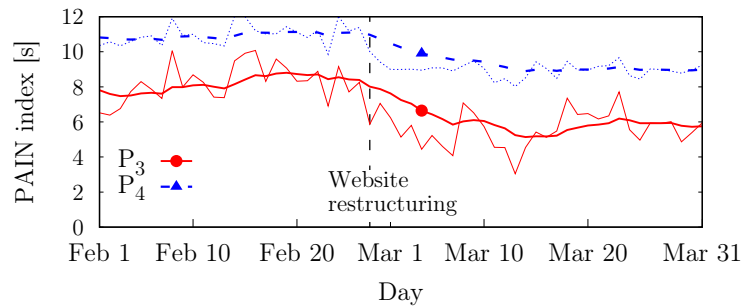

Figure 7.13: PAIN index trend for *www.repubblica.it* before and after website restructuring (ADSL dataset).

## 7.8   Conclusions

In this chapter I presented PAIN, an automatic and unsupervised system to monitor website performance using flow-level measurements, and release it as open source. PAIN builds a behavioral model for the websites' traffic, leveraging flows automatically opened by browsers to retrieve images, scripts etc. The model is used for assessing performance.

I validated PAIN by showing that PAIN metrics are strongly correlated with well-known objective metrics used as indication of users' QoE, i.e., `onLoad` time and `SpeedIn-dex`. I showed that PAIN performance is similar to supervised alternatives, which are however harder to be deployed in practical scenarios. Finally, my results show that PAIN metrics highlight sudden performance deterioration due to changes on network conditions, that may result in degrade web browsing performance.

Finally, I run PAIN on operational network traces for one full year. PAIN allowed me to quantify page load speed differences across customers with different access link capacities. Moreover, PAIN pinpointed sudden performance variations for websites that incurred restructuring.

# Chapter 8

# Conclusions

In my thesis, I presented several works addressing the problem of analyzing network traffic. Helped by big data and machine learning techniques, I provided several analyses of Internet traffic, and proposed innovative algorithms for web service traffic classification. I also faced the problems of per-service traffic management and web browsing quality.

In the first part of my thesis I characterized Internet traffic as seen from a nationwide ISP, and illustrated the trends emerging from 2013 to 2018. Then, I showed the challenges and complexity of understanding Internet traffic in the modern web, where encryption and convergences around few big players make the life of network analysts harder. To overcome the limitations of classical solutions, I proposed novel algorithms to address the problems of traffic classification and management. These approaches make use of machine learning and big data techniques to achieve visibility on the complicated traffic generated by modern web services. The key idea behind these work is to exploit temporal correlation among network events (i.e., network *flows*), mining common patterns from possibly large datasets of real traffic. In the last chapter, I used similar techniques to study the Quality of Experience of users accessing web services. Again, exploiting temporal correlation among events, I built a system that helps the ISP monitor the performance of web browsing of subscribers.

In general, I believe my thesis contains interesting advances in the field of network traffic analysis. The detailed analyses of ISP traffic provide useful insights to for operators, researchers and practitioners. The proposed algorithms are novel attempts to enhance network awareness, in a scenario where the size and complexity of traffic severely challenge its analysis.

Several research directions emerge from the encountered topics, and I am happy to face them in my future life as a researcher. First, the proposed algorithms are well-suited for traffic classification and management, but their accuracy should be further improved to be used in security environments, field in which data science could significantly help where classical approaches based on pure domain knowledge fail to scale or are loosing visibility. Moreover, the trend towards encryption is still modifying what is carried by computer networks. In particular, recent proposals for DNS encryption make the network to loose visibility also on the domain names contacted by clients, while URLs are already hidden by the widespread deployment of HTTPS. If encrypted DNS will be deployed, several monitoring techniques (including some of those presented in this thesis) will become ineffective. Indeed, this will claim for new research threads, in which different and more sophisticated techniques are needed to re-obtain, again, visibility on network traffic.

# Appendix A

# List of Publications

## Journal Publications

1. **Martino Trevisan**, Idilio Drago, Marco Mellia, "*PAIN: A Passive Web Performance Indicator for ISPs*", To appear in Computer Networks, ISSN: 1389-1286, January 2019

2. **Martino Trevisan**, Idilio Drago, "*Robust URL Classification With Generative Adversarial Networks*", ACM Performance Evaluation Review, Volume 46 (3), ISSN: 0163-5999, December 2018

3. **Martino Trevisan**, Stefano Traverso, Eleonora Bassi, Marco Mellia, "*4 Years of EU Cookie Law: Results and Lessons Learned*", To appear in Proceedings on Privacy Enhancing Technologies, ISSN: 2299-0984, April 2019

4. **Martino Trevisan**, Idilio Drago, Marco Mellia, Han Hee Song, Mario Baldi, "*AWESoME: Big Data for Automatic Web Service Management in SDN*", IEEE Transactions on Network and Service Management, pp.1-14, ISSN: 1932-4537, March 2018

5. Giuseppe Siracusano, Roberto Bifulco, **Martino Trevisan**, Tobias Jacobs, Simon Kuenzer, Stefano Salsano, Nicola Blefari-Melazzi, Felipe Huici, "*Re-designing Dynamic Content Delivery in the Light of a Virtualized Infrastructure*", IEEE Journal on Selected Areas in Communications, Vol.35, ISSN: 0733-8716, November 2017

6. **Martino Trevisan**, Alessandro Finamore, Marco Mellia, Maurizio Munafò, Dario Rossi, "*Traffic Analysis with Off-the-Shelf Hardware: Challenges and Lessons Learned*", IEEE Communications Magazine - Network Testing and Analytics Series, March 2017

# Conference and Workshop Publications

1. Andrea Morichetta, **Martino Trevisan**, Luca Vassio "*Characterizing Web Pornography Consumption From Passive Measurements*", to appear in the 2019 Passive and Active Measurement Conference (PAM 2019), Puerto Varas (Chile), March 27-29, 2019

2. Azadeh Faroughi, Reza Javidan, Marco Mella, Andrea Morichetta, Francesca Soro, **Martino Trevisan**, "*Achieving Horizontal Scalability in Density-based Clustering for URLs*", 2nd Workshop on Benchmarking, Performance Tuning and Optimization for Big Data Applications (BPOD) 2018, Seattle (USA), December 10, 2018

3. **Martino Trevisan**, Danilo Giordano, Idilio Drago, Marco Mella, Maurizio Munafò, "*Five Years at the Edge: Watching Internet from the ISP Network*", 14th International Conference on emerging Networking EXperiments and Technologies (CoNEXT 2018), Heraklion (Grece), December 5-7, 2018

4. **Martino Trevisan**, Idilio Drago, Marco Mella, "*Measuring Web Speed From Passive Traces*", ACM, IRTF & ISOC Applied Networking Research Workshop 2018 (ANRW 18), Montreal (Canada), July 16, 2018

5. **Martino Trevisan**, Idilio Drago, Marco Mella, Maurizio Munafò, "*Automatic Detection of DNS Manipulations*", IEEE Workshop Data Science for Networking (DS4N), Boston (USA), 14 December 2017

6. Ali Safari Khatouni, **Martino Trevisan**, Leonardo Regano, Alessio Viticchie, "*Privacy Issues of ISPs in the Modern Web*", IEEE IEMCON 2017, Vancouver (Canada), October 3-5

7. Luca Vassio, Danilo Giordano, **Martino Trevisan**, Marco Mella, Ana Paula Couto da Silva, "*Users' Fingerprinting Techniques from TCP Traffic*", ACM SIGCOMM Workshop on Big Data Analytics and Machine Learning for Data Communication Networks, Los Angeles (USA), August 2017

8. **Martino Trevisan**, Idilio Drago, Marco Mella, "*PAIN: A Passive Web Speed Indicator for ISPs*", ACM SIGCOMM Workshop on QoE-based Analysis and Management of Data Communication Networks, Los Angeles (USA), August 2017

9. Stefano Traverso, **Martino Trevisan**, Leonardo Giannantoni, Marco Mella, Hassan Metwalley, "*Benchmark and Comparison of Tracker-blockers: Should You Trust Them?*", Traffic Monitoring and Analysis workshop (TMA), Dublin (Ireland), 21-22 June 2017

10. **Martino Trevisan**, Idilio Drago, Marco Mella, Han Hee Song, Mario Baldi, "*WHAT: A Big Data Approach for Accounting of Modern Web Services*", IEEE Workshop on

Big Data and Machine Learning in Telecom (BMLIT), Washington DC (USA), December 2016

11. **Martino Trevisan**, Idilio Drago, Marco Mellia, Maurizio Munafo', "*Towards Web Service Classification using Addresses and DNS*", 7th International Workshop on TRaffic Analysis and Characterization, TRAC2016, Paphos (Cyprus), September, 2016

12. **Martino Trevisan**, Idilio Drago, Marco Mellia, "*Impact of Access Line Capacity on Adaptive Video Streaming Quality - A Passive Perspective*", ACM SIGCOMM Workshop on QoE-based Analysis and Management of Data Communication Networks, Florianopolis (Brasil), August 22 2016